\documentclass[twocolumn, prl, showpacs, superscriptaddress]{revtex4-1}

\usepackage{graphicx}
\usepackage{amsmath}
\usepackage{hyperref}
\usepackage{dcolumn}

\newcommand{\WFU}{Department of Physics, Wake Forest University,
Winston-Salem, NC 27106, USA.}
\newcommand{\MCtwo}{Microtechnology and Nanoscience, MC2, Chalmers
University of Technology, SE-412 96 G{\"o}teborg, Sweden.}
\newcommand{\Oslo}{Centre for Materials Science and Nanotechnology
(SMN), University of Oslo, 0316 Oslo, Norway.}

\newcommand{\rr}{{\bf r}}
\newcommand{\qq}{{\bf q}}
\newcommand{\nup}{\ensuremath{n_\uparrow}}

\newcommand{\ndown}{\ensuremath{n_\downarrow}}

\newcommand{\nl}{\text{nl}}

\begin{document}

\title{Spin signature of nonlocal-correlation binding in metal organic
frameworks}

\author{T. Thonhauser} \email{thonhauser@wfu.edu} \affiliation{\WFU}
\author{S. Zuluaga}                               \affiliation{\WFU}
\author{C. A. Arter}                              \affiliation{\WFU}
\author{K. Berland}           \affiliation{\MCtwo}\affiliation{\Oslo}
\author{E. Schr\"oder}                            \affiliation{\MCtwo}
\author{P. Hyldgaard}                             \affiliation{\MCtwo}

\date{\today}

\begin{abstract}
We develop a proper nonempirical spin-density formalism for the van der
Waals density functional (vdW-DF) method. We show that this
generalization, termed svdW-DF, is firmly rooted in the single-particle
nature of exchange and we test it on a range of spin systems.  We
investigate in detail the role of spin in the nonlocal-correlation
driven adsorption of H$_2$ and CO$_2$ in the linear magnets Mn-MOF74,
Fe-MOF74, Co-MOF74, and Ni-MOF74. In all cases, we find that spin plays
a significant role during the adsorption process despite the general
weakness of the molecular-magnetic responses. The case of CO$_2$
adsorption in Ni-MOF74 is particularly interesting, as the inclusion of
spin effects results in an increased attraction, opposite to what the
diamagnetic nature of CO$_2$ would suggest. We explain this
counter-intuitive result, tracking the behavior to a coincidental
hybridization of the O $p$ states with the Ni $d$ states in the
down-spin channel. More generally, by providing insight on nonlocal
correlation in concert with spin effects, our nonempirical svdW-DF
method opens the door for a deeper understanding of weak nonlocal
magnetic interactions.
\end{abstract}

\pacs{71.15.Mb, 31.15.ej, 81.05.Rm}
\maketitle

The modular building-block nature of metal organic frameworks (MOFs) and
their extraordinary affinity for adsorption of small molecules make
these nano-porous materials ideal for technologically important
applications. MOFs are used, for example, for gas storage and
sequestration \cite{Murray_2009:hydrogen_storage,
Li_2011:carbon_dioxide, Qiu_2009:molecular_engineering,
Nijem_2012:tuning_gate, Lee_2015:small-molecule_adsorption}, catalysis
\cite{Lee_2009:metal_organic, Luz_2010:bridging_homogeneous},
polymerization \cite{Uemura_2009:polymerization_reactions,
Vitorino_2009:lanthanide_metal}, luminescence
\cite{Allendorf_2009:luminescent_metal,
White_2009:near-infrared_luminescent}, non-linear optics
\cite{Bordiga_2004:electronic_vibrational}, magnetic networks
\cite{Kurmoo_2009:magnetic_metal-organic}, targeted drug delivery
\cite{Horcajada_2010:porous_metal-organic-framework}, multiferroics
\cite{Stroppa_2011:electric_control, Stroppa_2013:hybrid_improper,
Di-Sante_2013:tuning_ferroelectric}, and sensing
\cite{Serre_2007:role_solvent-host,
Allendorf_2008:stress-induced_chemical, Tan_2011:mechanical_properties,
Kreno_2012:metal-organic_framework}. The design of novel MOFs with
improved properties requires insight into the molecule/MOF interaction.
The large unit cells and periodic nature of MOFs make density functional
theory (DFT) the prospective tool for a theory exploration. However,
both the adsorbate molecule and the MOF's metal centers can carry spin,
giving rise to complex magnetic interactions and a molecular-spin
response. It is thus crucial that DFT can reliably capture van der Waals
(vdW) forces---which govern adsorption in MOFs---in concert with spin
effects.

Concerning the former, the last decade witnessed the development of DFT
descriptions for these forces \cite{Berland_2015:van_waals}. Here, the
vdW-DF versions \cite{Dion_2004:van_waals, Thonhauser_2007:van_waals,
Lee_2010:higher-accuracy_van, Berland_2014:exchange_functional} stand
out by being nonempirical exchange-correlation functionals that are
systematic and truly nonlocal extensions beyond LDA
\cite{Perdew_1992:accurate_simple} and GGA
\cite{Perdew_1996:generalized_gradient} in the electron-gas tradition
\cite{Berland_2015:van_waals, Berland_2014:van_waals,
Hyldgaard_2014:interpretation_van}. Subsequent developments include
variants which differ by their choice of the semi-local exchange
\cite{Cooper_2010:van_waals, Klimes_2010:chemical_accuracy,
Klimes_2011:van_waals, Wellendorff_2012:density_functionals,
Hamada_2014:van_waals} and related nonlocal correlation functionals that
rely on optimizing parameters
\cite{Vydrov_2008:self-consistent_implementation,
Vydrov_2009:nonlocal_van, Vydrov_2010:nonlocal_van}. The vdW-DF method
and relatives have been successfully applied to numerous materials in
general \cite{Langreth_2009:density_functional, Berland_2014:van_waals,
Berland_2015:van_waals}, and to small-molecule adsorption in MOFs in
particular \cite{Poloni_2014:understanding_trends,
Lee_2014:design_metal-organic, Lee_2015:small-molecule_adsorption,
Canepa_2013:diffusion_small, Nijem_2012:tuning_gate,
Canepa_2013:high-throughput_screening, Nijem_2013:water_cluster,
Tan_2014:water_reaction, Zuluaga_2014:study_van}.

Concerning the spin effects, however, a systematic description within
the vdW-DF framework is still missing. Such effects can play important
roles not only in MOFs, but in many systems, as Hund's rules reflect a
preference for spin-polarized ground states. For example, spin and vdW
effects are essential in organic spintronics
\cite{Dediu_2009:spin_routes}, dimer binding in excited states
\cite{Musial_2014:first_principle}, overlayer formation on magnetic
substrates \cite{Sipahi_2014:spin_polarization}, and correctly assessing
formation energies~\cite{Gunnarsson_1974:contribution_cohesive}. While
the nonlocal functional VV09 considers spin in its own way \cite{Troy,
real_space_code}, there have so far only been pragmatic approaches for
vdW-DF---ignoring the effect of spin on the nonlocal correlation
altogether~\cite{balancing} or estimating the effect
\cite{Ziambaras_2007:potassium_intercalation,
Obata_2013:implementation_van, Obata_2015:improving_description} using
the semi-local correlation of PBE
\cite{Perdew_1996:generalized_gradient}.

In this letter, we formulate a proper extension of vdW-DF to
spin-polarized systems, termed svdW-DF, following the design-logic of
the original functional. We apply svdW-DF to study the
nonlocal-correlation driven adsorption of H$_2$ and CO$_2$ in MOF74 and
find that spin plays a significant role, providing a detailed analysis
of spin signatures in such vdW bonding.  Beyond MOFs, we envision that
svdW-DF will lead to wider materials-theory progress in a stimulating
role like that of LSDA, i.e., LDA's spin extension
\cite{Perdew_1992:accurate_simple}. LSDA was introduced to describe
bulk-cohesive and molecular-binding energies
\cite{Gunnarsson_1974:contribution_cohesive,
Gunnarsson_1976:exchange_correlation,
Barth_1972:local_exchange-correlation} but also led DFT to important
successes in the study of magnetism
\cite{Jones_1989:density_functional}. The svdW-DF formulation enables a
robust exploration of systems where spin and nonlocal correlations are
both important and it makes vdW-DF a general purpose method
\cite{general_purpose}.

To design svdW-DF as the natural extension of vdW-DF to spin-polarized
systems, we revisit the derivation of its nonlocal correlation energy
functional. The starting point is the adiabatic-connection formula (ACF)
expressed in terms of a scalar dielectric function $\epsilon$ that
reflects a formal average of the coupling-constant integration over the
screened density-response function
\cite{Langreth_1977:exchange-correlation_energy,
Gunnarsson_1976:exchange_correlation, Rydberg_2003:van_waals,
Dion_2004:van_waals}. This provides a split-up of the total
exchange-correlation energy into a nonlocal and semi-local piece
$E_{xc}^\text{vdW-DF}=E_c^\text{nl}+E_{xc}^\text{int}$, defined by
\cite{Dion_2004:van_waals, Lee_2010:higher-accuracy_van,
Hyldgaard_2014:interpretation_van, Berland_2015:van_waals} in terms of
the Coulomb Green's function $G$ and an integral over the imaginary
frequency $u$:
\begin{eqnarray}
E_c^{\nl}[n]        &=& \int_0^\infty\frac{du}{2\pi}\;\text{Tr}\,
                        \big[\ln\big(\nabla\epsilon \cdot \nabla G\big) -
                        \ln\epsilon\big] \label{equ:ExcG}\\
E^{\rm int}_{xc}[n] &=& \int_0^\infty\frac{du}{2\pi}\;\text{Tr}\,
                        \ln\epsilon -E_{\rm self}\label{equ:Ex0}\;.
\end{eqnarray}
The general-geometry vdW-DF versions \cite{Dion_2004:van_waals,
Lee_2010:higher-accuracy_van,Berland_2014:exchange_functional} expand
the nonlocal correlation energy (\ref{equ:ExcG}) in terms of a
semi-local response function $S\equiv\ln \epsilon$ that is parameterized
via the choice of internal semi-local (GGA-type) functional $E_{xc}^{\rm
int}$ (\ref{equ:Ex0}). 

To obtain a computationally tractable approximation for $E_c^{\nl}$
\cite{Dion_2004:van_waals, Lee_2010:higher-accuracy_van,
Berland_2015:van_waals}, vdW-DF relies on a plasmon-pole approximation
of $S$ defined in plane-wave representation as
$S_{\qq,\qq'}=\frac{1}{2}\bar{S}_{\qq,\qq'}+
\frac{1}{2}\bar{S}_{-\qq',-\qq}$, with
\begin{equation}
\bar{S}_{\qq,\qq'} = \int d\rr\;e^{-i(\qq-\qq')\cdot\rr}
\frac{4\pi e^2\,n(\rr)/m}{[\omega_q(\rr)+\omega][\omega_{q'}(\rr)-\omega]}
\label{equ:S}\;.
\end{equation}
Here, $n(\rr)$ is the total electron density and $\omega_q(\rr)$ is the
effective local plasmon dispersion, parameterized by an effective
response parameter in the form of an inverse length scale
$q_0(\rr)=q_0[n]= q_0\big(n(\rr),\nabla n(\rr)\big)$; $m$ and $e$ are
the electronic mass and charge. The link between the energy-per-particle
of the internal functional $\varepsilon_{xc}^{\rm int}(\rr)$ and
$\omega_q(\rr)$ \cite{Dion_2004:van_waals, Berland_2015:van_waals,
Hyldgaard_2014:interpretation_van} follows from combining
Eqs.~(\ref{equ:Ex0})  and~(\ref{equ:S}) together with a plasmon
dispersion $\omega_{q}(\rr)= q^2/2h\big(q/q_0(\rr)\big)$ with a Gaussian
shape of $h(x)=1-\exp(-\gamma x^2)$, where  $\gamma$ is an arbitrary
constant set to $4\pi/9$.  Expanding Eq.~(\ref{equ:ExcG}) to second
order in $S$, one arrives at the well-known six-dimensional integral
over a universal kernel $\Phi_0(a,b)$,
\begin{equation}
\label{equ:functional}
E_c^\nl = \frac{1}{2}\int\!d\rr\,d\rr'\,\,
n(\rr)\;\Phi_0\big(q_0(\rr)|\rr-\rr'|,q_0(\rr')|\rr-\rr'|\big)\;n(\rr')\;,
\end{equation}
which defines the approximation for $E_c^{\nl}$
\cite{Dion_2004:van_waals}. The total exchange-correlation energy
$E_{xc}^\text{vdW-DF}$ also consists of the semi-local functional
$E_{xc}^\text{int}$ (\ref{equ:Ex0}).  This is in practice approximated
as $E_{xc}^\text{int}\approx E_{xc}^0 = E^{\text{GGA}}_x +
E^{\text{LDA}}_c $, based on a number of criteria
\cite{Dion_2004:van_waals, Lee_2010:higher-accuracy_van,
Klimes_2010:chemical_accuracy, Cooper_2010:van_waals,
Berland_2014:exchange_functional, Hamada_2014:van_waals} and differing
from the internal functional $E_{xc}^\text{int}$ to varying degrees.

The extension of the semi-local part $E_{xc}^0$ to spin-polarized
systems is straightforward. It is given by the exact spin scaling of
exchange~\cite{Perdew_1986:accurate_simple}, i.e.\ $E_x[\nup,\ndown]
=E_x[2\nup]/2+E_x[2\ndown]/2$, and the well-established spin-dependence
of the local correlation~\cite{Perdew_1992:accurate_simple}.  Here \nup\
and \ndown\ denote the spin-density components. Crucially, by applying
the very same criteria, we obtain a fully consistent extension of
$E_c^\nl$ for the spin case. 

The spin-scaling of exchange results in a spin-dependent semi-local
response $S$ in Eq.~(\ref{equ:S}), with spin entering exclusively in the
denominator through $\omega_q(\rr)$. The numerator is given by the
$f$-sum rule, specified as the classical plasmon frequency which depends
only on the total electron density $n(\rr)$.  The formulation of svdW-DF
can therefore be based on an universal-kernel evaluation using the exact
same function $\Phi_0(a,b)$ as in vdW-DF.  Nevertheless, the form of the
effective response parameter $q_0$---which acts as scaling parameter in
the arguments of $\Phi_0$---must be adjusted, $q_0[n]\rightarrow
\tilde{q}_0[\nup,\ndown]$, to reflect the explicit spin dependence of
the plasmon dispersion. 

Motivating our procedure for extending the original vdW-DF formulations
to spin-polarized system is the interpretation of the vdW-DF nonlocal
correlation energy as a formal summation of zero-point energy shifts
\cite{Mahan_1965:van_waals, Rapcewicz_1991:fluctuation_attraction,
Hyldgaard_2014:interpretation_van}.  The vdW-DF framework starts with a
description of the semilocal exchange-correlation holes corresponding to
the internal functional $E_{xc}^\text{int}$
\cite{Lee_2010:higher-accuracy_van, Berland_2014:van_waals,
Berland_2015:van_waals}, using a plasmon model to characterize the
associated response. The vdW-DF nonlocal correlation energy
(\ref{equ:ExcG}) is a rigorous summation of the plasmon-pole shifts that
result when such holes couple
electrodynamically~\cite{Hyldgaard_2014:interpretation_van}.  Spin
clearly affects the GGA-type internal hole and our svdW-DF formalism
represents a proper implementation of how such semilocal spin effects
impact the summation of zero-point energy shifts in Eq.
(\ref{equ:ExcG}).

To establish the updated form of $\tilde{q}_0[\nup,\ndown]$ it is
instructive to first revisit how $q_0[n]$ is specified in the
spin-neutral case, where it is given as scaling of the Fermi wave vector
$k_F(\rr)=(3\pi^2n)^{1/3}$ as follows
\begin{eqnarray}
q_0(\rr)  &=& \frac{\varepsilon_{xc}^{\rm int}(\rr)}{\varepsilon_x^{\text{LDA}}
              (\rr)}\,k_F(\rr) \equiv q_{0c}[n] + q_{0x}[n]\;,\label{equ:q_simple}\\
q_{0c}[n] &=& -\frac{4\pi}{3e^2}\,\varepsilon_c^\text{LDA}\;,\label{equ:q0c}\\
q_{0x}[n] &=& -\Big(1-\frac{Z_{ab}}{9}s^2\Big)\,
              \frac{4\pi}{3e^2}\varepsilon_x^{\text{LDA}}\;.\label{equ:q0x}
\end{eqnarray}
Here $\varepsilon_x^{\text{LDA}}=-3e^2k_F/4\pi$ and the exchange
gradient corrections are expressed in terms of the scaled gradient
$s=|\nabla n|/2k_Fn$.  These relations
(\ref{equ:q_simple})--(\ref{equ:q0x}) can be directly adapted to the
spin case. The correlation part $\tilde{q}_{0c}[\nup,\ndown]$ is
specified by the spin-dependent PW92 LDA correlation energy-per-particle
$\varepsilon_c^\text{LDA}$ \cite{Perdew_1992:accurate_simple}. For the
exchange part $\tilde{q}_{0x}[\nup,\ndown]$, the spin scaling relation
\cite{Perdew_1986:accurate_simple} gives the following form
\begin{equation}
\label{equ:q_spin_scaling}
\tilde{q}_{0x}[\nup,\ndown] = \frac{\nup  }{\nup+\ndown}q_{0x}[2\nup] +
\frac{\ndown}{\nup+\ndown}q_{0x}[2\ndown]\;.
\end{equation}
These equations fully specify the nonlocal correlation energy of
svdW-DF. We make svdW-DF self-consistent, implemented in \textsc{Quantum
Espresso} \cite{Giannozzi_2009:quantum_espresso}, by computing the
corresponding exchange-correlation potential
\cite{Thonhauser_2007:van_waals}. Further details on the implementation
and calculations are provided in the Supplemental Material.

\begin{table}[b]
\caption{\label{energy_tab} Binding energies [meV] of small molecules in
the system $\mathcal{M}$-MOF74+$\mathcal{A}$ with $\mathcal{M}$ = Mn,
Fe, Co, and Ni and $\mathcal{A}$ = H$_2$ and CO$_2$. In number triplets
the first number refers to the bare binding energy $\Delta E$, the
second one includes the zero-point correction $\Delta E_\text{ZPE}$, and
the third refers to the binding enthalpy at room temperature $\Delta
H_{298}$.}
\begin{tabular*}{\columnwidth}{@{}l@{\extracolsep{\fill}}lccr@{}}
\hline\hline
$\mathcal{M}$ & $\mathcal{A}$ & Exp. & no spin     & \multicolumn{1}{c}{spin}\\\hline
Mn & H$_2$  &  91 \cite{Zhou_2008:enhanced_h2}                  & $-473$/$-524$/$-524$  & $-133$/$-117$/$-117$ \\
Fe & H$_2$  & 104 \cite{Marcz_2012:iron_member}                 & $-134$/$-137$/$-137$  & $-122$/$-124$/$-124$ \\
Co & H$_2$  & 111 \cite{Zhou_2008:enhanced_h2}                  & $-177$/$-181$/$-181$  & $-111$/$-117$/$-117$ \\
Ni & H$_2$  & 134 \cite{Zhou_2008:enhanced_h2}                  & $-134$/$-137$/$-137$  & $-131$/$-133$/$-133$ \\
Mn & CO$_2$ & 331 \cite{Yu_2013:combined_experimental}          & $-528$/$-551$/$-550$  & $-337$/$-345$/$-344$ \\
Fe & CO$_2$ & 352 \cite{Marcz_2012:iron_member}                 & $-344$/$-353$/$-351$  & $-315$/$-323$/$-321$ \\
Co & CO$_2$ & 383 \cite{Caskey_2008:dramatic_tuning}            & $-377$/$-387$/$-385$  & $-350$/$-359$/$-357$ \\
Ni & CO$_2$ & 394 \cite{Dietzel_2009:application_metal-organic} & $-300$/$-311$/$-309$  & $-377$/$-390$/$-387$ \\\hline\hline
\end{tabular*}
\end{table}

We test svdW-DF on three cases of increasing complexity; results are
summarized here and details are in the Supplemental Material. For our
test cases we use the implementations svdW-DF1
\cite{Dion_2004:van_waals} and svdW-DF2
\cite{Lee_2010:higher-accuracy_van} (which are better suited for small
molecules) as well as svdW-DF-cx \cite{Berland_2014:exchange_functional}
(which is better suited for larger, extended systems). We start with the
Li-dimer in its triplet state $^3\Sigma$---an ideal test case that
critically balances vdW and spin effects. We find a dissociation energy
of 53~meV for svdW-DF1 and 70~meV for svdW-DF-cx, the former in good
agreement with the experimental value of 41~meV
\cite{Linton_1999:high-lying_vibrational}; VV10 and PBE find 77~meV.  A
second case is given by atomization energies for molecules from the G1
set \cite{Pople_1989:gaussian-1_theory}, where spin enters through
magnetic molecular ground states and the isolated atoms. We find a mean
absolute percentage error of 4.59\%  and 7.75\% for svdW-DF1 and
svdW-DF-cx; VV10 and PBE find 5.14\% and 7.11\%, respectively.  For a
third, extended-system test, we study the weak-chemisorption of graphene
on Ni(111) \cite{Gamo_1997:atomic_structure,
Varykhalov_2008:electronic_magnetic, Dedkov_2010:electronic_magnetic},
finding a binding separation for svdW-DF-cx of 2.12~\AA, in excellent
agreement with experiment (2.11 $\pm$ 0.07~\AA\
\cite{Gamo_1997:atomic_structure}). In contrast, svdW-DF1 finds
3.76~\AA, VV10 finds 3.37~\AA, and PBE essentially does not
bind---unlike svdW-DF-cx they all miss a significant chemical component
to the binding.

Table~\ref{energy_tab} summarizes the main point of this letter: that
svdW-DF provides insight on the nature of nonlocal spin effects in the
adsorption of H$_2$ and CO$_2$ in the linear magnets Mn-MOF74, Fe-MOF74,
Co-MOF74, and Ni-MOF74 \cite{Canepa_2013:when_metal}.  The table reports
raw svdW-DF binding energies $\Delta E$, as well as values $\Delta
E_\text{ZPE}$ corrected for zero-point vibrations of the adsorbates and
binding enthalpies at room temperature $\Delta H_{298}$.  We note that
the adhesion comes entirely from $E_c^\text{nl}$---without nonlocal
correlations, CO$_2$ would not bind at all and H$_2$ would only bind
with a binding energy of $\sim$5~meV.

Overall, we find very good agreement with experiment, which we partly
attribute to the ``cx'' version of svdW-DF; agreement with other vdW-DF
calculations is also good \cite{Lee_2015:small-molecule_adsorption}.  In
all cases we find that the inclusion of spin has an important effect on
the binding. Note that the case of Mn is somewhat artificial and the
``no spin'' numbers seem inflated---this is a result of the fact that
Mn-MOF74 itself requires spin for a proper description of its structure.

\begin{figure}
\hspace*{\fill}\includegraphics[width=0.45\columnwidth]{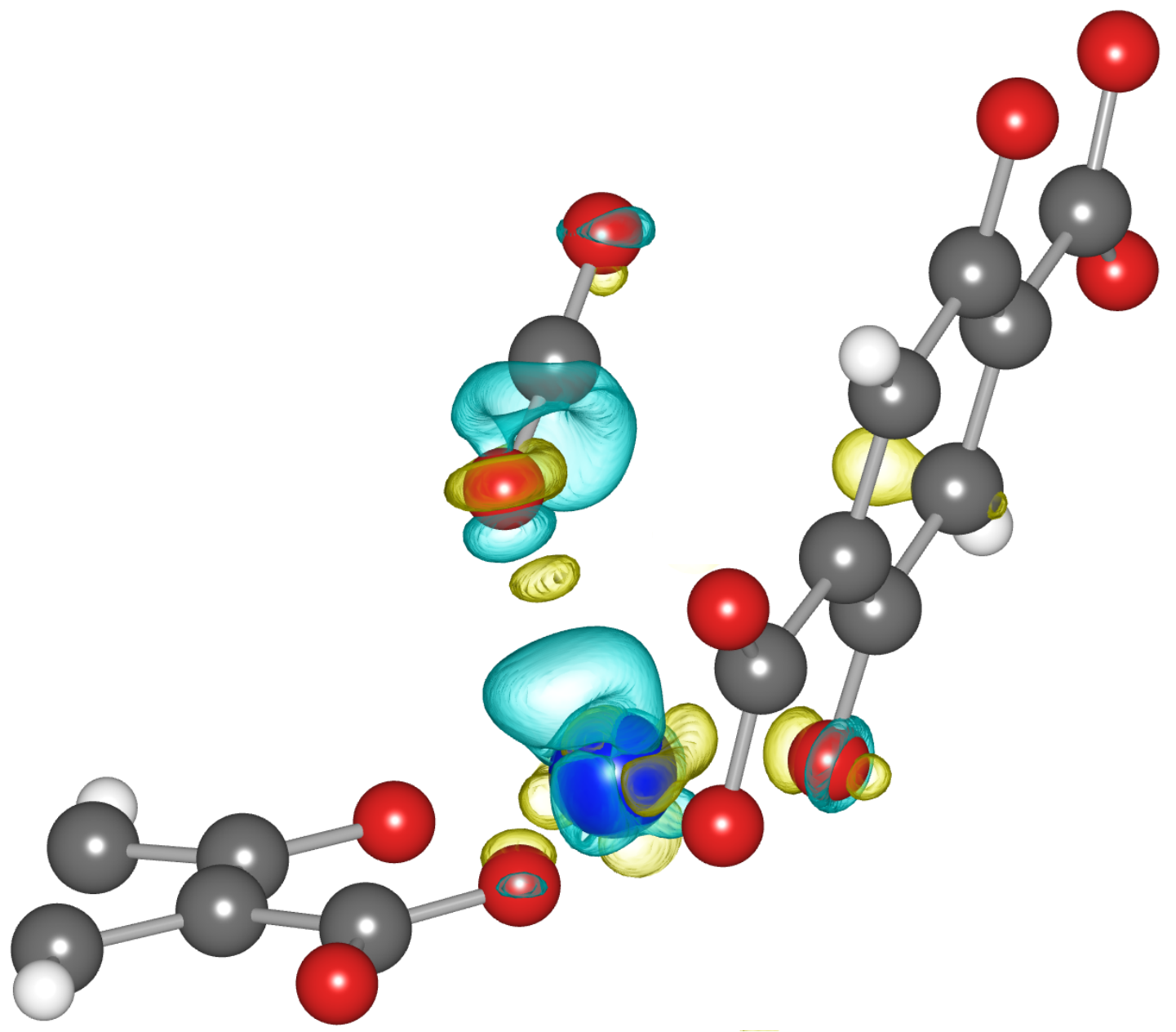}\hfill
\includegraphics[width=0.45\columnwidth]{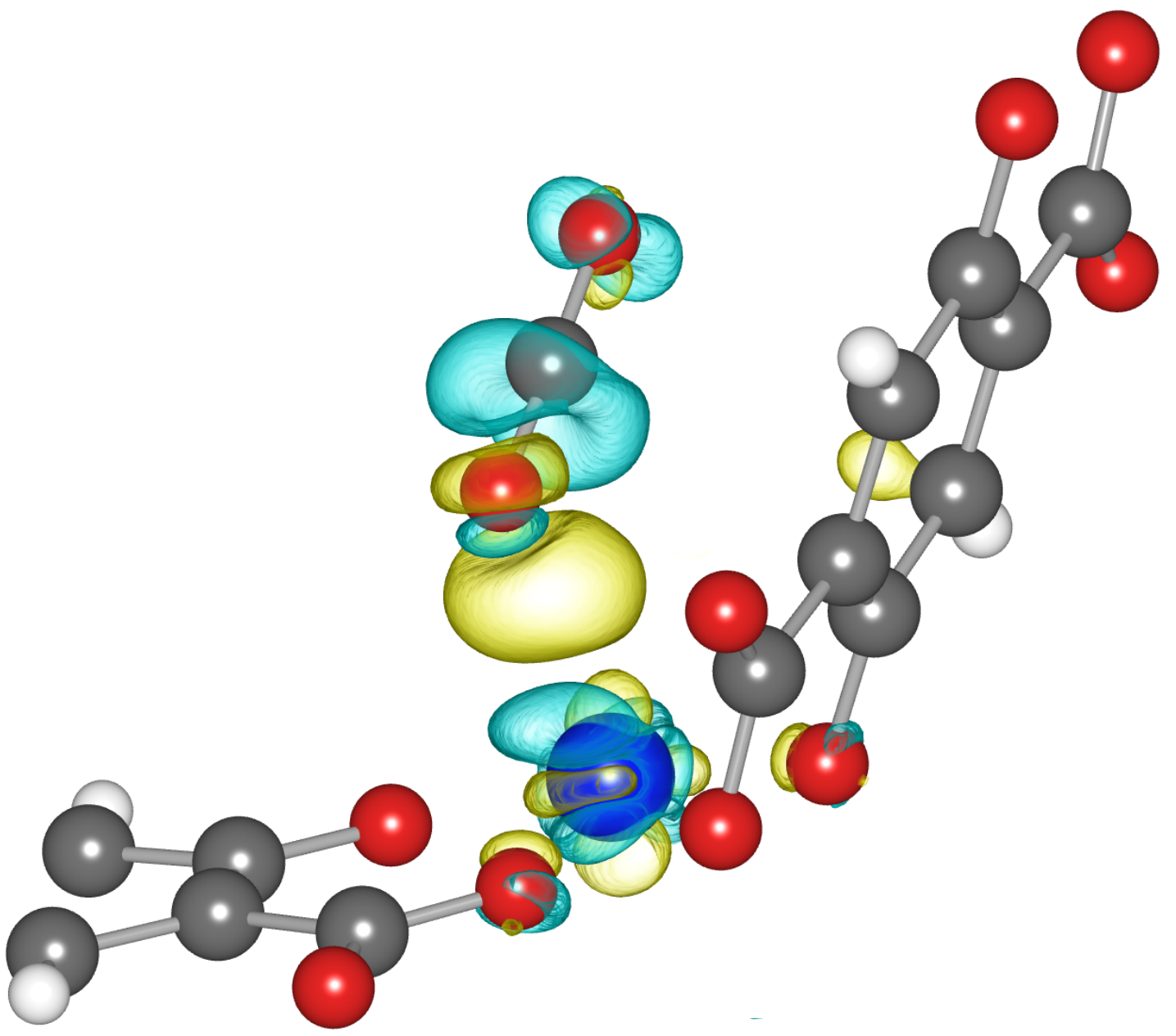}\hspace*{\fill}\mbox{}\\
\hspace*{\fill}no spin\hfill\hfill spin\hfill\mbox{}\\
\hspace*{\fill}\includegraphics[width=0.45\columnwidth]{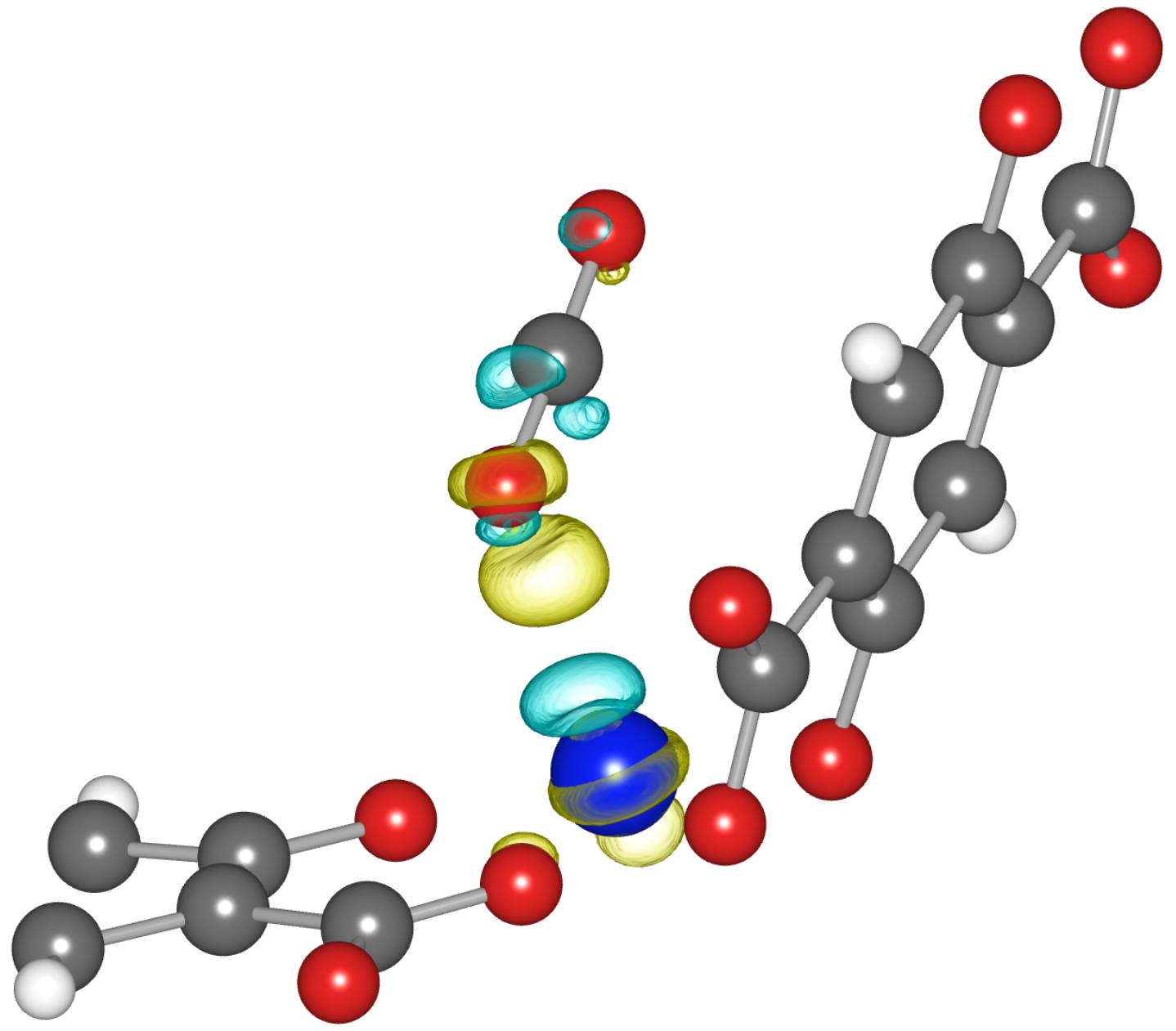}\hfill
\includegraphics[width=0.45\columnwidth]{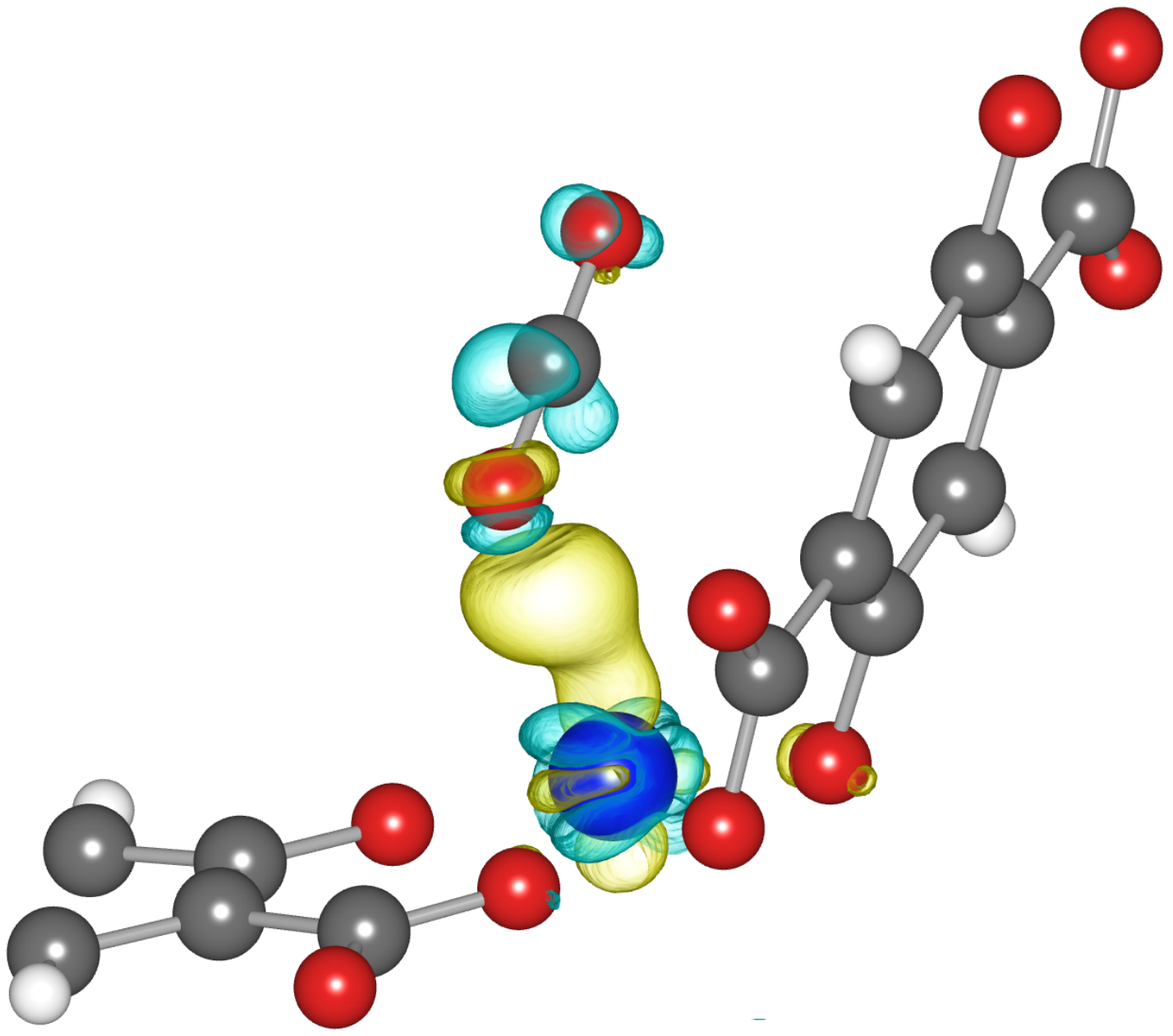}\hspace*{\fill}\mbox{}\\
\hspace*{\fill}spin up\hfill\hfill spin down\hfill\mbox{}
\caption{\label{fig:induced_density} (Upper panels) Induced charge
density upon CO$_2$ adsorption in Ni-MOF74. (Lower panels) Induced
charge density split into its up and down contribution. Blue (yellow)
areas show charge depletion (accumulation). Iso levels are 0.001
$e$/Bohr$^{3}$. See Fig.~\ref{fig:MOF} in the Supplemental Material for the
structure of the MOF.}
\end{figure}

\begin{figure}
\includegraphics[width=\columnwidth]{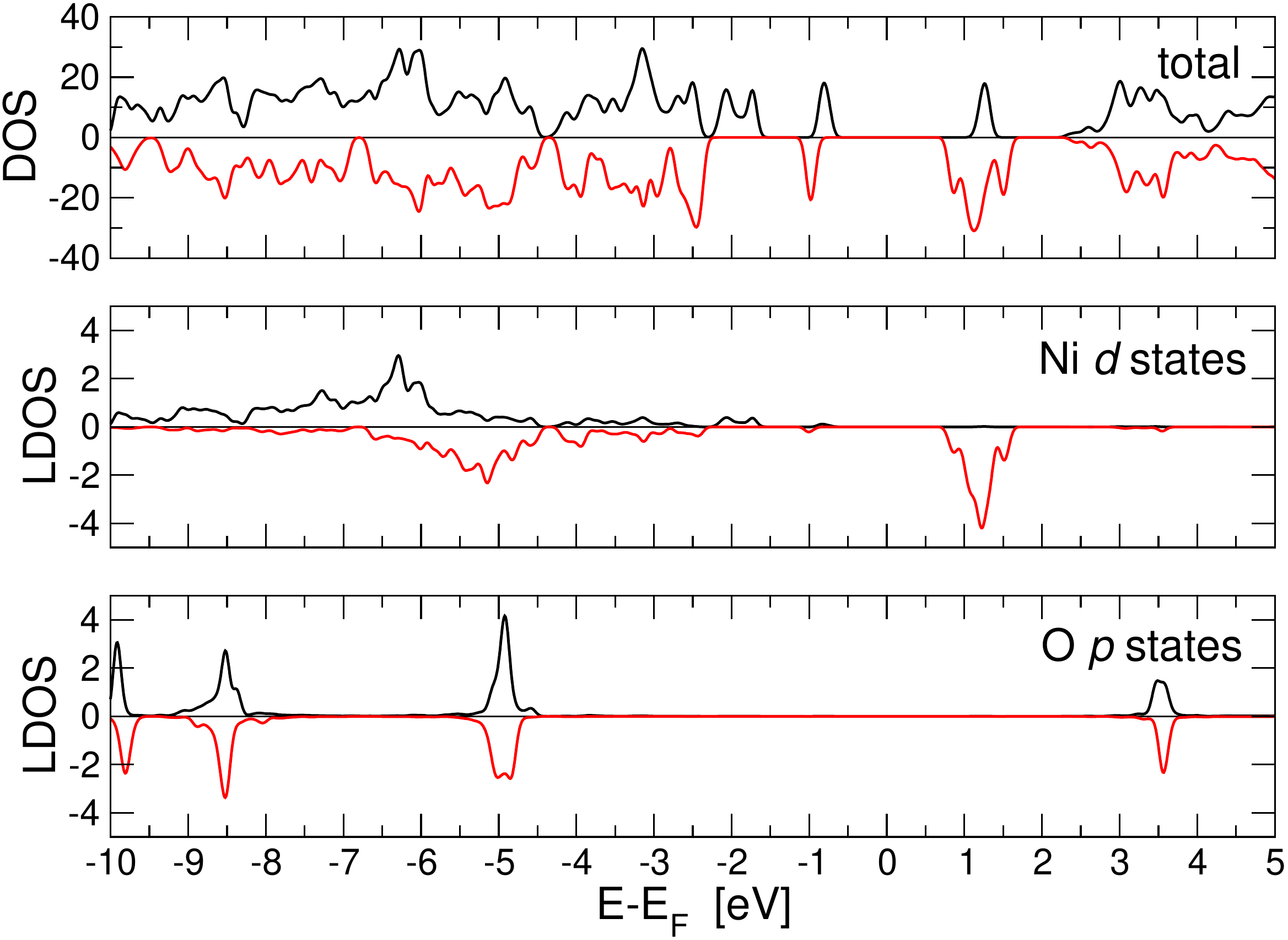}
\caption{\label{fig:DOS}Up (black) and down (red) total density of
states (top) and projected density of states on the Ni $d$ states
(middle) and O $p$ states (bottom) of the Ni-MOF74+CO$_2$ system.}
\end{figure}

Of particular interest is the binding of CO$_2$ in Ni-MOF74, where spin
effects play a tantalizing and unexpected role. The CO$_2$ molecule is
diamagnetic and should experience a slight repulsion and weaker binding
in the presence of the magnetic dipole of Ni---similarly to what is
observed in all the other cases in Table~\ref{energy_tab}. However, on
the contrary, when spin effects are included the binding increases and
the molecule experiences a stronger attraction, which warrants further
investigation. In the upper panels of Fig.~\ref{fig:induced_density} we
plot the induced charge density, i.e., the charge density redistribution
due to the formation of the bond. It is clearly visible that in the spin
case more charge is pulled in-between the CO$_2$ and the metal site,
resulting in the stronger binding. In the spin case, we can split this
induced charge density further into spin-up and spin-down contributions,
as shown in the lower panels of Fig.~\ref{fig:induced_density}. Here we
see the true spin effect: much more spin-down density is being pulled
into the bond, compared to spin-up density. This peculiar behavior can
be understood by analyzing the projected density of states in
Fig.~\ref{fig:DOS}. In particular, from the middle and bottom panel we
see that at $-5$~eV the O $p$ states show similar peaks in the spin up
and down channels. However, the projected Ni $d$ states at that point
have a large spin down density while the spin up density is much
smaller. Thus, the O {\it p} states hybridize with the Ni spin-down {\it
d} states, while the hybridization with the spin-up states is negligible
(see Fig.~\ref{charge_band} in the Supplemental Material for plots of the
corresponding orbitals). The interaction of the O $p$ states with the
down-spin Ni $d$ states is therefore responsible for the increased and
counter-intuitive strength of the bond.

\begin{figure}[h]
\hspace*{\fill}\includegraphics[width=0.45\columnwidth]{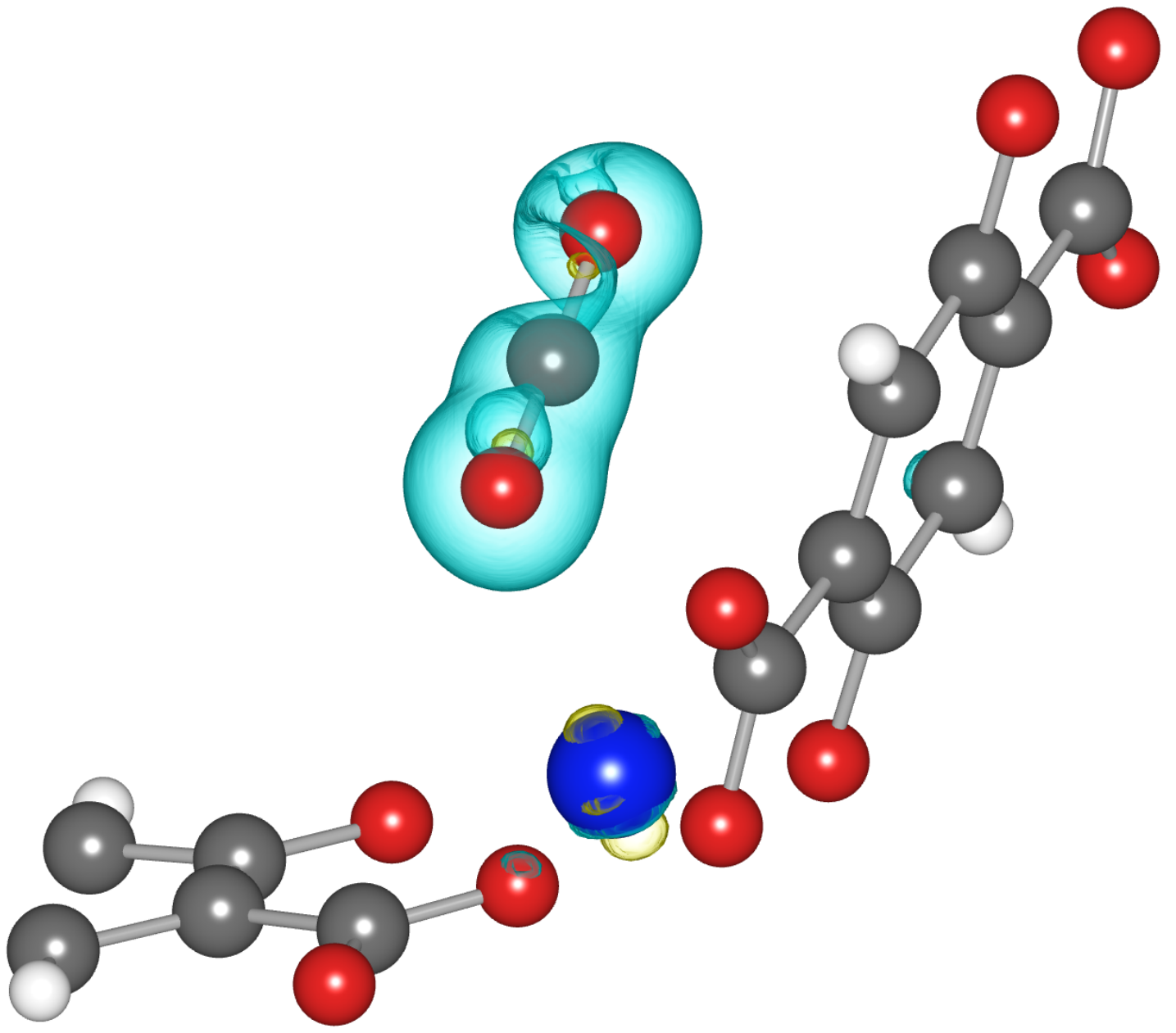}\hfill
\includegraphics[width=0.45\columnwidth]{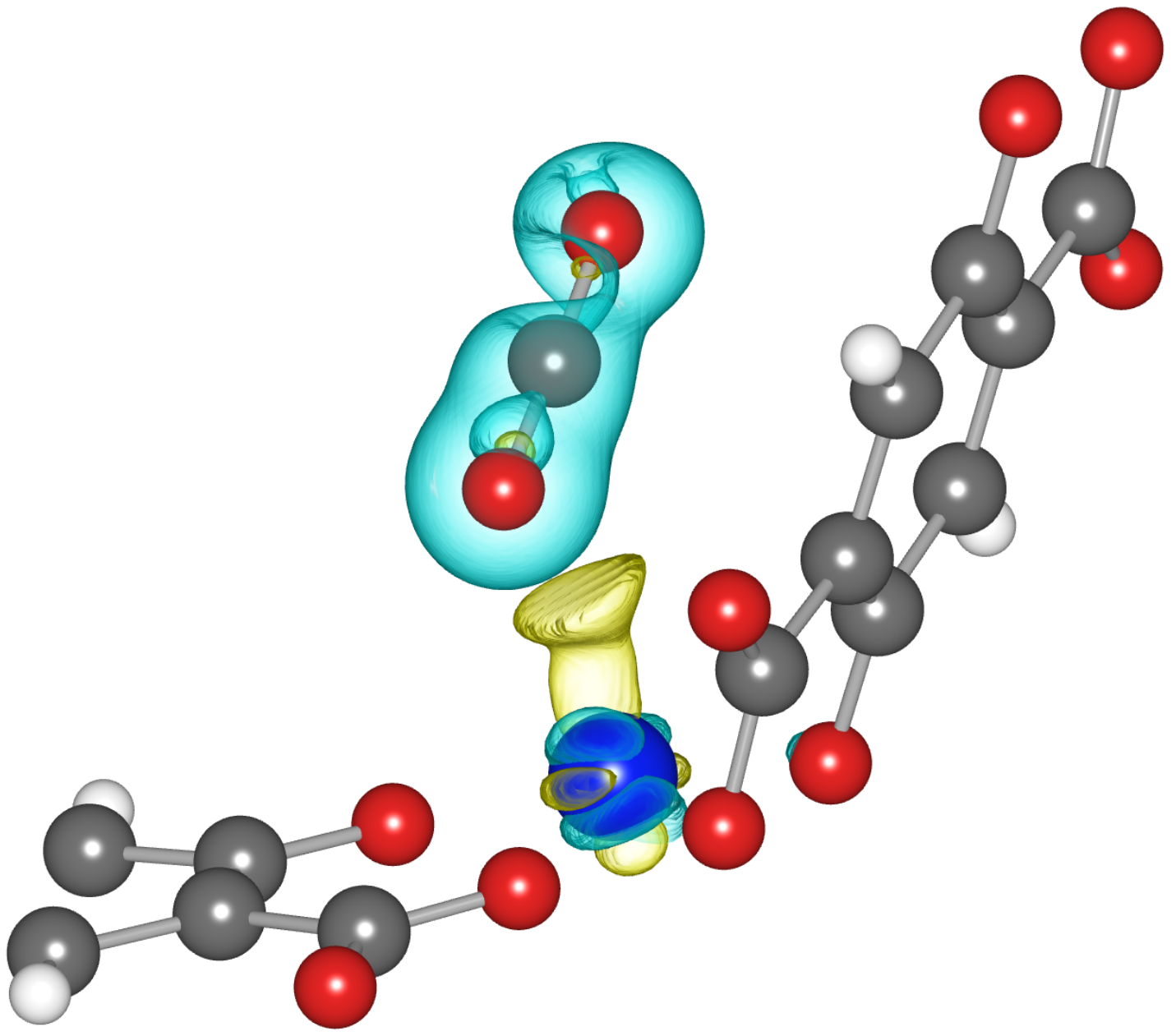}\hspace*{\fill}\mbox{}\\
\hspace*{\fill}spin up\hfill\hfill spin down\hfill\mbox{}
\caption{\label{induced_diff} Induced charge density originating from
the $E_c^\nl$ term upon CO$_2$ adsorption Ni-MOF74, split into its up
and down contribution. Blue (yellow) areas show charge depletion
(accumulation). Iso levels are 0.0001 $e$/Bohr$^{3}$.}
\end{figure}

As mentioned above, $E_c^\nl$ is responsible for the entire binding of
the small molecules. As such, it is at least indirectly responsible for
all the effects we see in our figures and tables. To examine the spin
effect of $E_c^\nl$ explicitly, we also calculate the difference between
the binding-induced density of a full svdW-DF calculation and the same
calculation without the $E_c^\nl$ term (Fig.~\ref{induced_diff}, Supplemental
Material). Although overall smaller in magnitude---as expected, since
the semi-local part $E_{xc}^0$ also contributes to the induced
density---we find the same behavior as in
Fig.~\ref{fig:induced_density}: more down density is being pulled into
the bond, strengthening the binding.

It is also revealing to partition the charge and magnetic moment of the
system, as detailed in Table~\ref{chrg-mgm-ba_tab} in the Supplemental Material.
Partitioning schemes are not unique, but one can still gain qualitative
information. Before adsorption, the CO$_2$ has no magnetic moment and
all six Ni atoms in the unitcell are equivalent. However, once
adsorption occurs, the up and down charge inside the CO$_2$ rearranges
differently and gives rise to a small but observable magnetic moment. At
the same time, the adsorption process leaves the total charge on the
CO$_2$ molecule unchanged, i.e.\ there is no net charge transfer. During
this adsorption process, the nearby Ni loses 0.03~$e$---our analysis
shows that it loses mostly down density---and thus its magnetic moment
increases, giving rise to a weak nonlocal-correlation induced magnetic
interaction.

Finally, in terms of absolute numbers, spin effects in the
Fe-MOF74+CO$_2$ and Co-MOF74+CO$_2$ systems seem to worsen agreement
with experiment---but, at the same time, they actually resolve a more
pressing issue. Experimentally, the CO$_2$ binding strength should
follow the order Mn $<$ Fe $<$ Co $<$ Ni. In calculations without spin
(not considering the artificial case of Mn), the order is reversed.
However, after including spin effects the correct order is restored.

In summary, we have developed a consistent spin-polarized version of the
nonlocal exchange-correlation functional vdW-DF, which we find to now
become an all-purpose functional.  We then apply this framework to study
small-molecule adsorption in MOF74 with magnetic open metal sites and
find that including nonlocal spin effects can significantly influence
the binding of the adsorbates. In the case of Ni-MOF-74+CO$_2$ we find a
counter-intuitive increase in binding due to nonlocal spin effects,
which we explain by a coincidental interaction of the Ni $d$ and O $p$
states. The additional degree of freedom from such unexpected magnetic
interactions can be used to tailor the specificity of MOFs in novel gas
storage, sequestration, and sensing applications.

\begin{acknowledgments}
Work in the US has been supported by NSF Grant No.\ DMR--1145968 and DOE
Grant No.\ DE--FG02--08ER46491. Work in Sweden has been supported by
grants from the Swedish Research Council (VR), the Swedish Foundation
for Strategic Research (SSF), the Chalmers e-Science Centre, and the
Chalmers Materials Area of Advance.
\end{acknowledgments}

\clearpage
\onecolumngrid
\begin{center}
\Large\bfseries ---\;\;Supplemental Material\;\;---
\end{center}
\vspace{5ex}
\twocolumngrid

\section{I.\quad Computational Details}

We have implemented svdW-DF in \textsc{PWscf} 5.1, which is part of the
\textsc{Quantum Espresso} package
\cite{Giannozzi_2009:quantum_espresso}.  We used ultrasoft
pseudopotentials and a plane wave cutoff for the wave functions of
40~Ryd together with the cx \cite{Berland_2014:exchange_functional}
implementation of svdW-DF.  Hubbard $U$ corrections are applied to the
$d$ electrons of Mn, Fe, Co, and Ni, using the well-established values
of 4.0, 4.0, 3.3, and 6.4~eV from
Ref.~\cite{Wang_2006:oxidation_energies}, where the $U$ values for each
metal $\mathcal{M}$ were determined so that they reproduce the
experimental oxidation energy of the metal monoxide to
$\mathcal{M}_2\text{O}_3$.  Note that these $U$ values have been shown
to give highly accurate results for small-molecule adsorption in
MOFs~\cite{Lee_2015:small-molecule_adsorption,
Poloni_2014:understanding_trends}. Due to the large unitcell size of the
MOF, we only sampled the $\Gamma$-point. In all our calculations, we
optimized the entire structure until the forces on all atoms reached
less than 1~meV/\AA.

\begin{figure}[h!]
\centering\includegraphics[width=0.9\columnwidth]{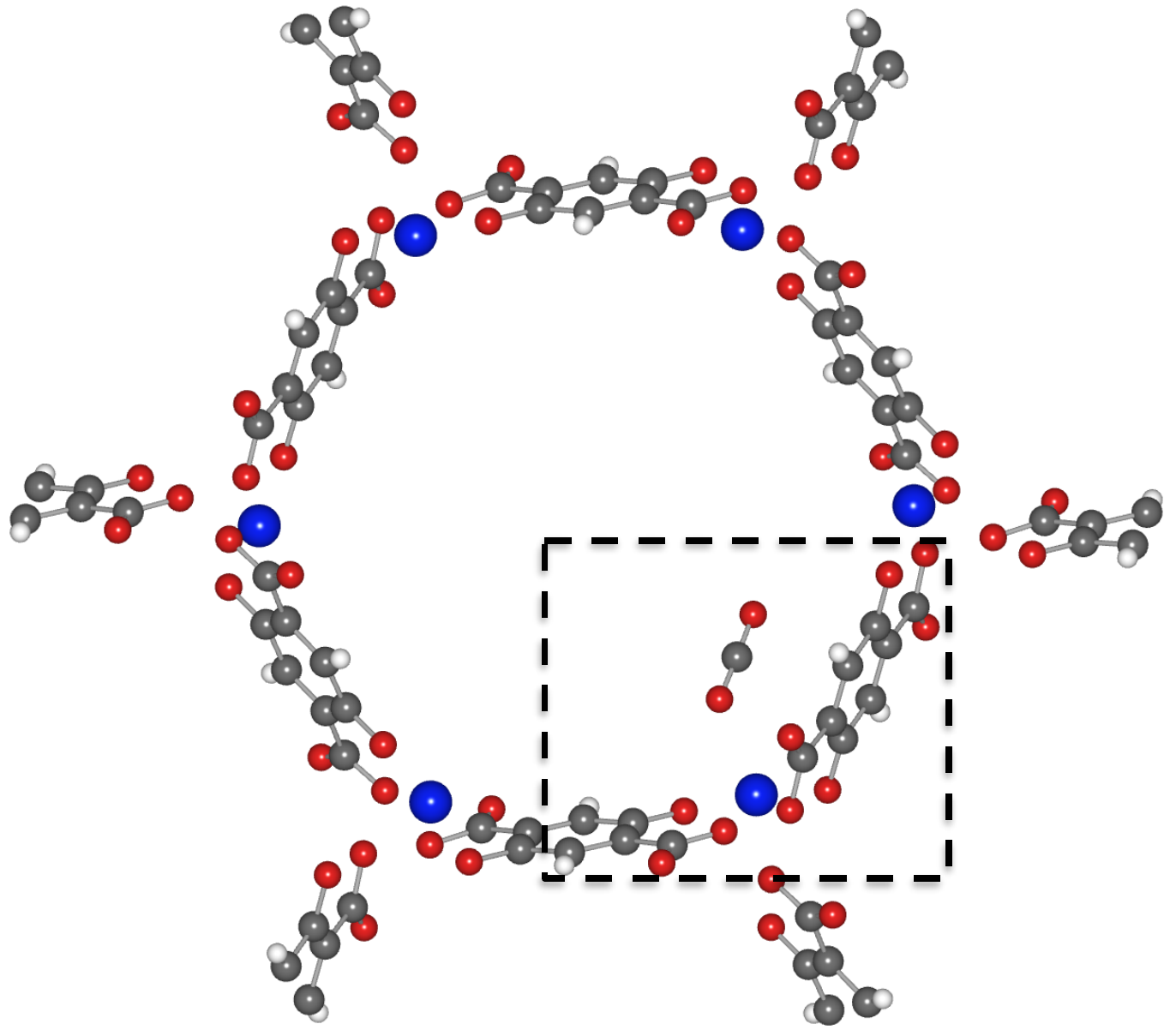}
\caption{\label{fig:MOF}MOF74 structure with a CO$_2$ molecule adsorbed
at one of the open metal sites. The hexagonal channel structure with its
six equivalent open metal sites per unit cell is clearly visible. Carbon
atoms are depicted as grey spheres, while oxygen, hydrogen, and metal
atoms are shown in red, white, and blue. The rectangle indicates the
portion of MOF74 shown in the main manuscript and
Figs.~\ref{charge_band} and \ref{induced_diff}.}
\end{figure}

The starting points of the calculations are the experimental
rhombohedral structures of Mn-, Fe-, Co-, and Ni-MOF74 with 54 atoms in
the primitive cell. Table~\ref{lattice_tab} shows the experimental
lattice constants, listed in the more convenient hexagonal lattice
\cite{Zhou_2008:enhanced_h2, Bloch_2011:Selective_Binding}. For the
magnetic ordering of the metal ions we adopted the ferromagnetic
arrangement, which has been found to be the ground-state configuration
\cite{Canepa_2013:when_metal}.  Figure~\ref{fig:MOF} provides a
graphical representation of the isostructural Mn-, Fe-, Co-, and
Ni-MOF74.

Vibrational properties were calculated using the finite difference
method with atomic displacements of 0.015~\AA\ in each direction. The
corresponding Hessian matrix was symmetrized and the acoustic sum rule
was enforced. Only atoms of the adsorbed molecule were allowed to
move---an approximation which we found to result in errors on the order
of merely 1 cm$^{-1}$.  The zero-point energy and the thermal correction
to the binding energy was calculated as
\begin{equation}
\Delta E_\text{ZPE}+\Delta H_T = \frac{1}{2}\sum_i v_i +
\sum_i \frac{v_i}{\exp({v_{i}/k_BT})-1}\;,
\label{equ:freq-corr}
\end{equation} 
where $k_B$ is Boltzmann's constant and $v_i$ are the vibrational
frequencies of the system in eV. The temperature was set to $T=298$~K.
Contributions from low frequency vibrations were not included.

\begin{table}[h!]
\caption{\label{lattice_tab}Experimental lattice constants used [\AA]
for our $\mathcal{M}$-MOF74 systems with $\mathcal{M}$ = Mn, Fe, Co, and
Ni.}
\begin{tabular*}{\columnwidth}{@{}l@{\extracolsep{\fill}}ccc@{}}\hline\hline
$\mathcal{M}$ & Ref.                                & $a=b$   & $c$   \\\hline
Mn            & \cite{Zhou_2008:enhanced_h2}        & 26.230  & 7.035 \\
Fe            & \cite{Bloch_2011:Selective_Binding} & 26.098  & 6.851 \\
Co            & \cite{Zhou_2008:enhanced_h2}        & 25.948  & 6.838 \\
Ni            & \cite{Zhou_2008:enhanced_h2}        & 25.719  & 6.741 \\\hline\hline
\end{tabular*}
\end{table}

\newpage\clearpage

\section{II.\quad Additional Tables and Figures}

\begin{figure}[h]
\raisebox{4ex}{\rotatebox{90}{Ni-MOF74+CO$_2$}}\hspace*{\fill}
\includegraphics[width=0.45\columnwidth]{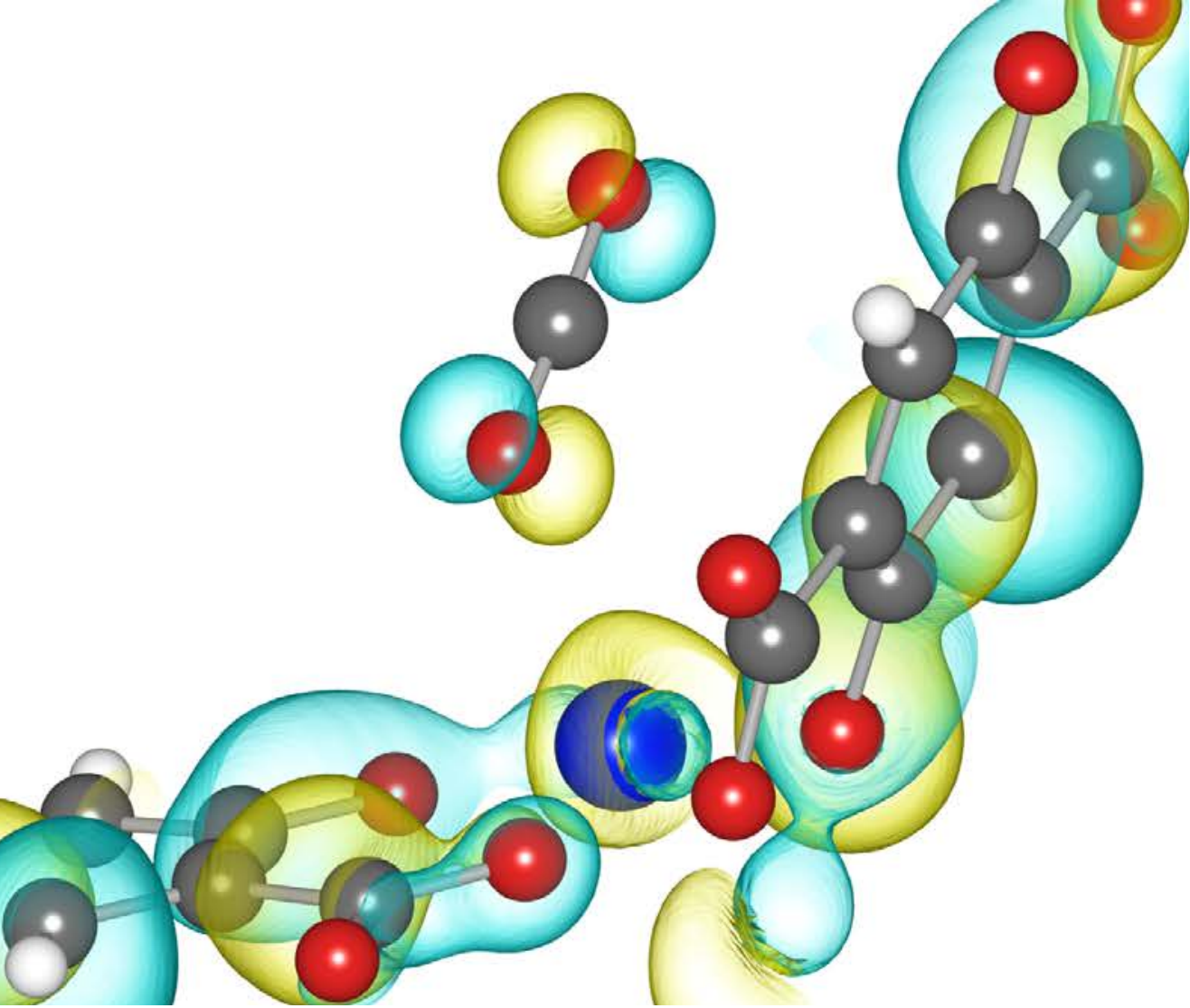}\hfill
\includegraphics[width=0.45\columnwidth]{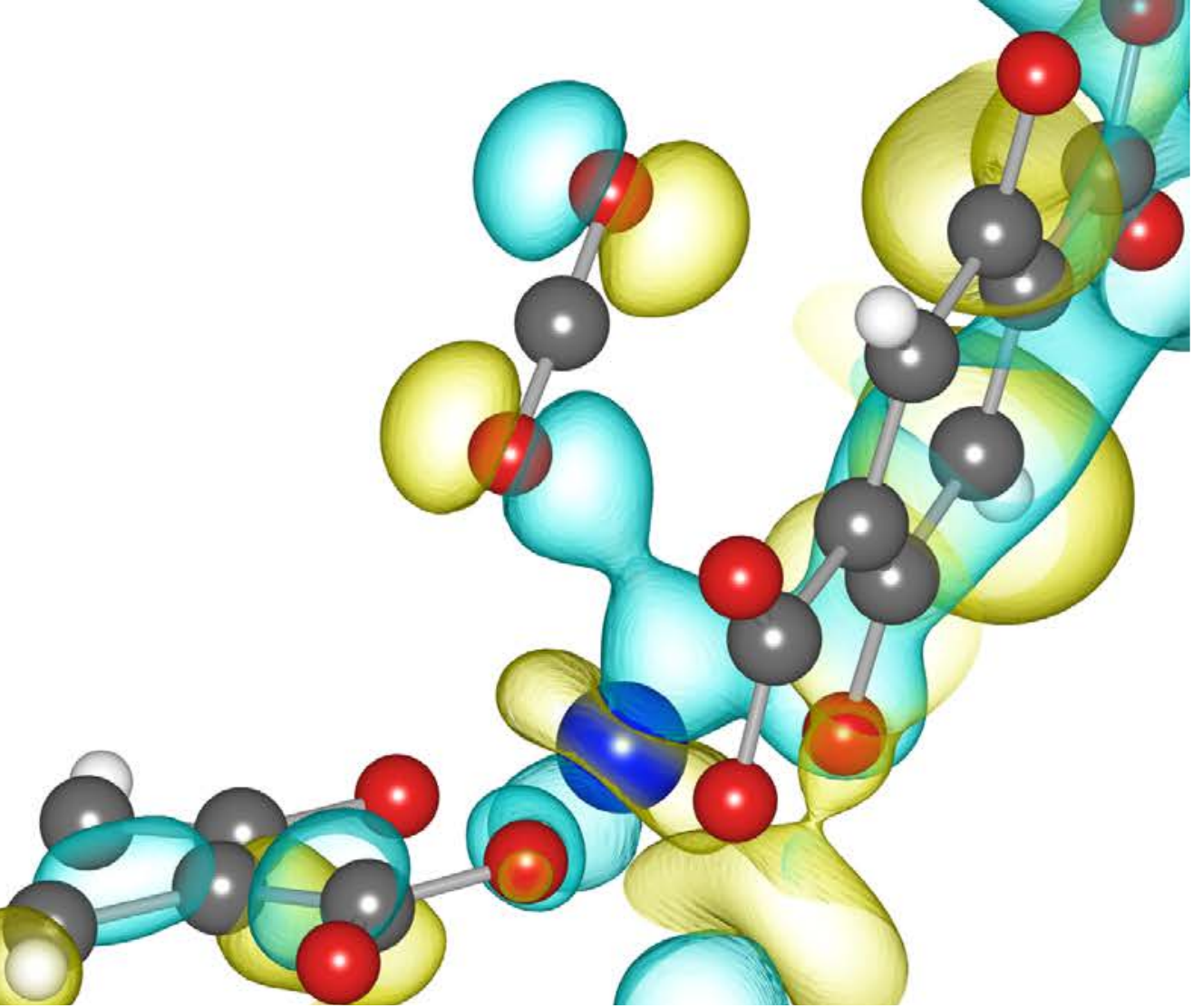}\hspace*{\fill}\mbox{}\\[4ex]
\raisebox{7ex}{\rotatebox{90}{Ni-MOF74}}\hspace*{\fill}
\includegraphics[width=0.45\columnwidth]{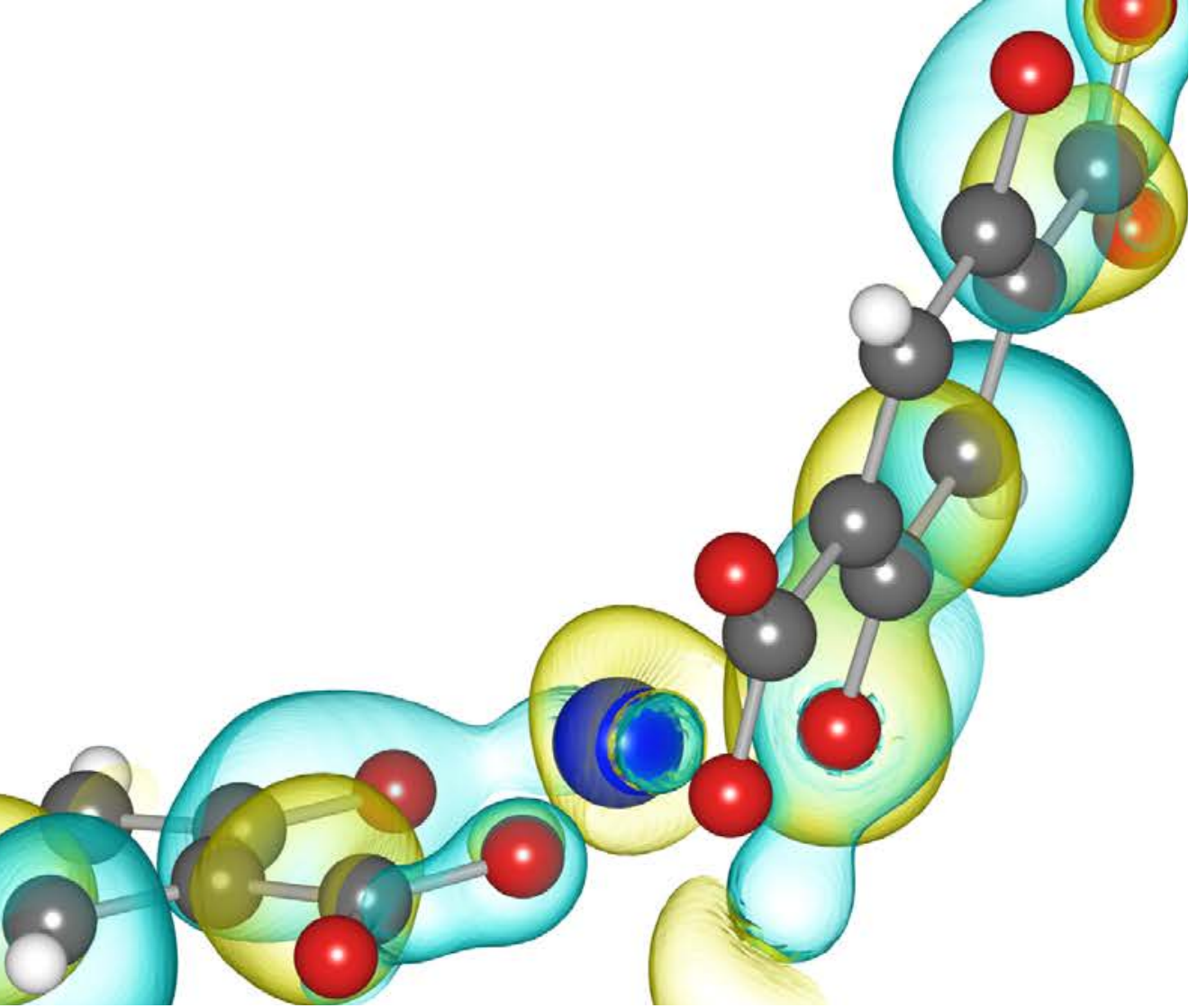}\hfill
\includegraphics[width=0.45\columnwidth]{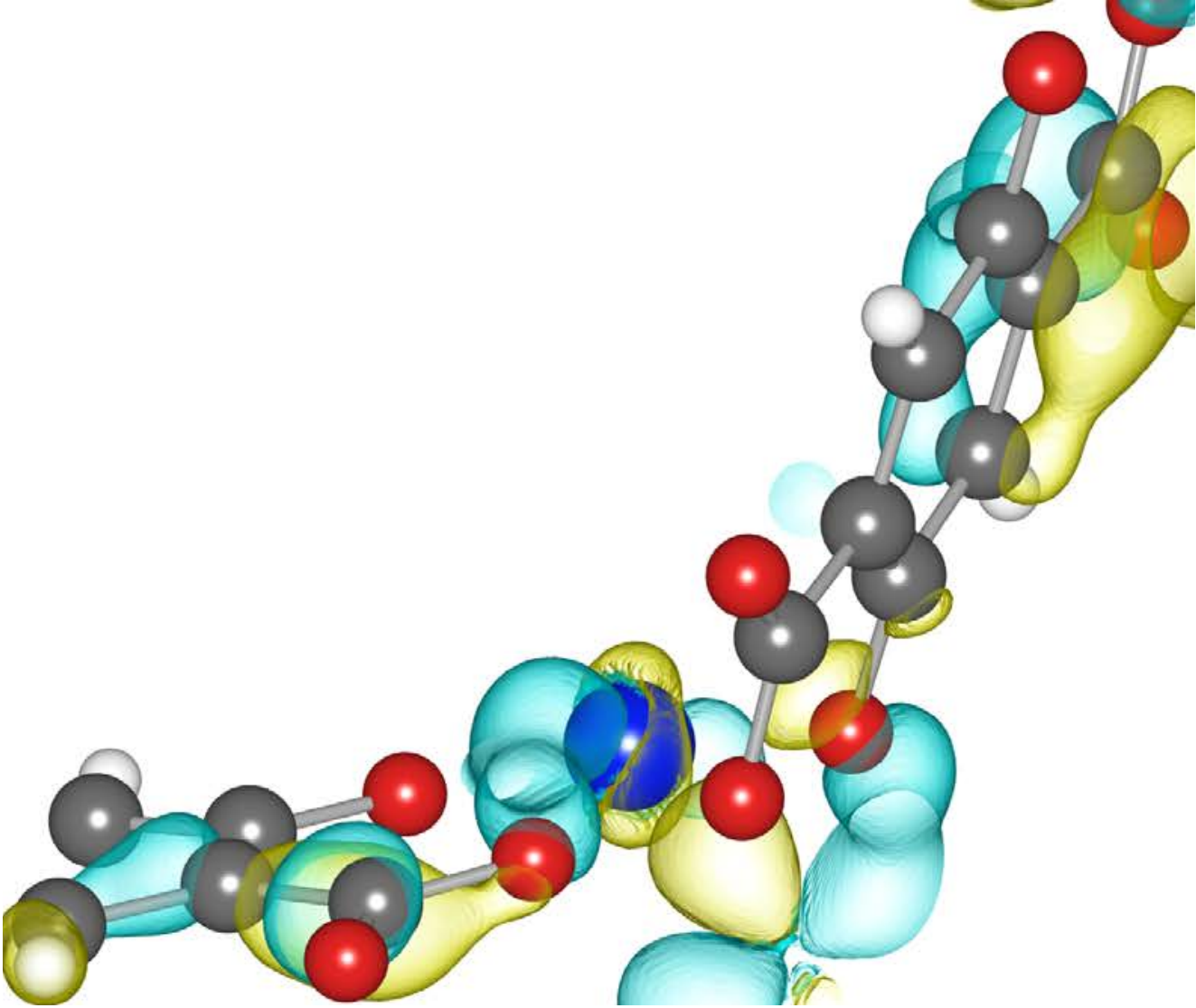}\hspace*{\fill}\mbox{}\\
\hspace*{\fill}spin up\hfill\hfill spin down\hfill\mbox{}
\caption{\label{charge_band} Contour plots of $|\psi|^2
\times\text{sign}(\psi)$ evaluated (at the $\Gamma$-point) for the
spin-up and spin-down Ni-MOF74 bands around $-5$~eV with (top row) and
without (bottom row) the adsorbed CO$_2$ molecule. Iso levels are
0.0002~$e$/Bohr$^{3}$.  In all cases, the $d_{z^2}$-like orbital on the
Ni is visible. In the spin-up channel, the adsorption of CO$_2$ leaves
this orbital almost unchanged. In contrast, in the spin-down channel,
the $d_{z^2}$-like orbital rotates towards the oxygen $p$ orbital to
form a new, hybridized orbital.}
\end{figure}

\begin{table}[h]
\caption{\label{chrg-mgm-ba_tab} Bader charge $q$ (see
Refs.~\cite{Bader_1990:atoms_molecules, Henkelman_2006:fast_robust}) [in
units of $e$, showing electron \emph{loss} relative to the neutral unit]
and magnetic moment $\mu$ [in units of $\mu_B$] during the CO$_2$
adsorption process in Ni-MOF74. $^*$ indicates the Ni atom participating
in the bond. Only one representative of the remaining 5 equivalent Ni is
given.} 
\begin{tabular*}{\columnwidth}{@{}l@{\extracolsep{\fill}}cccr@{}}\hline\hline
        & \multicolumn{2}{c}{before adsorption}
        & \multicolumn{2}{c}{after adsorption} \\\cline{2-3}\cline{4-5}
Unit    & $q$     & $\mu$    & $q$   & $\mu$   \\\hline
CO$_2$  &    0.00 & 0.00   & 0.00  & 0.01      \\
Ni$^*$  &    1.30 & 1.57   & 1.33  & 1.60      \\
Ni      &    1.30 & 1.57   & 1.30  & 1.57      \\\hline\hline
\end{tabular*}
\end{table}

\newpage\vspace*{3ex}
\section{III.\quad Self-Consistent Derivatives}

As for the evaluation of $\tilde{q}_{0c}[\nup,\ndown]$ and its
derivatives, they are completely defined through the choice of
correlation inside $\varepsilon_{c}^\text{LDA}$ and are given elsewhere
\cite{Perdew_1992:accurate_simple}. For $\tilde{q}_{0x}[\nup,\ndown]$ it
follows that
\begin{subequations}
\begin{eqnarray}
\frac{d\tilde{q}_{0x}[\nup,\ndown]}{d\nup}   &=& \frac{\ndown}{(\nup+\ndown)^2}
       \big(q_{0x}[2\nup] - q_{0x}[2\ndown]\big)\nonumber\\
   &+& \frac{2\nup}{\nup+\ndown}\frac{dq_{0x}[n]}{dn}\Big|_{n=2\nup}\\[3ex]
\frac{d\tilde{q}_{0x}[\nup,\ndown]}{d\ndown} &=& \frac{\nup}{(\nup+\ndown)^2}
       \big(q_{0x}[2\ndown] - q_{0x}[2\nup]\big)\nonumber\\
   &+& \frac{2\ndown}{\nup+\ndown}\frac{dq_{0x}[n]}{dn}\Big|_{n=2\ndown}
\end{eqnarray}
\end{subequations}
and
\begin{subequations}
\begin{eqnarray}
\frac{d\tilde{q}_{0x}[\nup,\ndown]}{d|\nabla \nup|}   &=& \frac{\nup  }{\nup+\ndown}
   \frac{dq_{0x}[2\nup]}{d|\nabla\nup|}\\[3ex]
\frac{d\tilde{q}_{0x}[\nup,\ndown]}{d|\nabla \ndown|} &=& \frac{\ndown}{\nup+\ndown}
   \frac{dq_{0x}[2\ndown]}{d|\nabla\ndown|}\;.
\end{eqnarray}
\end{subequations}

\clearpage\onecolumngrid
\section{IV.\quad Testing svdW-DF}

Testing and benchmarking has been performed on three model systems,
i.e.\ the Li dimer in its triplet state $^3\Sigma$, atomization energies
of a set of small molecules, and graphene on the Ni(111) surface. We
used our implementation in
\textsc{PWscf}~\cite{Giannozzi_2009:quantum_espresso} and calculated
binding curves and energies using svdW-DF1 \cite{Dion_2004:van_waals},
svdW-DF2 \cite{Lee_2010:higher-accuracy_van}, svdW-DF-cx
\cite{Berland_2014:exchange_functional}, VV10
\cite{Vydrov_2010:nonlocal_van, Sabatini_2013:nonlocal_van}, and PBE
\cite{Perdew_1996:generalized_gradient}; these studies were performed
using pseudopotentials from the Rutgers database with the recommended
cutoffs \cite{Garrity_2014:pseudopotentials_high-throughput}.  Note that
svdW-DF1 and svdW-DF2 are better suited for small molecules, while
svdW-DF-cx is better suited for larger, extended systems.  We compare
our \textsc{PWscf} calculations to high-level quantum chemistry (QC),
diffusion Monte Carlo (DMC), and experiment where appropriate. Please
see the footnotes \cite{Troy} and \cite{balancing} in the main text
concerning the neglect of spin-polarization effects in the nonlocal
part of VV10.

\vspace{0.5in}
\subsection{A.\quad Li Dimer in the Triplet State $^3\Sigma$}
\vspace*{-0.25in}

\begin{table}[h]
\caption{\label{tab:Lidimer}Dissociation energy $D_e$ [meV] and
fundamental frequency $\omega_e$ [meV] for Li$_{2}$ in the triplet state
$^3\Sigma$. The experimental value is $D_e^\text{exp}=41$~meV
\cite{Linton_1999:high-lying_vibrational, dsepDFTdisc}. Early papers
using quantum chemistry methods report a considerable spread in $D_{e}$
values \cite{Konowalow_1984:molecular_electronic,
Poteau_1995:calculation_electronic, Jasik_2006:calculation_adiabatic},
but the most recent high-level quantum chemistry (QC) results show in
essence perfect agreement with
experiment~\cite{Musial_2014:first_principle}.}
\begin{tabular*}{\columnwidth}{@{\extracolsep{\fill}}lrrrrrr@{}}\hline\hline
\rule[-1.1ex]{0ex}{3.5ex}                         &
\multicolumn{1}{c}{svdW-DF1}          \cite{Dion_2004:van_waals}&
\multicolumn{1}{c}{svdW-DF2}          \cite{Lee_2010:higher-accuracy_van}&
\multicolumn{1}{c}{svdW-DF-cx}        \cite{Berland_2014:exchange_functional}&
\multicolumn{1}{c}{VV10}              \cite{Vydrov_2010:nonlocal_van}&
\multicolumn{1}{c}{PBE}               \cite{Perdew_1996:generalized_gradient}&
QC                                    \cite{Musial_2014:first_principle}\\\hline
$D_e$      &  53 & 55 & 70 & 77 & 77 & 41 \\
$\omega_e$ &   8 &  8 &  8 &  8 &  8 &  8 \\\hline\hline
\end{tabular*}
\end{table}

\clearpage
\subsection{B.\quad Atomization Energy of Small Molecules}
\vspace*{-0.25in}

\begin{table}[h]
\caption{\label{tab:atomizationenergies} Atomization energies
$E_\text{at}$ [eV] of small molecules from the G1 set
\cite{Pople_1989:gaussian-1_theory}. Spin expectation values $\langle
S(S+1)\rangle$ different from zero indicate a magnetic ground state.
Diffusion Monte Carlo (DMC) values are taken from
\cite{Grossman_2002:benchmark_quantum}; experimental numbers from
\cite{Chase_1998:nist-janaf_thermochemical}. An error analysis follows
in Table~\ref{tab:erroratomization}. Spin enters through the isolated
atoms and/or magnetic molecular ground states.}
\renewcommand{\arraystretch}{1.2}
\begin{tabular*}{1.0\textwidth}{@{\extracolsep{\fill}}lcddddddr@{}}\hline\hline
\rule[-1.1ex]{0ex}{3.5ex}                   &
\multicolumn{1}{c}{$\langle S(S+1)\rangle$} &
\multicolumn{1}{c}{svdW-DF1}                \cite{Dion_2004:van_waals}&
\multicolumn{1}{c}{svdW-DF2}                \cite{Lee_2010:higher-accuracy_van}&
\multicolumn{1}{c}{svdW-DF-cx}              \cite{Berland_2014:exchange_functional}&
\multicolumn{1}{c}{VV10}                    \cite{Vydrov_2010:nonlocal_van}&
\multicolumn{1}{c}{PBE}                     \cite{Perdew_1996:generalized_gradient}&
\multicolumn{1}{c}{DMC}                     \cite{Grossman_2002:benchmark_quantum}&
Exp.\                                       \cite{Chase_1998:nist-janaf_thermochemical}\\\hline
CH$_3$       & 3/4  &  13.421   &  13.363  &  13.639  & 13.373  &  13.507   &  12.615  &  12.545 \\
CH$_4$       & 0    &  18.074   &  17.937  &  18.502  & 17.963  &  18.212   &  17.129  &  17.020 \\
NH           & 2    &  3.763    &  3.810   &  3.758   & 3.795   &  3.778    &  3.393   &  3.426  \\
NH$_2$       & 3/4  &  7.942    &  7.934   &  8.100   & 7.957   &  8.042    &  7.339   &  7.372  \\
NH$_3$       & 0    &  12.648   &  12.516  &  13.088  & 12.620  &  12.880   &  11.991  &  11.999 \\
OH           & 3/4  &  4.236    &  4.098   &  4.557   & 4.195   &  4.408    &  4.390   &  4.397  \\
H$_2$O       & 0    &  9.660    &  9.449   &  10.194  & 9.612   &  9.953    &  9.514   &  9.512  \\
SiH$_3$      & 3/4  &  9.887    &  9.863   &  10.015  & 9.594   &  9.668    &  9.328   &  9.280  \\
SiH$_4$      & 0    &  13.921   &  13.941  &  14.086  & 13.528  &  13.595   &  13.261  &  13.122 \\
PH$_2$       & 3/4  &  6.716    &  6.694   &  6.899   & 6.572   &  6.656    &  6.232   &  6.275  \\
H$_2$S       & 0    &  7.901    &  7.786   &  8.190   & 7.871   &  8.045    &  7.463   &  7.506  \\
HCl          & 0    &  4.575    &  4.457   &  4.785   & 4.575   &  4.697    &  4.486   &  4.432  \\
LiF          & 0    &  5.885    &  5.990   &  5.984   & 5.897   &  6.209    &  6.292   &  5.984  \\
C$_2$H$_4$   & 0    &  24.272   &  24.042  &  25.020  & 24.439  &  24.819   &  23.136  &  23.065 \\
CO           & 0    &  10.940   &  10.732  &  11.489  & 11.179  &  11.501   &  10.981  &  11.110 \\
N$_2$        & 0    &  9.590    &  9.494   &  10.017  & 9.972   &  10.164   &  9.587   &  9.761  \\
NO           & 3/4  &  6.373    &  6.133   &  7.043   & 6.725   &  7.079    &  6.198   &  6.507  \\
O$_2$        & 2    &  5.171    &  4.805   &  6.017   & 5.434   &  5.925    &  4.844   &  5.115  \\
CO$_2$       & 0    &  16.526   &  16.028  &  17.768  & 17.103  &  17.796   &  16.458  &  16.562 \\
Na$_2$       & 0    &  0.794    &  0.767   &  0.740   & 0.671   &  0.707    &  0.750   &  0.729  \\
S$_2$        & 2    &  4.820    &  4.650   &  5.197   & 5.161   &  5.334    &  4.264   &  4.365  \\
NaCl         & 0    &  4.085    &  3.993   &  4.190   & 4.129   &  4.157    &  4.284   &  4.219  \\
SiO          & 0    &  8.045    &  7.971   &  8.508   & 8.229   &  8.429    &  8.096   &  8.239  \\
CS           & 0    &  7.504    &  7.374   &  7.861   & 7.801   &  7.980    &  7.172   &  7.328  \\
SO           & 2    &  5.589    &  5.357   &  6.156   & 5.866   &  6.178    &  5.100   &  5.351  \\
\hline\hline
\end{tabular*}
\end{table}

\begin{table}[h]
\caption{\label{tab:erroratomization} Error analysis for the datasets
given in Table~\ref{tab:atomizationenergies}. Given are the mean
(signed) percentage error (MPE), mean absolute percentage error (MAPE),
and mean absolute deviation (MAD) values when comparing with experiment.
Note that svdW-DF-cx, due to its design, is not necessarily expected to
perform well for small molecules, but still gives good results,
comparable to PBE.}
\renewcommand{\arraystretch}{1.2}
\begin{tabular*}{\textwidth}{@{\extracolsep{\fill}}ldddddr@{}}\hline\hline
\rule[-1.1ex]{0ex}{3.5ex}       &
\multicolumn{1}{c}{svdW-DF1}          \cite{Dion_2004:van_waals}&
\multicolumn{1}{c}{svdW-DF2}          \cite{Lee_2010:higher-accuracy_van}&
\multicolumn{1}{c}{svdW-DF-cx}        \cite{Berland_2014:exchange_functional}&
\multicolumn{1}{c}{VV10}              \cite{Vydrov_2010:nonlocal_van}&
\multicolumn{1}{c}{PBE}               \cite{Perdew_1996:generalized_gradient}&
DMC                                   \cite{Grossman_2002:benchmark_quantum}\\\hline
MPE [\%]               &  -3.28 &  -1.52  & -7.69  &  -3.84 &  -6.75 &  0.54  \\
MAPE [\%]              &   4.59 &   4.50  &  7.75  &   5.14 &   7.11 &  1.65  \\
MAD [eV]               &   0.37 &   0.38  &  0.67  &   0.40 &   0.59 &  0.11  \\
\hline\hline
\end{tabular*}
\end{table}

\clearpage\twocolumngrid
\subsection{C.\quad Graphene on Ni(111)}

\begin{figure}[h]
\includegraphics[width=0.6\columnwidth]{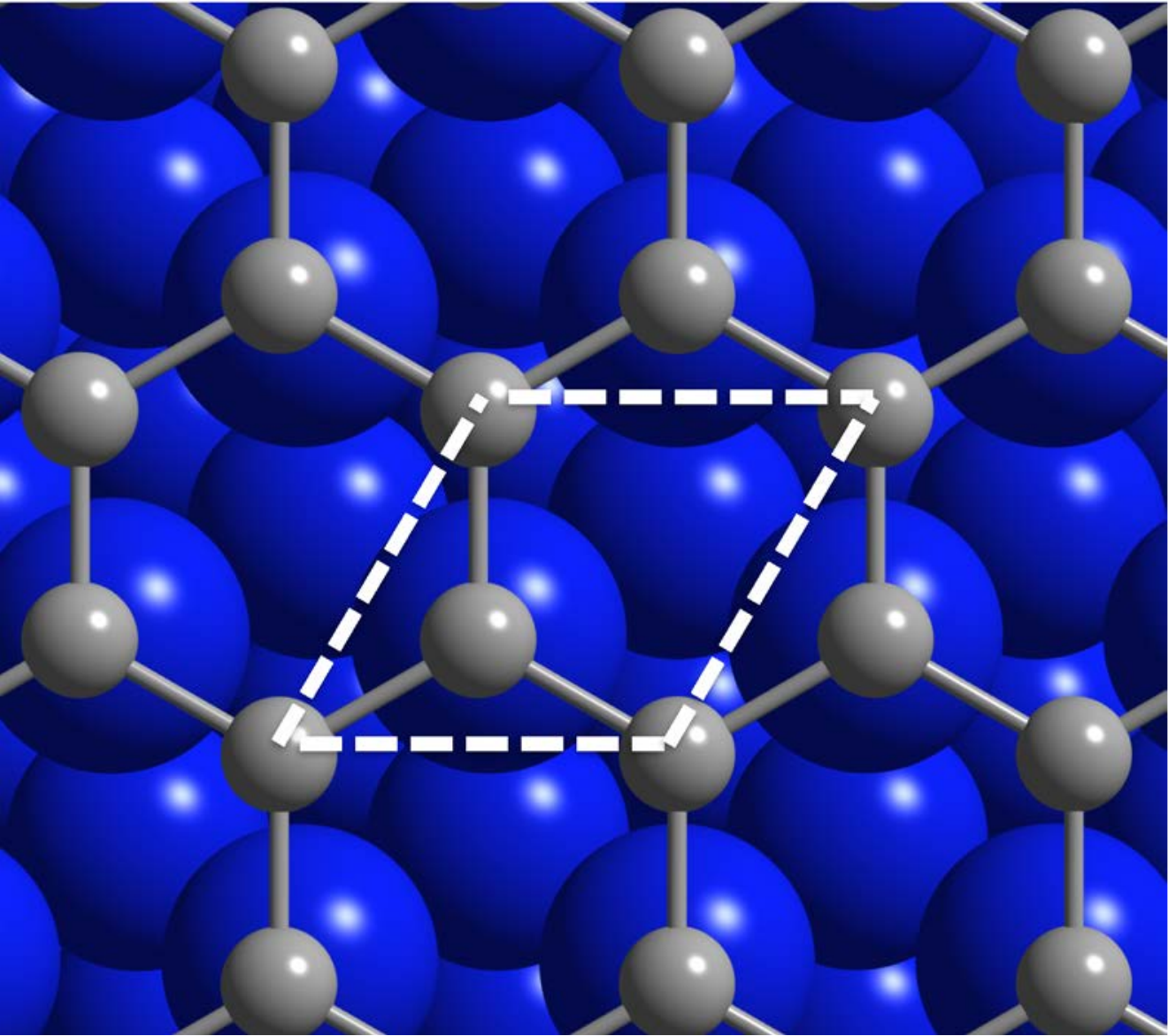}
\caption{\label{graphene_on_Ni_surface} Unit cell used for the
adsorption of graphene on the Ni(111) surface. Ni was modeled as a slab
consisting of 6 layers of atoms, the bottom three of which were fixed to
their bulk positions. Blue and grey spheres denote Ni and C atoms,
respectively.}
\end{figure}

\begin{figure}[h]
\includegraphics[width=\columnwidth]{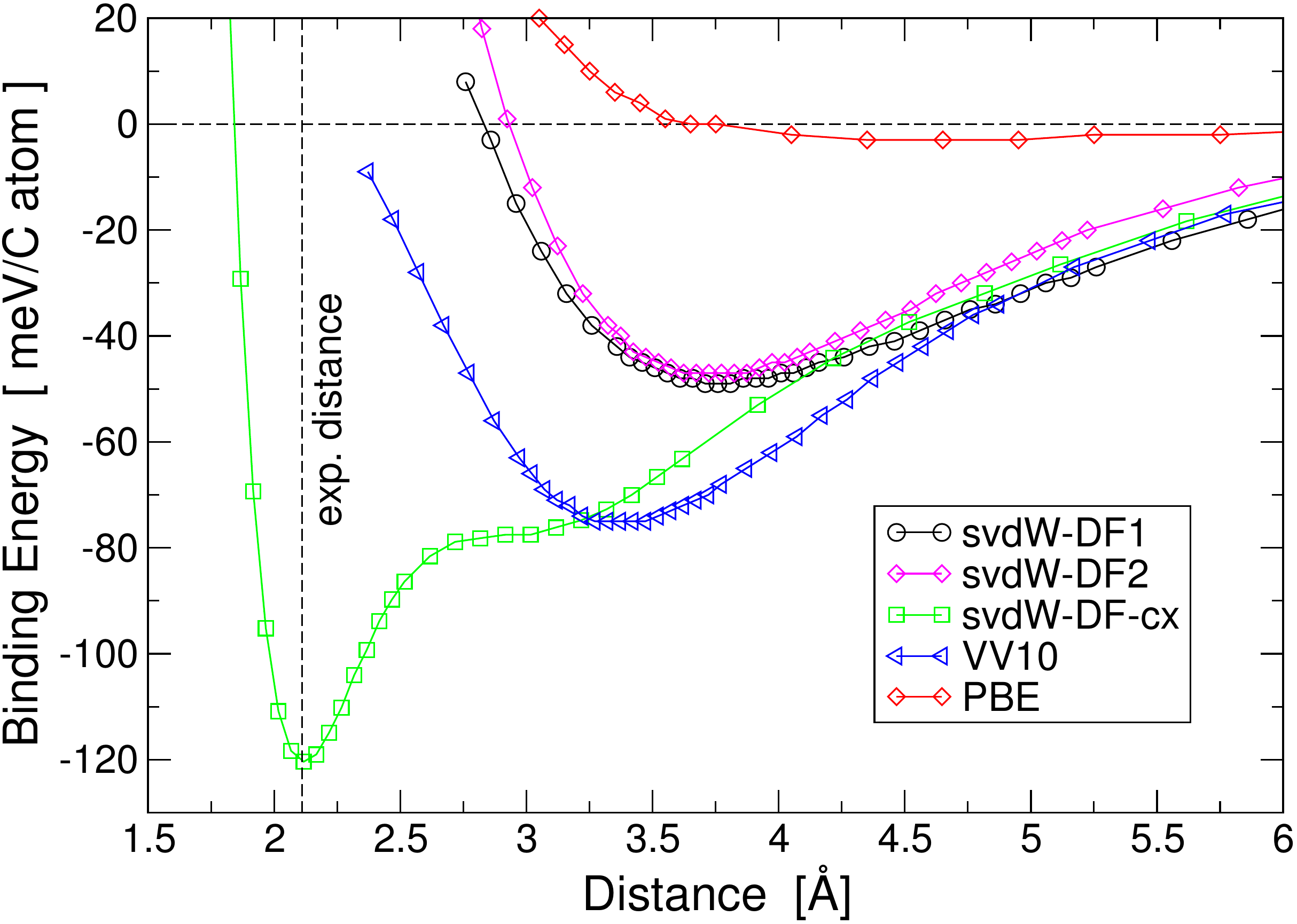}
\caption{\label{graphene_on_Ni_results} Binding energy curve of graphene
on the Ni(111) surface. The dashed vertical line denotes the
experimental value for the distance between graphene and the Ni(111)
surface \cite{Gamo_1997:atomic_structure}.}
\end{figure}

\clearpage
%


\begin{thebibliography}{90}%
\makeatletter
\providecommand \@ifxundefined [1]{%
 \@ifx{#1\undefined}
}%
\providecommand \@ifnum [1]{%
 \ifnum #1\expandafter \@firstoftwo
 \else \expandafter \@secondoftwo
 \fi
}%
\providecommand \@ifx [1]{%
 \ifx #1\expandafter \@firstoftwo
 \else \expandafter \@secondoftwo
 \fi
}%
\providecommand \natexlab [1]{#1}%
\providecommand \enquote  [1]{``#1''}%
\providecommand \bibnamefont  [1]{#1}%
\providecommand \bibfnamefont [1]{#1}%
\providecommand \citenamefont [1]{#1}%
\providecommand \href@noop [0]{\@secondoftwo}%
\providecommand \href [0]{\begingroup \@sanitize@url \@href}%
\providecommand \@href[1]{\@@startlink{#1}\@@href}%
\providecommand \@@href[1]{\endgroup#1\@@endlink}%
\providecommand \@sanitize@url [0]{\catcode `\\12\catcode `\$12\catcode
  `\&12\catcode `\#12\catcode `\^12\catcode `\_12\catcode `\%12\relax}%
\providecommand \@@startlink[1]{}%
\providecommand \@@endlink[0]{}%
\providecommand \url  [0]{\begingroup\@sanitize@url \@url }%
\providecommand \@url [1]{\endgroup\@href {#1}{\urlprefix }}%
\providecommand \urlprefix  [0]{URL }%
\providecommand \Eprint [0]{\href }%
\providecommand \doibase [0]{http://dx.doi.org/}%
\providecommand \selectlanguage [0]{\@gobble}%
\providecommand \bibinfo  [0]{\@secondoftwo}%
\providecommand \bibfield  [0]{\@secondoftwo}%
\providecommand \translation [1]{[#1]}%
\providecommand \BibitemOpen [0]{}%
\providecommand \bibitemStop [0]{}%
\providecommand \bibitemNoStop [0]{.\EOS\space}%
\providecommand \EOS [0]{\spacefactor3000\relax}%
\providecommand \BibitemShut  [1]{\csname bibitem#1\endcsname}%
\let\auto@bib@innerbib\@empty
\bibitem [{\citenamefont {Murray}\ \emph {et~al.}(2009)\citenamefont {Murray},
  \citenamefont {Dinca},\ and\ \citenamefont
  {Long}}]{Murray_2009:hydrogen_storage}%
  \BibitemOpen
  \bibfield  {author} {\bibinfo {author} {\bibfnamefont {L.~J.}\ \bibnamefont
  {Murray}}, \bibinfo {author} {\bibfnamefont {M.}~\bibnamefont {Dinca}}, \
  and\ \bibinfo {author} {\bibfnamefont {J.~R.}\ \bibnamefont {Long}},\ }\href
  {\doibase 10.1039/b802256a} {\bibfield  {journal} {\bibinfo  {journal} {Chem.
  Soc. Rev.}\ }\textbf {\bibinfo {volume} {38}},\ \bibinfo {pages} {1294}
  (\bibinfo {year} {2009})}\BibitemShut {NoStop}%
\bibitem [{\citenamefont {Li}\ \emph {et~al.}(2011)\citenamefont {Li},
  \citenamefont {Ma}, \citenamefont {McCarthy}, \citenamefont {Sculley},
  \citenamefont {Yu}, \citenamefont {Jeong}, \citenamefont {Balbuena},\ and\
  \citenamefont {Zhou}}]{Li_2011:carbon_dioxide}%
  \BibitemOpen
  \bibfield  {author} {\bibinfo {author} {\bibfnamefont {J.-R.}\ \bibnamefont
  {Li}}, \bibinfo {author} {\bibfnamefont {Y.}~\bibnamefont {Ma}}, \bibinfo
  {author} {\bibfnamefont {M.~C.}\ \bibnamefont {McCarthy}}, \bibinfo {author}
  {\bibfnamefont {J.}~\bibnamefont {Sculley}}, \bibinfo {author} {\bibfnamefont
  {J.}~\bibnamefont {Yu}}, \bibinfo {author} {\bibfnamefont {H.-K.}\
  \bibnamefont {Jeong}}, \bibinfo {author} {\bibfnamefont {P.~B.}\ \bibnamefont
  {Balbuena}}, \ and\ \bibinfo {author} {\bibfnamefont {H.-C.}\ \bibnamefont
  {Zhou}},\ }\href {\doibase 10.1016/j.ccr.2011.02.012} {\bibfield  {journal}
  {\bibinfo  {journal} {Coord. Chem. Rev.}\ }\textbf {\bibinfo {volume}
  {255}},\ \bibinfo {pages} {1791} (\bibinfo {year} {2011})}\BibitemShut
  {NoStop}%
\bibitem [{\citenamefont {Qiu}\ and\ \citenamefont
  {Zhu}(2009)}]{Qiu_2009:molecular_engineering}%
  \BibitemOpen
  \bibfield  {author} {\bibinfo {author} {\bibfnamefont {S.}~\bibnamefont
  {Qiu}}\ and\ \bibinfo {author} {\bibfnamefont {G.}~\bibnamefont {Zhu}},\
  }\href {\doibase 10.1016/j.ccr.2009.07.020} {\bibfield  {journal} {\bibinfo
  {journal} {Coord. Chem. Rev.}\ }\textbf {\bibinfo {volume} {253}},\ \bibinfo
  {pages} {2891} (\bibinfo {year} {2009})}\BibitemShut {NoStop}%
\bibitem [{\citenamefont {Nijem}\ \emph {et~al.}(2012)\citenamefont {Nijem},
  \citenamefont {Wu}, \citenamefont {Canepa}, \citenamefont {Marti},
  \citenamefont {Balkus}, \citenamefont {Thonhauser}, \citenamefont {Li},\ and\
  \citenamefont {Chabal}}]{Nijem_2012:tuning_gate}%
  \BibitemOpen
  \bibfield  {author} {\bibinfo {author} {\bibfnamefont {N.}~\bibnamefont
  {Nijem}}, \bibinfo {author} {\bibfnamefont {H.}~\bibnamefont {Wu}}, \bibinfo
  {author} {\bibfnamefont {P.}~\bibnamefont {Canepa}}, \bibinfo {author}
  {\bibfnamefont {A.}~\bibnamefont {Marti}}, \bibinfo {author} {\bibfnamefont
  {K.~J.}\ \bibnamefont {Balkus}}, \bibinfo {author} {\bibfnamefont
  {T.}~\bibnamefont {Thonhauser}}, \bibinfo {author} {\bibfnamefont
  {J.}~\bibnamefont {Li}}, \ and\ \bibinfo {author} {\bibfnamefont {Y.~J.}\
  \bibnamefont {Chabal}},\ }\href {\doibase 10.1021/ja305754f} {\bibfield
  {journal} {\bibinfo  {journal} {J. Am. Chem. Soc.}\ }\textbf {\bibinfo
  {volume} {134}},\ \bibinfo {pages} {15201} (\bibinfo {year}
  {2012})}\BibitemShut {NoStop}%
\bibitem [{\citenamefont {Lee}\ \emph {et~al.}(2015)\citenamefont {Lee},
  \citenamefont {Howe}, \citenamefont {Lin}, \citenamefont {Smit},\ and\
  \citenamefont {Neaton}}]{Lee_2015:small-molecule_adsorption}%
  \BibitemOpen
  \bibfield  {author} {\bibinfo {author} {\bibfnamefont {K.}~\bibnamefont
  {Lee}}, \bibinfo {author} {\bibfnamefont {J.~D.}\ \bibnamefont {Howe}},
  \bibinfo {author} {\bibfnamefont {L.-C.}\ \bibnamefont {Lin}}, \bibinfo
  {author} {\bibfnamefont {B.}~\bibnamefont {Smit}}, \ and\ \bibinfo {author}
  {\bibfnamefont {J.~B.}\ \bibnamefont {Neaton}},\ }\href {\doibase 10.1021/cm502760q} {\bibfield  {journal} {\bibinfo  {journal} {Chem. Mater.}\
  }\textbf {\bibinfo {volume} {27}},\ \bibinfo {pages} {668} (\bibinfo {year}
  {2015})}\BibitemShut {NoStop}%
\bibitem [{\citenamefont {Lee}\ \emph {et~al.}(2009)\citenamefont {Lee},
  \citenamefont {Farha}, \citenamefont {Roberts}, \citenamefont {Scheidt},
  \citenamefont {Nguyen},\ and\ \citenamefont {Hupp}}]{Lee_2009:metal_organic}%
  \BibitemOpen
  \bibfield  {author} {\bibinfo {author} {\bibfnamefont {J.~Y.}\ \bibnamefont
  {Lee}}, \bibinfo {author} {\bibfnamefont {O.}~\bibnamefont {Farha}}, \bibinfo
  {author} {\bibfnamefont {J.}~\bibnamefont {Roberts}}, \bibinfo {author}
  {\bibfnamefont {K.~A.}\ \bibnamefont {Scheidt}}, \bibinfo {author}
  {\bibfnamefont {S.~T.}\ \bibnamefont {Nguyen}}, \ and\ \bibinfo {author}
  {\bibfnamefont {J.~T.}\ \bibnamefont {Hupp}},\ }\href {\doibase 10.1039/B807080F} {\bibfield  {journal} {\bibinfo  {journal} {Chem. Soc.
  Rev.}\ }\textbf {\bibinfo {volume} {38}},\ \bibinfo {pages} {1450} (\bibinfo
  {year} {2009})}\BibitemShut {NoStop}%
\bibitem [{\citenamefont {Luz}\ \emph {et~al.}(2010)\citenamefont {Luz},
  \citenamefont {Llabr\'{e}s~i Xamena},\ and\ \citenamefont
  {Corma}}]{Luz_2010:bridging_homogeneous}%
  \BibitemOpen
  \bibfield  {author} {\bibinfo {author} {\bibfnamefont {I.}~\bibnamefont
  {Luz}}, \bibinfo {author} {\bibfnamefont {F.~X.}\ \bibnamefont {Llabr\'{e}s~i
  Xamena}}, \ and\ \bibinfo {author} {\bibfnamefont {A.}~\bibnamefont
  {Corma}},\ }\href {\doibase 10.1016/j.jcat.2010.09.010} {\bibfield  {journal}
  {\bibinfo  {journal} {J. Catal.}\ }\textbf {\bibinfo {volume} {276}},\
  \bibinfo {pages} {134} (\bibinfo {year} {2010})}\BibitemShut {NoStop}%
\bibitem [{\citenamefont {Uemura}\ \emph {et~al.}(2009)\citenamefont {Uemura},
  \citenamefont {Yanai},\ and\ \citenamefont
  {Kitagawa}}]{Uemura_2009:polymerization_reactions}%
  \BibitemOpen
  \bibfield  {author} {\bibinfo {author} {\bibfnamefont {T.}~\bibnamefont
  {Uemura}}, \bibinfo {author} {\bibfnamefont {N.}~\bibnamefont {Yanai}}, \
  and\ \bibinfo {author} {\bibfnamefont {S.}~\bibnamefont {Kitagawa}},\ }\href
  {\doibase 10.1039/B802583P} {\bibfield  {journal} {\bibinfo  {journal} {Chem.
  Soc. Rev.}\ }\textbf {\bibinfo {volume} {38}},\ \bibinfo {pages} {1228}
  (\bibinfo {year} {2009})}\BibitemShut {NoStop}%
\bibitem [{\citenamefont {Vitorino}\ \emph {et~al.}(2009)\citenamefont
  {Vitorino}, \citenamefont {Devic}, \citenamefont {Tromp}, \citenamefont
  {F\'{e}rey},\ and\ \citenamefont
  {Visseaux}}]{Vitorino_2009:lanthanide_metal}%
  \BibitemOpen
  \bibfield  {author} {\bibinfo {author} {\bibfnamefont {M.~J.}\ \bibnamefont
  {Vitorino}}, \bibinfo {author} {\bibfnamefont {T.}~\bibnamefont {Devic}},
  \bibinfo {author} {\bibfnamefont {M.}~\bibnamefont {Tromp}}, \bibinfo
  {author} {\bibfnamefont {G.}~\bibnamefont {F\'{e}rey}}, \ and\ \bibinfo
  {author} {\bibfnamefont {M.}~\bibnamefont {Visseaux}},\ }\href {\doibase 10.1002/macp.200900354} {\bibfield  {journal} {\bibinfo  {journal} {Macromol.
  Chem. Phys.}\ }\textbf {\bibinfo {volume} {210}},\ \bibinfo {pages} {1923}
  (\bibinfo {year} {2009})}\BibitemShut {NoStop}%
\bibitem [{\citenamefont {Allendorf}\ \emph {et~al.}(2009)\citenamefont
  {Allendorf}, \citenamefont {Bauer}, \citenamefont {Bhakta},\ and\
  \citenamefont {Houk}}]{Allendorf_2009:luminescent_metal}%
  \BibitemOpen
  \bibfield  {author} {\bibinfo {author} {\bibfnamefont {M.~D.}\ \bibnamefont
  {Allendorf}}, \bibinfo {author} {\bibfnamefont {C.~A.}\ \bibnamefont
  {Bauer}}, \bibinfo {author} {\bibfnamefont {R.~K.}\ \bibnamefont {Bhakta}}, \
  and\ \bibinfo {author} {\bibfnamefont {R.}~\bibnamefont {Houk}},\ }\href
  {\doibase 10.1039/B802352M} {\bibfield  {journal} {\bibinfo  {journal} {Chem.
  Soc. Rev.}\ }\textbf {\bibinfo {volume} {38}},\ \bibinfo {pages} {1330}
  (\bibinfo {year} {2009})}\BibitemShut {NoStop}%
\bibitem [{\citenamefont {White}\ \emph {et~al.}(2009)\citenamefont {White},
  \citenamefont {Chengelis}, \citenamefont {Gogick}, \citenamefont {Stehman},
  \citenamefont {Rosi},\ and\ \citenamefont
  {Petoud}}]{White_2009:near-infrared_luminescent}%
  \BibitemOpen
  \bibfield  {author} {\bibinfo {author} {\bibfnamefont {K.~A.}\ \bibnamefont
  {White}}, \bibinfo {author} {\bibfnamefont {D.~A.}\ \bibnamefont
  {Chengelis}}, \bibinfo {author} {\bibfnamefont {K.~A.}\ \bibnamefont
  {Gogick}}, \bibinfo {author} {\bibfnamefont {J.}~\bibnamefont {Stehman}},
  \bibinfo {author} {\bibfnamefont {N.~L.}\ \bibnamefont {Rosi}}, \ and\
  \bibinfo {author} {\bibfnamefont {S.}~\bibnamefont {Petoud}},\ }\href
  {\doibase 10.1021/ja907885m} {\bibfield  {journal} {\bibinfo  {journal} {J.
  Am. Chem. Soc.}\ }\textbf {\bibinfo {volume} {131}},\ \bibinfo {pages}
  {18069} (\bibinfo {year} {2009})}\BibitemShut {NoStop}%
\bibitem [{\citenamefont {Bordiga}\ \emph {et~al.}(2004)\citenamefont
  {Bordiga}, \citenamefont {Lamberti}, \citenamefont {Ricchiardi},
  \citenamefont {Regli}, \citenamefont {Bonino}, \citenamefont {Damin},
  \citenamefont {Lillerud}, \citenamefont {Bjorgen},\ and\ \citenamefont
  {Zecchina}}]{Bordiga_2004:electronic_vibrational}%
  \BibitemOpen
  \bibfield  {author} {\bibinfo {author} {\bibfnamefont {S.}~\bibnamefont
  {Bordiga}}, \bibinfo {author} {\bibfnamefont {C.}~\bibnamefont {Lamberti}},
  \bibinfo {author} {\bibfnamefont {G.}~\bibnamefont {Ricchiardi}}, \bibinfo
  {author} {\bibfnamefont {L.}~\bibnamefont {Regli}}, \bibinfo {author}
  {\bibfnamefont {F.}~\bibnamefont {Bonino}}, \bibinfo {author} {\bibfnamefont
  {A.}~\bibnamefont {Damin}}, \bibinfo {author} {\bibfnamefont {K.-P.}\
  \bibnamefont {Lillerud}}, \bibinfo {author} {\bibfnamefont {M.}~\bibnamefont
  {Bjorgen}}, \ and\ \bibinfo {author} {\bibfnamefont {A.}~\bibnamefont
  {Zecchina}},\ }\href {\doibase 10.1039/B407246D} {\bibfield  {journal}
  {\bibinfo  {journal} {Chem. Commun.}\ ,\ \bibinfo {pages} {2300}} (\bibinfo
  {year} {2004})}\BibitemShut {NoStop}%
\bibitem [{\citenamefont {Kurmoo}(2009)}]{Kurmoo_2009:magnetic_metal-organic}%
  \BibitemOpen
  \bibfield  {author} {\bibinfo {author} {\bibfnamefont {M.}~\bibnamefont
  {Kurmoo}},\ }\href {\doibase 10.1039/B804757J} {\bibfield  {journal}
  {\bibinfo  {journal} {Chem. Soc. Rev.}\ }\textbf {\bibinfo {volume} {38}},\
  \bibinfo {pages} {1353} (\bibinfo {year} {2009})}\BibitemShut {NoStop}%
\bibitem [{\citenamefont {Horcajada}\ \emph {et~al.}(2010)\citenamefont
  {Horcajada}, \citenamefont {Chalati}, \citenamefont {Serre}, \citenamefont
  {Gillet}, \citenamefont {Sebrie}, \citenamefont {Baati}, \citenamefont
  {Eubank}, \citenamefont {Heurtaux}, \citenamefont {Clayette}, \citenamefont
  {Kreuz}, \citenamefont {Chang}, \citenamefont {Hwang}, \citenamefont
  {Marsaud}, \citenamefont {Bories}, \citenamefont {Cynober}, \citenamefont
  {Gil}, \citenamefont {F\'{e}rey}, \citenamefont {Couvreur},\ and\
  \citenamefont {Gref}}]{Horcajada_2010:porous_metal-organic-framework}%
  \BibitemOpen
  \bibfield  {author} {\bibinfo {author} {\bibfnamefont {P.}~\bibnamefont
  {Horcajada}}, \bibinfo {author} {\bibfnamefont {T.}~\bibnamefont {Chalati}},
  \bibinfo {author} {\bibfnamefont {C.}~\bibnamefont {Serre}}, \bibinfo
  {author} {\bibfnamefont {B.}~\bibnamefont {Gillet}}, \bibinfo {author}
  {\bibfnamefont {C.}~\bibnamefont {Sebrie}}, \bibinfo {author} {\bibfnamefont
  {T.}~\bibnamefont {Baati}}, \bibinfo {author} {\bibfnamefont
  {J.}~\bibnamefont {Eubank}}, \bibinfo {author} {\bibfnamefont
  {D.}~\bibnamefont {Heurtaux}}, \bibinfo {author} {\bibfnamefont
  {P.}~\bibnamefont {Clayette}}, \bibinfo {author} {\bibfnamefont
  {C.}~\bibnamefont {Kreuz}}, \bibinfo {author} {\bibfnamefont {J.-S.}\
  \bibnamefont {Chang}}, \bibinfo {author} {\bibfnamefont {Y.}~\bibnamefont
  {Hwang}}, \bibinfo {author} {\bibfnamefont {V.}~\bibnamefont {Marsaud}},
  \bibinfo {author} {\bibfnamefont {P.-N.}\ \bibnamefont {Bories}}, \bibinfo
  {author} {\bibfnamefont {L.}~\bibnamefont {Cynober}}, \bibinfo {author}
  {\bibfnamefont {S.}~\bibnamefont {Gil}}, \bibinfo {author} {\bibfnamefont
  {G.}~\bibnamefont {F\'{e}rey}}, \bibinfo {author} {\bibfnamefont
  {P.}~\bibnamefont {Couvreur}}, \ and\ \bibinfo {author} {\bibfnamefont
  {R.}~\bibnamefont {Gref}},\ }\href {\doibase 10.1038/nmat2608} {\bibfield
  {journal} {\bibinfo  {journal} {Nat. Mater.}\ }\textbf {\bibinfo {volume}
  {9}},\ \bibinfo {pages} {172} (\bibinfo {year} {2010})}\BibitemShut {NoStop}%
\bibitem [{\citenamefont {Stroppa}\ \emph {et~al.}(2011)\citenamefont
  {Stroppa}, \citenamefont {Jain}, \citenamefont {Barone}, \citenamefont
  {Marsman}, \citenamefont {Perez-Mato}, \citenamefont {Cheetham},
  \citenamefont {Kroto},\ and\ \citenamefont
  {Picozzi}}]{Stroppa_2011:electric_control}%
  \BibitemOpen
  \bibfield  {author} {\bibinfo {author} {\bibfnamefont {A.}~\bibnamefont
  {Stroppa}}, \bibinfo {author} {\bibfnamefont {P.}~\bibnamefont {Jain}},
  \bibinfo {author} {\bibfnamefont {P.}~\bibnamefont {Barone}}, \bibinfo
  {author} {\bibfnamefont {M.}~\bibnamefont {Marsman}}, \bibinfo {author}
  {\bibfnamefont {J.~M.}\ \bibnamefont {Perez-Mato}}, \bibinfo {author}
  {\bibfnamefont {A.~K.}\ \bibnamefont {Cheetham}}, \bibinfo {author}
  {\bibfnamefont {H.~W.}\ \bibnamefont {Kroto}}, \ and\ \bibinfo {author}
  {\bibfnamefont {S.}~\bibnamefont {Picozzi}},\ }\href {\doibase 10.1002/anie.201101405} {\bibfield  {journal} {\bibinfo  {journal} {Angew.
  Chem., Int. Ed.}\ }\textbf {\bibinfo {volume} {50}},\ \bibinfo {pages} {5847}
  (\bibinfo {year} {2011})}\BibitemShut {NoStop}%
\bibitem [{\citenamefont {Stroppa}\ \emph {et~al.}(2013)\citenamefont
  {Stroppa}, \citenamefont {Barone}, \citenamefont {Jain}, \citenamefont
  {Perez-Mato},\ and\ \citenamefont {Picozzi}}]{Stroppa_2013:hybrid_improper}%
  \BibitemOpen
  \bibfield  {author} {\bibinfo {author} {\bibfnamefont {A.}~\bibnamefont
  {Stroppa}}, \bibinfo {author} {\bibfnamefont {P.}~\bibnamefont {Barone}},
  \bibinfo {author} {\bibfnamefont {P.}~\bibnamefont {Jain}}, \bibinfo {author}
  {\bibfnamefont {J.~M.}\ \bibnamefont {Perez-Mato}}, \ and\ \bibinfo {author}
  {\bibfnamefont {S.}~\bibnamefont {Picozzi}},\ }\href {\doibase 10.1002/adma.201204738} {\bibfield  {journal} {\bibinfo  {journal} {Adv.
  Mater.}\ }\textbf {\bibinfo {volume} {25}},\ \bibinfo {pages} {2284}
  (\bibinfo {year} {2013})}\BibitemShut {NoStop}%
\bibitem [{\citenamefont {Di~Sante}\ \emph {et~al.}(2013)\citenamefont
  {Di~Sante}, \citenamefont {Stroppa}, \citenamefont {Jain},\ and\
  \citenamefont {Picozzi}}]{Di-Sante_2013:tuning_ferroelectric}%
  \BibitemOpen
  \bibfield  {author} {\bibinfo {author} {\bibfnamefont {D.}~\bibnamefont
  {Di~Sante}}, \bibinfo {author} {\bibfnamefont {A.}~\bibnamefont {Stroppa}},
  \bibinfo {author} {\bibfnamefont {P.}~\bibnamefont {Jain}}, \ and\ \bibinfo
  {author} {\bibfnamefont {S.}~\bibnamefont {Picozzi}},\ }\href {\doibase 10.1021/ja408283a} {\bibfield  {journal} {\bibinfo  {journal} {J. Am. Chem.
  Soc.}\ }\textbf {\bibinfo {volume} {135}},\ \bibinfo {pages} {18126}
  (\bibinfo {year} {2013})}\BibitemShut {NoStop}%
\bibitem [{\citenamefont {Serre}\ \emph {et~al.}(2007)\citenamefont {Serre},
  \citenamefont {Mellot-Draznieks}, \citenamefont {Surbl\'{e}}, \citenamefont
  {Audebrand}, \citenamefont {Filinchuck},\ and\ \citenamefont
  {F\'{e}rey}}]{Serre_2007:role_solvent-host}%
  \BibitemOpen
  \bibfield  {author} {\bibinfo {author} {\bibfnamefont {C.}~\bibnamefont
  {Serre}}, \bibinfo {author} {\bibfnamefont {C.}~\bibnamefont
  {Mellot-Draznieks}}, \bibinfo {author} {\bibfnamefont {S.}~\bibnamefont
  {Surbl\'{e}}}, \bibinfo {author} {\bibfnamefont {N.}~\bibnamefont
  {Audebrand}}, \bibinfo {author} {\bibfnamefont {Y.}~\bibnamefont
  {Filinchuck}}, \ and\ \bibinfo {author} {\bibfnamefont {G.}~\bibnamefont
  {F\'{e}rey}},\ }\href {\doibase 10.1126/science.1137975} {\bibfield
  {journal} {\bibinfo  {journal} {Science}\ }\textbf {\bibinfo {volume}
  {315}},\ \bibinfo {pages} {1828} (\bibinfo {year} {2007})}\BibitemShut
  {NoStop}%
\bibitem [{\citenamefont {Allendorf}\ \emph {et~al.}(2008)\citenamefont
  {Allendorf}, \citenamefont {Houk}, \citenamefont {Andruskiewicz},
  \citenamefont {Talin}, \citenamefont {Pikarsky}, \citenamefont {Choundhury},
  \citenamefont {Gall},\ and\ \citenamefont
  {Hensketh}}]{Allendorf_2008:stress-induced_chemical}%
  \BibitemOpen
  \bibfield  {author} {\bibinfo {author} {\bibfnamefont {M.~D.}\ \bibnamefont
  {Allendorf}}, \bibinfo {author} {\bibfnamefont {R.~J.~T.}\ \bibnamefont
  {Houk}}, \bibinfo {author} {\bibfnamefont {L.}~\bibnamefont {Andruskiewicz}},
  \bibinfo {author} {\bibfnamefont {A.~A.}\ \bibnamefont {Talin}}, \bibinfo
  {author} {\bibfnamefont {J.}~\bibnamefont {Pikarsky}}, \bibinfo {author}
  {\bibfnamefont {A.}~\bibnamefont {Choundhury}}, \bibinfo {author}
  {\bibfnamefont {K.~A.}\ \bibnamefont {Gall}}, \ and\ \bibinfo {author}
  {\bibfnamefont {P.~J.}\ \bibnamefont {Hensketh}},\ }\href {\doibase 10.1021/ja805235k} {\bibfield  {journal} {\bibinfo  {journal} {J. Am. Chem.
  Soc.}\ }\textbf {\bibinfo {volume} {130}},\ \bibinfo {pages} {14404}
  (\bibinfo {year} {2008})}\BibitemShut {NoStop}%
\bibitem [{\citenamefont {Tan}\ and\ \citenamefont
  {Cheetham}(2011)}]{Tan_2011:mechanical_properties}%
  \BibitemOpen
  \bibfield  {author} {\bibinfo {author} {\bibfnamefont {J.-C.}\ \bibnamefont
  {Tan}}\ and\ \bibinfo {author} {\bibfnamefont {A.~K.}\ \bibnamefont
  {Cheetham}},\ }\href {\doibase 10.1039/C0CS00163E} {\bibfield  {journal}
  {\bibinfo  {journal} {Chem. Soc. Rev.}\ }\textbf {\bibinfo {volume} {40}},\
  \bibinfo {pages} {1059} (\bibinfo {year} {2011})}\BibitemShut {NoStop}%
\bibitem [{\citenamefont {Kreno}\ \emph {et~al.}(2012)\citenamefont {Kreno},
  \citenamefont {Leong}, \citenamefont {Farha}, \citenamefont {Allendorf},
  \citenamefont {Van~Duyne},\ and\ \citenamefont
  {Hupp}}]{Kreno_2012:metal-organic_framework}%
  \BibitemOpen
  \bibfield  {author} {\bibinfo {author} {\bibfnamefont {L.}~\bibnamefont
  {Kreno}}, \bibinfo {author} {\bibfnamefont {K.}~\bibnamefont {Leong}},
  \bibinfo {author} {\bibfnamefont {O.}~\bibnamefont {Farha}}, \bibinfo
  {author} {\bibfnamefont {M.}~\bibnamefont {Allendorf}}, \bibinfo {author}
  {\bibfnamefont {R.}~\bibnamefont {Van~Duyne}}, \ and\ \bibinfo {author}
  {\bibfnamefont {J.}~\bibnamefont {Hupp}},\ }\href {\doibase 10.1021/cr200324t} {\bibfield  {journal} {\bibinfo  {journal} {Chem. Rev.}\
  }\textbf {\bibinfo {volume} {112}},\ \bibinfo {pages} {1105} (\bibinfo {year}
  {2012})}\BibitemShut {NoStop}%
\bibitem [{\citenamefont {Berland}\ \emph {et~al.}(2015)\citenamefont
  {Berland}, \citenamefont {Cooper}, \citenamefont {Lee}, \citenamefont
  {Schr\"{o}der}, \citenamefont {Thonhauser}, \citenamefont {Hyldgaard},\ and\
  \citenamefont {Lundqvist}}]{Berland_2015:van_waals}%
  \BibitemOpen
  \bibfield  {author} {\bibinfo {author} {\bibfnamefont {K.}~\bibnamefont
  {Berland}}, \bibinfo {author} {\bibfnamefont {V.~R.}\ \bibnamefont {Cooper}},
  \bibinfo {author} {\bibfnamefont {K.}~\bibnamefont {Lee}}, \bibinfo {author}
  {\bibfnamefont {E.}~\bibnamefont {Schr\"{o}der}}, \bibinfo {author}
  {\bibfnamefont {T.}~\bibnamefont {Thonhauser}}, \bibinfo {author}
  {\bibfnamefont {P.}~\bibnamefont {Hyldgaard}}, \ and\ \bibinfo {author}
  {\bibfnamefont {B.~I.}\ \bibnamefont {Lundqvist}},\ }\href {\doibase 10.1088/0034-4885/78/6/066501} {\bibfield  {journal} {\bibinfo  {journal}
  {Rep. Prog. Phys.}\ }\textbf {\bibinfo {volume} {78}},\ \bibinfo {pages}
  {066501} (\bibinfo {year} {2015})}\BibitemShut {NoStop}%
\bibitem [{\citenamefont {Dion}\ \emph {et~al.}(2004)\citenamefont {Dion},
  \citenamefont {Rydberg}, \citenamefont {Schr\"{o}der}, \citenamefont
  {Langreth},\ and\ \citenamefont {Lundqvist}}]{Dion_2004:van_waals}%
  \BibitemOpen
  \bibfield  {author} {\bibinfo {author} {\bibfnamefont {M.}~\bibnamefont
  {Dion}}, \bibinfo {author} {\bibfnamefont {H.}~\bibnamefont {Rydberg}},
  \bibinfo {author} {\bibfnamefont {E.}~\bibnamefont {Schr\"{o}der}}, \bibinfo
  {author} {\bibfnamefont {D.~C.}\ \bibnamefont {Langreth}}, \ and\ \bibinfo
  {author} {\bibfnamefont {B.~I.}\ \bibnamefont {Lundqvist}},\ }\href {\doibase 10.1103/PhysRevLett.92.246401} {\bibfield  {journal} {\bibinfo  {journal}
  {Phys. Rev. Lett.}\ }\textbf {\bibinfo {volume} {92}},\ \bibinfo {pages}
  {246401} (\bibinfo {year} {2004})}\BibitemShut {NoStop}%
\bibitem [{\citenamefont {Thonhauser}\ \emph {et~al.}(2007)\citenamefont
  {Thonhauser}, \citenamefont {Cooper}, \citenamefont {Li}, \citenamefont
  {Puzder}, \citenamefont {Hyldgaard},\ and\ \citenamefont
  {Langreth}}]{Thonhauser_2007:van_waals}%
  \BibitemOpen
  \bibfield  {author} {\bibinfo {author} {\bibfnamefont {T.}~\bibnamefont
  {Thonhauser}}, \bibinfo {author} {\bibfnamefont {V.~R.}\ \bibnamefont
  {Cooper}}, \bibinfo {author} {\bibfnamefont {S.}~\bibnamefont {Li}}, \bibinfo
  {author} {\bibfnamefont {A.}~\bibnamefont {Puzder}}, \bibinfo {author}
  {\bibfnamefont {P.}~\bibnamefont {Hyldgaard}}, \ and\ \bibinfo {author}
  {\bibfnamefont {D.~C.}\ \bibnamefont {Langreth}},\ }\href {\doibase 10.1103/PhysRevB.76.125112} {\bibfield  {journal} {\bibinfo  {journal} {Phys.
  Rev. B}\ }\textbf {\bibinfo {volume} {76}},\ \bibinfo {pages} {125112}
  (\bibinfo {year} {2007})}\BibitemShut {NoStop}%
\bibitem [{\citenamefont {Lee}\ \emph {et~al.}(2010)\citenamefont {Lee},
  \citenamefont {Murray}, \citenamefont {Kong}, \citenamefont {Lundqvist},\
  and\ \citenamefont {Langreth}}]{Lee_2010:higher-accuracy_van}%
  \BibitemOpen
  \bibfield  {author} {\bibinfo {author} {\bibfnamefont {K.}~\bibnamefont
  {Lee}}, \bibinfo {author} {\bibfnamefont {E.~D.}\ \bibnamefont {Murray}},
  \bibinfo {author} {\bibfnamefont {L.}~\bibnamefont {Kong}}, \bibinfo {author}
  {\bibfnamefont {B.~I.}\ \bibnamefont {Lundqvist}}, \ and\ \bibinfo {author}
  {\bibfnamefont {D.~C.}\ \bibnamefont {Langreth}},\ }\href {\doibase 10.1103/PhysRevB.82.081101} {\bibfield  {journal} {\bibinfo  {journal} {Phys.
  Rev. B}\ }\textbf {\bibinfo {volume} {82}},\ \bibinfo {pages} {081101}
  (\bibinfo {year} {2010})}\BibitemShut {NoStop}%
\bibitem [{\citenamefont {Berland}\ and\ \citenamefont
  {Hyldgaard}(2014)}]{Berland_2014:exchange_functional}%
  \BibitemOpen
  \bibfield  {author} {\bibinfo {author} {\bibfnamefont {K.}~\bibnamefont
  {Berland}}\ and\ \bibinfo {author} {\bibfnamefont {P.}~\bibnamefont
  {Hyldgaard}},\ }\href {\doibase 10.1103/PhysRevB.89.035412} {\bibfield
  {journal} {\bibinfo  {journal} {Phys. Rev. B}\ }\textbf {\bibinfo {volume}
  {89}},\ \bibinfo {pages} {035412} (\bibinfo {year} {2014})}\BibitemShut
  {NoStop}%
\bibitem [{\citenamefont {Perdew}\ and\ \citenamefont
  {Wang}(1992)}]{Perdew_1992:accurate_simple}%
  \BibitemOpen
  \bibfield  {author} {\bibinfo {author} {\bibfnamefont {J.~P.}\ \bibnamefont
  {Perdew}}\ and\ \bibinfo {author} {\bibfnamefont {Y.}~\bibnamefont {Wang}},\
  }\href {\doibase 10.1103/PhysRevB.45.13244} {\bibfield  {journal} {\bibinfo
  {journal} {Phys. Rev. B}\ }\textbf {\bibinfo {volume} {45}},\ \bibinfo
  {pages} {13244} (\bibinfo {year} {1992})}\BibitemShut {NoStop}%
\bibitem [{\citenamefont {Perdew}\ \emph {et~al.}(1996)\citenamefont {Perdew},
  \citenamefont {Burke},\ and\ \citenamefont
  {Ernzerhof}}]{Perdew_1996:generalized_gradient}%
  \BibitemOpen
  \bibfield  {author} {\bibinfo {author} {\bibfnamefont {J.~P.}\ \bibnamefont
  {Perdew}}, \bibinfo {author} {\bibfnamefont {K.}~\bibnamefont {Burke}}, \
  and\ \bibinfo {author} {\bibfnamefont {M.}~\bibnamefont {Ernzerhof}},\ }\href
  {\doibase 10.1103/PhysRevLett.77.3865} {\bibfield  {journal} {\bibinfo
  {journal} {Phys. Rev. Lett.}\ }\textbf {\bibinfo {volume} {77}},\ \bibinfo
  {pages} {3865} (\bibinfo {year} {1996})}\BibitemShut {NoStop}%
\bibitem [{\citenamefont {Berland}\ \emph {et~al.}(2014)\citenamefont
  {Berland}, \citenamefont {Arter}, \citenamefont {Cooper}, \citenamefont
  {Lee}, \citenamefont {Lundqvist}, \citenamefont {Schr\"{o}der}, \citenamefont
  {Thonhauser},\ and\ \citenamefont {Hyldgaard}}]{Berland_2014:van_waals}%
  \BibitemOpen
  \bibfield  {author} {\bibinfo {author} {\bibfnamefont {K.}~\bibnamefont
  {Berland}}, \bibinfo {author} {\bibfnamefont {C.~A.}\ \bibnamefont {Arter}},
  \bibinfo {author} {\bibfnamefont {V.~R.}\ \bibnamefont {Cooper}}, \bibinfo
  {author} {\bibfnamefont {K.}~\bibnamefont {Lee}}, \bibinfo {author}
  {\bibfnamefont {B.~I.}\ \bibnamefont {Lundqvist}}, \bibinfo {author}
  {\bibfnamefont {E.}~\bibnamefont {Schr\"{o}der}}, \bibinfo {author}
  {\bibfnamefont {T.}~\bibnamefont {Thonhauser}}, \ and\ \bibinfo {author}
  {\bibfnamefont {P.}~\bibnamefont {Hyldgaard}},\ }\href {\doibase 10.1063/1.4871731} {\bibfield  {journal} {\bibinfo  {journal} {J. Chem.
  Phys.}\ }\textbf {\bibinfo {volume} {140}},\ \bibinfo {pages} {18A539}
  (\bibinfo {year} {2014})}\BibitemShut {NoStop}%
\bibitem [{\citenamefont {Hyldgaard}\ \emph {et~al.}(2014)\citenamefont
  {Hyldgaard}, \citenamefont {Berland},\ and\ \citenamefont
  {Schr\"oder}}]{Hyldgaard_2014:interpretation_van}%
  \BibitemOpen
  \bibfield  {author} {\bibinfo {author} {\bibfnamefont {P.}~\bibnamefont
  {Hyldgaard}}, \bibinfo {author} {\bibfnamefont {K.}~\bibnamefont {Berland}},
  \ and\ \bibinfo {author} {\bibfnamefont {E.}~\bibnamefont {Schr\"oder}},\
  }\href {\doibase 10.1103/PhysRevB.90.075148} {\bibfield  {journal} {\bibinfo
  {journal} {Phys. Rev. B}\ }\textbf {\bibinfo {volume} {90}},\ \bibinfo
  {pages} {075148} (\bibinfo {year} {2014})}\BibitemShut {NoStop}%
\bibitem [{\citenamefont {Cooper}(2010)}]{Cooper_2010:van_waals}%
  \BibitemOpen
  \bibfield  {author} {\bibinfo {author} {\bibfnamefont {V.~R.}\ \bibnamefont
  {Cooper}},\ }\href {\doibase 10.1103/PhysRevB.81.161104} {\bibfield
  {journal} {\bibinfo  {journal} {Phys. Rev. B}\ }\textbf {\bibinfo {volume}
  {81}},\ \bibinfo {pages} {161104} (\bibinfo {year} {2010})}\BibitemShut
  {NoStop}%
\bibitem [{\citenamefont {Klime\v{s}}\ \emph {et~al.}(2010)\citenamefont
  {Klime\v{s}}, \citenamefont {Bowler},\ and\ \citenamefont
  {Michaelides}}]{Klimes_2010:chemical_accuracy}%
  \BibitemOpen
  \bibfield  {author} {\bibinfo {author} {\bibfnamefont {J.}~\bibnamefont
  {Klime\v{s}}}, \bibinfo {author} {\bibfnamefont {D.~R.}\ \bibnamefont
  {Bowler}}, \ and\ \bibinfo {author} {\bibfnamefont {A.}~\bibnamefont
  {Michaelides}},\ }\href {\doibase 10.1088/0953-8984/22/2/022201} {\bibfield
  {journal} {\bibinfo  {journal} {J. Phys. Condens. Matter}\ }\textbf {\bibinfo
  {volume} {22}},\ \bibinfo {pages} {022201} (\bibinfo {year}
  {2010})}\BibitemShut {NoStop}%
\bibitem [{\citenamefont {Klime\v{s}}\ \emph {et~al.}(2011)\citenamefont
  {Klime\v{s}}, \citenamefont {Bowler},\ and\ \citenamefont
  {Michaelides}}]{Klimes_2011:van_waals}%
  \BibitemOpen
  \bibfield  {author} {\bibinfo {author} {\bibfnamefont {J.}~\bibnamefont
  {Klime\v{s}}}, \bibinfo {author} {\bibfnamefont {D.~R.}\ \bibnamefont
  {Bowler}}, \ and\ \bibinfo {author} {\bibfnamefont {A.}~\bibnamefont
  {Michaelides}},\ }\href {\doibase 10.1103/PhysRevB.83.195131} {\bibfield
  {journal} {\bibinfo  {journal} {Phys. Rev. B}\ }\textbf {\bibinfo {volume}
  {83}},\ \bibinfo {pages} {195131} (\bibinfo {year} {2011})}\BibitemShut
  {NoStop}%
\bibitem [{\citenamefont {Wellendorff}\ \emph {et~al.}(2012)\citenamefont
  {Wellendorff}, \citenamefont {Lundgaard}, \citenamefont {M{\o}gelh{\o}j},
  \citenamefont {Petzold}, \citenamefont {Landis}, \citenamefont {N{\o}rskov},
  \citenamefont {Bligaard},\ and\ \citenamefont
  {Jacobsen}}]{Wellendorff_2012:density_functionals}%
  \BibitemOpen
  \bibfield  {author} {\bibinfo {author} {\bibfnamefont {J.}~\bibnamefont
  {Wellendorff}}, \bibinfo {author} {\bibfnamefont {K.~T.}\ \bibnamefont
  {Lundgaard}}, \bibinfo {author} {\bibfnamefont {A.}~\bibnamefont
  {M{\o}gelh{\o}j}}, \bibinfo {author} {\bibfnamefont {V.}~\bibnamefont
  {Petzold}}, \bibinfo {author} {\bibfnamefont {D.~D.}\ \bibnamefont {Landis}},
  \bibinfo {author} {\bibfnamefont {J.~K.}\ \bibnamefont {N{\o}rskov}},
  \bibinfo {author} {\bibfnamefont {T.}~\bibnamefont {Bligaard}}, \ and\
  \bibinfo {author} {\bibfnamefont {K.~W.}\ \bibnamefont {Jacobsen}},\ }\href
  {\doibase 10.1103/PhysRevB.85.235149} {\bibfield  {journal} {\bibinfo
  {journal} {Phys. Rev. B}\ }\textbf {\bibinfo {volume} {85}},\ \bibinfo
  {pages} {235149} (\bibinfo {year} {2012})}\BibitemShut {NoStop}%
\bibitem [{\citenamefont {Hamada}(2014)}]{Hamada_2014:van_waals}%
  \BibitemOpen
  \bibfield  {author} {\bibinfo {author} {\bibfnamefont {I.}~\bibnamefont
  {Hamada}},\ }\href {\doibase 10.1103/PhysRevB.89.121103} {\bibfield
  {journal} {\bibinfo  {journal} {Phys. Rev. B}\ }\textbf {\bibinfo {volume}
  {89}},\ \bibinfo {pages} {121103} (\bibinfo {year} {2014})}\BibitemShut
  {NoStop}%
\bibitem [{\citenamefont {Vydrov}\ \emph {et~al.}(2008)\citenamefont {Vydrov},
  \citenamefont {Wu},\ and\ \citenamefont {{Van
  Voorhis}}}]{Vydrov_2008:self-consistent_implementation}%
  \BibitemOpen
  \bibfield  {author} {\bibinfo {author} {\bibfnamefont {O.~A.}\ \bibnamefont
  {Vydrov}}, \bibinfo {author} {\bibfnamefont {Q.}~\bibnamefont {Wu}}, \ and\
  \bibinfo {author} {\bibfnamefont {T.}~\bibnamefont {{Van Voorhis}}},\ }\href
  {\doibase 10.1063/1.2948400} {\bibfield  {journal} {\bibinfo  {journal} {J.
  Chem. Phys.}\ }\textbf {\bibinfo {volume} {129}},\ \bibinfo {pages} {014106}
  (\bibinfo {year} {2008})}\BibitemShut {NoStop}%
\bibitem [{\citenamefont {Vydrov}\ and\ \citenamefont {{Van
  Voorhis}}(2009)}]{Vydrov_2009:nonlocal_van}%
  \BibitemOpen
  \bibfield  {author} {\bibinfo {author} {\bibfnamefont {O.~A.}\ \bibnamefont
  {Vydrov}}\ and\ \bibinfo {author} {\bibfnamefont {T.}~\bibnamefont {{Van
  Voorhis}}},\ }\href {\doibase 10.1103/PhysRevLett.103.063004} {\bibfield
  {journal} {\bibinfo  {journal} {Phys. Rev. Lett.}\ }\textbf {\bibinfo
  {volume} {103}},\ \bibinfo {pages} {063004} (\bibinfo {year}
  {2009})}\BibitemShut {NoStop}%
\bibitem [{\citenamefont {Vydrov}\ and\ \citenamefont {{Van
  Voorhis}}(2010)}]{Vydrov_2010:nonlocal_van}%
  \BibitemOpen
  \bibfield  {author} {\bibinfo {author} {\bibfnamefont {O.~A.}\ \bibnamefont
  {Vydrov}}\ and\ \bibinfo {author} {\bibfnamefont {T.}~\bibnamefont {{Van
  Voorhis}}},\ }\href {\doibase 10.1063/1.3521275} {\bibfield  {journal}
  {\bibinfo  {journal} {J. Chem. Phys.}\ }\textbf {\bibinfo {volume} {133}},\
  \bibinfo {pages} {244103} (\bibinfo {year} {2010})}\BibitemShut {NoStop}%
\bibitem [{\citenamefont {Langreth}\ \emph {et~al.}(2009)\citenamefont
  {Langreth}, \citenamefont {Lundqvist}, \citenamefont {Chakarova-K\"{a}ck},
  \citenamefont {Cooper}, \citenamefont {Dion}, \citenamefont {Hyldgaard},
  \citenamefont {Kelkkanen}, \citenamefont {Kleis}, \citenamefont {Kong},
  \citenamefont {Li}, \citenamefont {Moses}, \citenamefont {Murray},
  \citenamefont {Puzder}, \citenamefont {Rydberg}, \citenamefont
  {Schr\"{o}der},\ and\ \citenamefont
  {Thonhauser}}]{Langreth_2009:density_functional}%
  \BibitemOpen
  \bibfield  {author} {\bibinfo {author} {\bibfnamefont {D.~C.}\ \bibnamefont
  {Langreth}}, \bibinfo {author} {\bibfnamefont {B.~I.}\ \bibnamefont
  {Lundqvist}}, \bibinfo {author} {\bibfnamefont {S.~D.}\ \bibnamefont
  {Chakarova-K\"{a}ck}}, \bibinfo {author} {\bibfnamefont {V.~R.}\ \bibnamefont
  {Cooper}}, \bibinfo {author} {\bibfnamefont {M.}~\bibnamefont {Dion}},
  \bibinfo {author} {\bibfnamefont {P.}~\bibnamefont {Hyldgaard}}, \bibinfo
  {author} {\bibfnamefont {A.}~\bibnamefont {Kelkkanen}}, \bibinfo {author}
  {\bibfnamefont {J.}~\bibnamefont {Kleis}}, \bibinfo {author} {\bibfnamefont
  {L.}~\bibnamefont {Kong}}, \bibinfo {author} {\bibfnamefont {S.}~\bibnamefont
  {Li}}, \bibinfo {author} {\bibfnamefont {P.~G.}\ \bibnamefont {Moses}},
  \bibinfo {author} {\bibfnamefont {E.~D.}\ \bibnamefont {Murray}}, \bibinfo
  {author} {\bibfnamefont {A.}~\bibnamefont {Puzder}}, \bibinfo {author}
  {\bibfnamefont {H.}~\bibnamefont {Rydberg}}, \bibinfo {author} {\bibfnamefont
  {E.}~\bibnamefont {Schr\"{o}der}}, \ and\ \bibinfo {author} {\bibfnamefont
  {T.}~\bibnamefont {Thonhauser}},\ }\href {\doibase 10.1088/0953-8984/21/8/084203} {\bibfield  {journal} {\bibinfo  {journal} {J.
  Phys. Condens. Matter}\ }\textbf {\bibinfo {volume} {21}},\ \bibinfo {pages}
  {084203} (\bibinfo {year} {2009})}\BibitemShut {NoStop}%
\bibitem [{\citenamefont {Poloni}\ \emph {et~al.}(2014)\citenamefont {Poloni},
  \citenamefont {Lee}, \citenamefont {Berger}, \citenamefont {Smit},\ and\
  \citenamefont {Neaton}}]{Poloni_2014:understanding_trends}%
  \BibitemOpen
  \bibfield  {author} {\bibinfo {author} {\bibfnamefont {R.}~\bibnamefont
  {Poloni}}, \bibinfo {author} {\bibfnamefont {K.}~\bibnamefont {Lee}},
  \bibinfo {author} {\bibfnamefont {R.~F.}\ \bibnamefont {Berger}}, \bibinfo
  {author} {\bibfnamefont {B.}~\bibnamefont {Smit}}, \ and\ \bibinfo {author}
  {\bibfnamefont {J.~B.}\ \bibnamefont {Neaton}},\ }\href {\doibase 10.1021/jz500202x} {\bibfield  {journal} {\bibinfo  {journal} {J. Phys. Chem.
  Lett.}\ }\textbf {\bibinfo {volume} {5}},\ \bibinfo {pages} {861} (\bibinfo
  {year} {2014})}\BibitemShut {NoStop}%
\bibitem [{\citenamefont {Lee}\ \emph {et~al.}(2014)\citenamefont {Lee},
  \citenamefont {Isley}, \citenamefont {Dzubak}, \citenamefont {Verma},
  \citenamefont {Stoneburner}, \citenamefont {Lin}, \citenamefont {Howe},
  \citenamefont {Bloch}, \citenamefont {Reed}, \citenamefont {Hudson},
  \citenamefont {Brown}, \citenamefont {Long}, \citenamefont {Neaton},
  \citenamefont {Smit}, \citenamefont {Cramer}, \citenamefont {Truhlar},\ and\
  \citenamefont {Gagliardi}}]{Lee_2014:design_metal-organic}%
  \BibitemOpen
  \bibfield  {author} {\bibinfo {author} {\bibfnamefont {K.}~\bibnamefont
  {Lee}}, \bibinfo {author} {\bibfnamefont {W.~C.}\ \bibnamefont {Isley}},
  \bibinfo {author} {\bibfnamefont {A.~L.}\ \bibnamefont {Dzubak}}, \bibinfo
  {author} {\bibfnamefont {P.}~\bibnamefont {Verma}}, \bibinfo {author}
  {\bibfnamefont {S.~J.}\ \bibnamefont {Stoneburner}}, \bibinfo {author}
  {\bibfnamefont {L.~C.}\ \bibnamefont {Lin}}, \bibinfo {author} {\bibfnamefont
  {J.~D.}\ \bibnamefont {Howe}}, \bibinfo {author} {\bibfnamefont {E.~D.}\
  \bibnamefont {Bloch}}, \bibinfo {author} {\bibfnamefont {D.~A.}\ \bibnamefont
  {Reed}}, \bibinfo {author} {\bibfnamefont {M.~R.}\ \bibnamefont {Hudson}},
  \bibinfo {author} {\bibfnamefont {C.~M.}\ \bibnamefont {Brown}}, \bibinfo
  {author} {\bibfnamefont {J.~R.}\ \bibnamefont {Long}}, \bibinfo {author}
  {\bibfnamefont {J.~B.}\ \bibnamefont {Neaton}}, \bibinfo {author}
  {\bibfnamefont {B.}~\bibnamefont {Smit}}, \bibinfo {author} {\bibfnamefont
  {C.~J.}\ \bibnamefont {Cramer}}, \bibinfo {author} {\bibfnamefont {D.~G.}\
  \bibnamefont {Truhlar}}, \ and\ \bibinfo {author} {\bibfnamefont
  {L.}~\bibnamefont {Gagliardi}},\ }\href {\doibase 10.1021/ja4102979}
  {\bibfield  {journal} {\bibinfo  {journal} {J. Am. Chem. Soc.}\ }\textbf
  {\bibinfo {volume} {136}},\ \bibinfo {pages} {698} (\bibinfo {year}
  {2014})}\BibitemShut {NoStop}%
\bibitem [{\citenamefont {Canepa}\ \emph
  {et~al.}(2013{\natexlab{a}})\citenamefont {Canepa}, \citenamefont {Nijem},
  \citenamefont {Chabal},\ and\ \citenamefont
  {Thonhauser}}]{Canepa_2013:diffusion_small}%
  \BibitemOpen
  \bibfield  {author} {\bibinfo {author} {\bibfnamefont {P.}~\bibnamefont
  {Canepa}}, \bibinfo {author} {\bibfnamefont {N.}~\bibnamefont {Nijem}},
  \bibinfo {author} {\bibfnamefont {Y.~J.}\ \bibnamefont {Chabal}}, \ and\
  \bibinfo {author} {\bibfnamefont {T.}~\bibnamefont {Thonhauser}},\ }\href
  {\doibase 10.1103/PhysRevLett.110.026102} {\bibfield  {journal} {\bibinfo
  {journal} {Phys. Rev. Lett.}\ }\textbf {\bibinfo {volume} {110}},\ \bibinfo
  {pages} {026102} (\bibinfo {year} {2013}{\natexlab{a}})}\BibitemShut
  {NoStop}%
\bibitem [{\citenamefont {Canepa}\ \emph
  {et~al.}(2013{\natexlab{b}})\citenamefont {Canepa}, \citenamefont {Arter},
  \citenamefont {Conwill}, \citenamefont {Johnson}, \citenamefont {Shoemaker},
  \citenamefont {Soliman},\ and\ \citenamefont
  {Thonhauser}}]{Canepa_2013:high-throughput_screening}%
  \BibitemOpen
  \bibfield  {author} {\bibinfo {author} {\bibfnamefont {P.}~\bibnamefont
  {Canepa}}, \bibinfo {author} {\bibfnamefont {C.~A.}\ \bibnamefont {Arter}},
  \bibinfo {author} {\bibfnamefont {E.~M.}\ \bibnamefont {Conwill}}, \bibinfo
  {author} {\bibfnamefont {D.~H.}\ \bibnamefont {Johnson}}, \bibinfo {author}
  {\bibfnamefont {B.~A.}\ \bibnamefont {Shoemaker}}, \bibinfo {author}
  {\bibfnamefont {K.~Z.}\ \bibnamefont {Soliman}}, \ and\ \bibinfo {author}
  {\bibfnamefont {T.}~\bibnamefont {Thonhauser}},\ }\href {\doibase 10.1039/c3ta12395b} {\bibfield  {journal} {\bibinfo  {journal} {J. Mater.
  Chem. A}\ }\textbf {\bibinfo {volume} {1}},\ \bibinfo {pages} {13597}
  (\bibinfo {year} {2013}{\natexlab{b}})}\BibitemShut {NoStop}%
\bibitem [{\citenamefont {Nijem}\ \emph {et~al.}(2013)\citenamefont {Nijem},
  \citenamefont {Canepa}, \citenamefont {Kaipa}, \citenamefont {Tan},
  \citenamefont {Roodenko}, \citenamefont {Tekarli}, \citenamefont {Halbert},
  \citenamefont {Oswald}, \citenamefont {Arvapally}, \citenamefont {Yang},
  \citenamefont {Thonhauser}, \citenamefont {Omary},\ and\ \citenamefont
  {Chabal}}]{Nijem_2013:water_cluster}%
  \BibitemOpen
  \bibfield  {author} {\bibinfo {author} {\bibfnamefont {N.}~\bibnamefont
  {Nijem}}, \bibinfo {author} {\bibfnamefont {P.}~\bibnamefont {Canepa}},
  \bibinfo {author} {\bibfnamefont {U.}~\bibnamefont {Kaipa}}, \bibinfo
  {author} {\bibfnamefont {K.}~\bibnamefont {Tan}}, \bibinfo {author}
  {\bibfnamefont {K.}~\bibnamefont {Roodenko}}, \bibinfo {author}
  {\bibfnamefont {S.}~\bibnamefont {Tekarli}}, \bibinfo {author} {\bibfnamefont
  {J.}~\bibnamefont {Halbert}}, \bibinfo {author} {\bibfnamefont {I.~W.~H.}\
  \bibnamefont {Oswald}}, \bibinfo {author} {\bibfnamefont {R.~K.}\
  \bibnamefont {Arvapally}}, \bibinfo {author} {\bibfnamefont {C.}~\bibnamefont
  {Yang}}, \bibinfo {author} {\bibfnamefont {T.}~\bibnamefont {Thonhauser}},
  \bibinfo {author} {\bibfnamefont {M.~A.}\ \bibnamefont {Omary}}, \ and\
  \bibinfo {author} {\bibfnamefont {Y.~J.}\ \bibnamefont {Chabal}},\ }\href
  {\doibase 10.1021/ja400754p} {\bibfield  {journal} {\bibinfo  {journal} {J.
  Am. Chem. Soc.}\ }\textbf {\bibinfo {volume} {135}},\ \bibinfo {pages}
  {12615} (\bibinfo {year} {2013})}\BibitemShut {NoStop}%
\bibitem [{\citenamefont {Tan}\ \emph {et~al.}(2014)\citenamefont {Tan},
  \citenamefont {Zuluaga}, \citenamefont {Gong}, \citenamefont {Canepa},
  \citenamefont {Wang}, \citenamefont {Li}, \citenamefont {Chabal},\ and\
  \citenamefont {Thonhauser}}]{Tan_2014:water_reaction}%
  \BibitemOpen
  \bibfield  {author} {\bibinfo {author} {\bibfnamefont {K.}~\bibnamefont
  {Tan}}, \bibinfo {author} {\bibfnamefont {S.}~\bibnamefont {Zuluaga}},
  \bibinfo {author} {\bibfnamefont {Q.}~\bibnamefont {Gong}}, \bibinfo {author}
  {\bibfnamefont {P.}~\bibnamefont {Canepa}}, \bibinfo {author} {\bibfnamefont
  {H.}~\bibnamefont {Wang}}, \bibinfo {author} {\bibfnamefont {J.}~\bibnamefont
  {Li}}, \bibinfo {author} {\bibfnamefont {Y.~J.}\ \bibnamefont {Chabal}}, \
  and\ \bibinfo {author} {\bibfnamefont {T.}~\bibnamefont {Thonhauser}},\
  }\href {\doibase 10.1021/cm5038183} {\bibfield  {journal} {\bibinfo
  {journal} {Chem. Mater.}\ }\textbf {\bibinfo {volume} {26}},\ \bibinfo
  {pages} {6886} (\bibinfo {year} {2014})}\BibitemShut {NoStop}%
\bibitem [{\citenamefont {Zuluaga}\ \emph {et~al.}(2014)\citenamefont
  {Zuluaga}, \citenamefont {Canepa}, \citenamefont {Tan}, \citenamefont
  {Chabal},\ and\ \citenamefont {Thonhauser}}]{Zuluaga_2014:study_van}%
  \BibitemOpen
  \bibfield  {author} {\bibinfo {author} {\bibfnamefont {S.}~\bibnamefont
  {Zuluaga}}, \bibinfo {author} {\bibfnamefont {P.}~\bibnamefont {Canepa}},
  \bibinfo {author} {\bibfnamefont {K.}~\bibnamefont {Tan}}, \bibinfo {author}
  {\bibfnamefont {Y.~J.}\ \bibnamefont {Chabal}}, \ and\ \bibinfo {author}
  {\bibfnamefont {T.}~\bibnamefont {Thonhauser}},\ }\href {\doibase 10.1088/0953-8984/26/13/133002} {\bibfield  {journal} {\bibinfo  {journal}
  {J. Phys. Condens. Matter}\ }\textbf {\bibinfo {volume} {26}},\ \bibinfo
  {pages} {133002} (\bibinfo {year} {2014})}\BibitemShut {NoStop}%
\bibitem [{\citenamefont {Dediu}\ \emph {et~al.}(2009)\citenamefont {Dediu},
  \citenamefont {Hueso}, \citenamefont {Bergenti},\ and\ \citenamefont
  {Taliani}}]{Dediu_2009:spin_routes}%
  \BibitemOpen
  \bibfield  {author} {\bibinfo {author} {\bibfnamefont {V.~A.}\ \bibnamefont
  {Dediu}}, \bibinfo {author} {\bibfnamefont {L.~E.}\ \bibnamefont {Hueso}},
  \bibinfo {author} {\bibfnamefont {I.}~\bibnamefont {Bergenti}}, \ and\
  \bibinfo {author} {\bibfnamefont {C.}~\bibnamefont {Taliani}},\ }\href
  {\doibase 10.1038/nmat2510} {\bibfield  {journal} {\bibinfo  {journal} {Nat.
  Mater.}\ }\textbf {\bibinfo {volume} {8}},\ \bibinfo {pages} {707} (\bibinfo
  {year} {2009})}\BibitemShut {NoStop}%
\bibitem [{\citenamefont {Musia\l{}}\ and\ \citenamefont
  {Kucharski}(2014)}]{Musial_2014:first_principle}%
  \BibitemOpen
  \bibfield  {author} {\bibinfo {author} {\bibfnamefont {M.}~\bibnamefont
  {Musia\l{}}}\ and\ \bibinfo {author} {\bibfnamefont {S.~A.}\ \bibnamefont
  {Kucharski}},\ }\href {\doibase 10.1021/ct401076e} {\bibfield  {journal}
  {\bibinfo  {journal} {J. Chem. Theory Comp.}\ }\textbf {\bibinfo {volume}
  {10}},\ \bibinfo {pages} {1200} (\bibinfo {year} {2014})}\BibitemShut
  {NoStop}%
\bibitem [{\citenamefont {Sipahi}\ \emph {et~al.}(2014)\citenamefont {Sipahi},
  \citenamefont {Zutic}, \citenamefont {Atodiresei}, \citenamefont {Kawakami},\
  and\ \citenamefont {Lazic}}]{Sipahi_2014:spin_polarization}%
  \BibitemOpen
  \bibfield  {author} {\bibinfo {author} {\bibfnamefont {G.~M.}\ \bibnamefont
  {Sipahi}}, \bibinfo {author} {\bibfnamefont {I.}~\bibnamefont {Zutic}},
  \bibinfo {author} {\bibfnamefont {N.}~\bibnamefont {Atodiresei}}, \bibinfo
  {author} {\bibfnamefont {R.~W.}\ \bibnamefont {Kawakami}}, \ and\ \bibinfo
  {author} {\bibfnamefont {P.}~\bibnamefont {Lazic}},\ }\href {\doibase 10.1088/0953-8984/26/10/104204} {\bibfield  {journal} {\bibinfo  {journal}
  {J. Phys.: Condens. Matter}\ }\textbf {\bibinfo {volume} {26}},\ \bibinfo
  {pages} {104204} (\bibinfo {year} {2014})}\BibitemShut {NoStop}%
\bibitem [{\citenamefont {Gunnarsson}\ \emph {et~al.}(1974)\citenamefont
  {Gunnarsson}, \citenamefont {Lundqvist},\ and\ \citenamefont
  {Wilkins}}]{Gunnarsson_1974:contribution_cohesive}%
  \BibitemOpen
  \bibfield  {author} {\bibinfo {author} {\bibfnamefont {O.}~\bibnamefont
  {Gunnarsson}}, \bibinfo {author} {\bibfnamefont {B.~I.}\ \bibnamefont
  {Lundqvist}}, \ and\ \bibinfo {author} {\bibfnamefont {J.~W.}\ \bibnamefont
  {Wilkins}},\ }\href {\doibase 10.1103/PhysRevB.10.1319} {\bibfield  {journal}
  {\bibinfo  {journal} {Phys. Rev. B}\ }\textbf {\bibinfo {volume} {10}},\
  \bibinfo {pages} {1319} (\bibinfo {year} {1974})}\BibitemShut {NoStop}%
\bibitem [{Tro()}]{Troy}%
  \BibitemOpen
  \href@noop {} {}\bibinfo {note} {The functionals VV09
  \cite{Vydrov_2009:nonlocal_van} and VV10 \cite{Vydrov_2010:nonlocal_van},
  while related to vdW-DF, follow a different, simpler philosophy in defining
  the nonlocal correlation kernel. Spin does not affect the plasmon dispersion
  in VV09, and the VV09 strategy cannot be used for including spin in the
  vdW-DF method. Also note that no proper spin formulation exists for VV10 and
  the VV10 implementations use a spin-balancing procedure \cite{balancing} to
  compute $E_c^\nl$.}\BibitemShut {Stop}%
\bibitem [{rea()}]{real_space_code}%
  \BibitemOpen
  \href@noop {} {}\bibinfo {note} {VV09 \cite{Vydrov_2009:nonlocal_van} is only
  implemented in the real-space code Q-Chem \cite{Shao_2006:advances_methods}
  and a direct comparison to svdW-DF results for periodic crystals such as MOFs
  is thus not possible.}\BibitemShut {Stop}%
\bibitem [{bal()}]{balancing}%
  \BibitemOpen
  \href@noop {} {}\bibinfo {note} {Some DFT codes permit a pragmatic adaption
  of nonlocal functionals to spin systems, namely computing $E_c^\nl$ based on
  the total electron density, leaving $E_{xc}^0[\nup,\ndown]$ to reflect all
  spin effects. This spin-balanced $E_c^\nl$ evaluation constitutes an
  uncontrolled approximation, ignoring that spin changes the plasmon
  dispersion.}\BibitemShut {Stop}%
\bibitem [{\citenamefont {Ziambaras}\ \emph {et~al.}(2007)\citenamefont
  {Ziambaras}, \citenamefont {Kleis}, \citenamefont {Schr\"oder},\ and\
  \citenamefont {Hyldgaard}}]{Ziambaras_2007:potassium_intercalation}%
  \BibitemOpen
  \bibfield  {author} {\bibinfo {author} {\bibfnamefont {E.}~\bibnamefont
  {Ziambaras}}, \bibinfo {author} {\bibfnamefont {J.}~\bibnamefont {Kleis}},
  \bibinfo {author} {\bibfnamefont {E.}~\bibnamefont {Schr\"oder}}, \ and\
  \bibinfo {author} {\bibfnamefont {P.}~\bibnamefont {Hyldgaard}},\ }\href
  {\doibase 10.1103/PhysRevB.76.155425} {\bibfield  {journal} {\bibinfo
  {journal} {Phys. Rev. B}\ }\textbf {\bibinfo {volume} {76}},\ \bibinfo
  {pages} {155425} (\bibinfo {year} {2007})}\BibitemShut {NoStop}%
\bibitem [{\citenamefont {Obata}\ \emph {et~al.}(2013)\citenamefont {Obata},
  \citenamefont {Nakamura}, \citenamefont {Hamada},\ and\ \citenamefont
  {Oda}}]{Obata_2013:implementation_van}%
  \BibitemOpen
  \bibfield  {author} {\bibinfo {author} {\bibfnamefont {M.}~\bibnamefont
  {Obata}}, \bibinfo {author} {\bibfnamefont {M.}~\bibnamefont {Nakamura}},
  \bibinfo {author} {\bibfnamefont {I.}~\bibnamefont {Hamada}}, \ and\ \bibinfo
  {author} {\bibfnamefont {T.}~\bibnamefont {Oda}},\ }\href {\doibase 10.7566/JPSJ.82.093701} {\bibfield  {journal} {\bibinfo  {journal} {J. Phys.
  Soc. Jpn.}\ }\textbf {\bibinfo {volume} {82}},\ \bibinfo {pages} {093701}
  (\bibinfo {year} {2013})}\BibitemShut {NoStop}%
\bibitem [{\citenamefont {Obata}\ \emph {et~al.}(2015)\citenamefont {Obata},
  \citenamefont {Nakamura}, \citenamefont {Hamada},\ and\ \citenamefont
  {Oda}}]{Obata_2015:improving_description}%
  \BibitemOpen
  \bibfield  {author} {\bibinfo {author} {\bibfnamefont {M.}~\bibnamefont
  {Obata}}, \bibinfo {author} {\bibfnamefont {M.}~\bibnamefont {Nakamura}},
  \bibinfo {author} {\bibfnamefont {I.}~\bibnamefont {Hamada}}, \ and\ \bibinfo
  {author} {\bibfnamefont {T.}~\bibnamefont {Oda}},\ }\href@noop {} {\
  (\bibinfo {year} {2015})},\ \bibinfo {note}
  {http://arxiv.org/abs/1501.05081}\BibitemShut {NoStop}%
\bibitem [{\citenamefont {Gunnarsson}\ and\ \citenamefont
  {Lundqvist}(1976)}]{Gunnarsson_1976:exchange_correlation}%
  \BibitemOpen
  \bibfield  {author} {\bibinfo {author} {\bibfnamefont {O.}~\bibnamefont
  {Gunnarsson}}\ and\ \bibinfo {author} {\bibfnamefont {B.~I.}\ \bibnamefont
  {Lundqvist}},\ }\href {\doibase 10.1103/PhysRevB.13.4274} {\bibfield
  {journal} {\bibinfo  {journal} {Phys. Rev. B}\ }\textbf {\bibinfo {volume}
  {13}},\ \bibinfo {pages} {4274} (\bibinfo {year} {1976})}\BibitemShut
  {NoStop}%
\bibitem [{\citenamefont {von Barth}\ and\ \citenamefont
  {Hedin}(1972)}]{Barth_1972:local_exchange-correlation}%
  \BibitemOpen
  \bibfield  {author} {\bibinfo {author} {\bibfnamefont {U.}~\bibnamefont {von
  Barth}}\ and\ \bibinfo {author} {\bibfnamefont {L.}~\bibnamefont {Hedin}},\
  }\href {\doibase 10.1088/0022-3719/5/13/012} {\bibfield  {journal} {\bibinfo
  {journal} {J. Physics C: Solid State Phys.}\ }\textbf {\bibinfo {volume}
  {5}},\ \bibinfo {pages} {1629} (\bibinfo {year} {1972})}\BibitemShut
  {NoStop}%
\bibitem [{\citenamefont {Jones}\ and\ \citenamefont
  {Gunnarsson}(1989)}]{Jones_1989:density_functional}%
  \BibitemOpen
  \bibfield  {author} {\bibinfo {author} {\bibfnamefont {R.~O.}\ \bibnamefont
  {Jones}}\ and\ \bibinfo {author} {\bibfnamefont {O.}~\bibnamefont
  {Gunnarsson}},\ }\href {\doibase 10.1103/RevModPhys.61.689} {\bibfield
  {journal} {\bibinfo  {journal} {Rev. Mod. Phys.}\ }\textbf {\bibinfo {volume}
  {61}},\ \bibinfo {pages} {689} (\bibinfo {year} {1989})}\BibitemShut
  {NoStop}%
\bibitem [{gen()}]{general_purpose}%
  \BibitemOpen
  \href@noop {} {}\bibinfo {note} {Note that svdW-DF reduces to vdW-DF in the
  absence of spin polarization.}\BibitemShut {Stop}%
\bibitem [{\citenamefont {Langreth}\ and\ \citenamefont
  {Perdew}(1977)}]{Langreth_1977:exchange-correlation_energy}%
  \BibitemOpen
  \bibfield  {author} {\bibinfo {author} {\bibfnamefont {D.~C.}\ \bibnamefont
  {Langreth}}\ and\ \bibinfo {author} {\bibfnamefont {J.~P.}\ \bibnamefont
  {Perdew}},\ }\href {\doibase 10.1103/PhysRevB.15.2884} {\bibfield  {journal}
  {\bibinfo  {journal} {Phys. Rev. B}\ }\textbf {\bibinfo {volume} {15}},\
  \bibinfo {pages} {2884} (\bibinfo {year} {1977})}\BibitemShut {NoStop}%
\bibitem [{\citenamefont {Rydberg}\ \emph {et~al.}(2003)\citenamefont
  {Rydberg}, \citenamefont {Dion}, \citenamefont {Jacobson}, \citenamefont
  {Schr\"{o}der}, \citenamefont {Hyldgaard}, \citenamefont {Simak},
  \citenamefont {Langreth},\ and\ \citenamefont
  {Lundqvist}}]{Rydberg_2003:van_waals}%
  \BibitemOpen
  \bibfield  {author} {\bibinfo {author} {\bibfnamefont {H.}~\bibnamefont
  {Rydberg}}, \bibinfo {author} {\bibfnamefont {M.}~\bibnamefont {Dion}},
  \bibinfo {author} {\bibfnamefont {N.}~\bibnamefont {Jacobson}}, \bibinfo
  {author} {\bibfnamefont {E.}~\bibnamefont {Schr\"{o}der}}, \bibinfo {author}
  {\bibfnamefont {P.}~\bibnamefont {Hyldgaard}}, \bibinfo {author}
  {\bibfnamefont {S.~I.}\ \bibnamefont {Simak}}, \bibinfo {author}
  {\bibfnamefont {D.~C.}\ \bibnamefont {Langreth}}, \ and\ \bibinfo {author}
  {\bibfnamefont {B.~I.}\ \bibnamefont {Lundqvist}},\ }\href {\doibase 10.1103/PhysRevLett.91.126402} {\bibfield  {journal} {\bibinfo  {journal}
  {Phys. Rev. Lett.}\ }\textbf {\bibinfo {volume} {91}},\ \bibinfo {pages}
  {126402} (\bibinfo {year} {2003})}\BibitemShut {NoStop}%
\bibitem [{\citenamefont {Perdew}\ and\ \citenamefont
  {Wang}(1986)}]{Perdew_1986:accurate_simple}%
  \BibitemOpen
  \bibfield  {author} {\bibinfo {author} {\bibfnamefont {J.~P.}\ \bibnamefont
  {Perdew}}\ and\ \bibinfo {author} {\bibfnamefont {Y.}~\bibnamefont {Wang}},\
  }\href {\doibase 10.1103/PhysRevB.33.8800} {\bibfield  {journal} {\bibinfo
  {journal} {Phys. Rev. B}\ }\textbf {\bibinfo {volume} {33}},\ \bibinfo
  {pages} {8800} (\bibinfo {year} {1986})}\BibitemShut {NoStop}%
\bibitem [{\citenamefont {Mahan}(1965)}]{Mahan_1965:van_waals}%
  \BibitemOpen
  \bibfield  {author} {\bibinfo {author} {\bibfnamefont {G.~D.}\ \bibnamefont
  {Mahan}},\ }\href {\doibase 10.1063/1.1696973} {\bibfield  {journal}
  {\bibinfo  {journal} {J. Chem. Phys.}\ }\textbf {\bibinfo {volume} {43}},\
  \bibinfo {pages} {1569} (\bibinfo {year} {1965})}\BibitemShut {NoStop}%
\bibitem [{\citenamefont {Rapcewicz}\ and\ \citenamefont
  {Ashcroft}(1991)}]{Rapcewicz_1991:fluctuation_attraction}%
  \BibitemOpen
  \bibfield  {author} {\bibinfo {author} {\bibfnamefont {K.}~\bibnamefont
  {Rapcewicz}}\ and\ \bibinfo {author} {\bibfnamefont {N.~W.}\ \bibnamefont
  {Ashcroft}},\ }\href {\doibase 10.1103/PhysRevB.44.4032} {\bibfield
  {journal} {\bibinfo  {journal} {Phys. Rev. B}\ }\textbf {\bibinfo {volume}
  {44}},\ \bibinfo {pages} {4032} (\bibinfo {year} {1991})}\BibitemShut
  {NoStop}%
\bibitem [{\citenamefont {Giannozzi}\ \emph {et~al.}(2009)\citenamefont
  {Giannozzi}, \citenamefont {Baroni}, \citenamefont {Bonini}, \citenamefont
  {Calandra}, \citenamefont {Car}, \citenamefont {Cavazzoni}, \citenamefont
  {Ceresoli}, \citenamefont {Chiarotti}, \citenamefont {Cococcioni},
  \citenamefont {Dabo}, \citenamefont {{Dal Corso}}, \citenamefont
  {de~Gironcoli}, \citenamefont {Fabris}, \citenamefont {Fratesi},
  \citenamefont {Gebauer}, \citenamefont {Gerstmann}, \citenamefont
  {Gougoussis}, \citenamefont {Kokalj}, \citenamefont {Lazzeri}, \citenamefont
  {Martin-Samos}, \citenamefont {Marzari}, \citenamefont {Mauri}, \citenamefont
  {Mazzarello}, \citenamefont {Paolini}, \citenamefont {Pasquarello},
  \citenamefont {Paulatto}, \citenamefont {Sbraccia}, \citenamefont {Scandolo},
  \citenamefont {Sclauzero}, \citenamefont {Seitsonen}, \citenamefont
  {Smogunov}, \citenamefont {Umari},\ and\ \citenamefont
  {Wentzcovitch}}]{Giannozzi_2009:quantum_espresso}%
  \BibitemOpen
  \bibfield  {author} {\bibinfo {author} {\bibfnamefont {P.}~\bibnamefont
  {Giannozzi}}, \bibinfo {author} {\bibfnamefont {S.}~\bibnamefont {Baroni}},
  \bibinfo {author} {\bibfnamefont {N.}~\bibnamefont {Bonini}}, \bibinfo
  {author} {\bibfnamefont {M.}~\bibnamefont {Calandra}}, \bibinfo {author}
  {\bibfnamefont {R.}~\bibnamefont {Car}}, \bibinfo {author} {\bibfnamefont
  {C.}~\bibnamefont {Cavazzoni}}, \bibinfo {author} {\bibfnamefont
  {D.}~\bibnamefont {Ceresoli}}, \bibinfo {author} {\bibfnamefont {G.~L.}\
  \bibnamefont {Chiarotti}}, \bibinfo {author} {\bibfnamefont {M.}~\bibnamefont
  {Cococcioni}}, \bibinfo {author} {\bibfnamefont {I.}~\bibnamefont {Dabo}},
  \bibinfo {author} {\bibfnamefont {A.}~\bibnamefont {{Dal Corso}}}, \bibinfo
  {author} {\bibfnamefont {S.}~\bibnamefont {de~Gironcoli}}, \bibinfo {author}
  {\bibfnamefont {S.}~\bibnamefont {Fabris}}, \bibinfo {author} {\bibfnamefont
  {G.}~\bibnamefont {Fratesi}}, \bibinfo {author} {\bibfnamefont
  {R.}~\bibnamefont {Gebauer}}, \bibinfo {author} {\bibfnamefont
  {U.}~\bibnamefont {Gerstmann}}, \bibinfo {author} {\bibfnamefont
  {C.}~\bibnamefont {Gougoussis}}, \bibinfo {author} {\bibfnamefont
  {A.}~\bibnamefont {Kokalj}}, \bibinfo {author} {\bibfnamefont
  {M.}~\bibnamefont {Lazzeri}}, \bibinfo {author} {\bibfnamefont
  {L.}~\bibnamefont {Martin-Samos}}, \bibinfo {author} {\bibfnamefont
  {N.}~\bibnamefont {Marzari}}, \bibinfo {author} {\bibfnamefont
  {F.}~\bibnamefont {Mauri}}, \bibinfo {author} {\bibfnamefont
  {R.}~\bibnamefont {Mazzarello}}, \bibinfo {author} {\bibfnamefont
  {S.}~\bibnamefont {Paolini}}, \bibinfo {author} {\bibfnamefont
  {A.}~\bibnamefont {Pasquarello}}, \bibinfo {author} {\bibfnamefont
  {L.}~\bibnamefont {Paulatto}}, \bibinfo {author} {\bibfnamefont
  {C.}~\bibnamefont {Sbraccia}}, \bibinfo {author} {\bibfnamefont
  {S.}~\bibnamefont {Scandolo}}, \bibinfo {author} {\bibfnamefont
  {G.}~\bibnamefont {Sclauzero}}, \bibinfo {author} {\bibfnamefont {A.~P.}\
  \bibnamefont {Seitsonen}}, \bibinfo {author} {\bibfnamefont {A.}~\bibnamefont
  {Smogunov}}, \bibinfo {author} {\bibfnamefont {P.}~\bibnamefont {Umari}}, \
  and\ \bibinfo {author} {\bibfnamefont {R.~M.}\ \bibnamefont {Wentzcovitch}},\
  }\href {\doibase 10.1088/0953-8984/21/39/395502} {\bibfield  {journal}
  {\bibinfo  {journal} {J. Phys. Condens. Matter}\ }\textbf {\bibinfo {volume}
  {21}},\ \bibinfo {pages} {395502} (\bibinfo {year} {2009})}\BibitemShut
  {NoStop}%
\bibitem [{\citenamefont {Zhou}\ \emph {et~al.}(2008)\citenamefont {Zhou},
  \citenamefont {Wu},\ and\ \citenamefont {Yildirim}}]{Zhou_2008:enhanced_h2}%
  \BibitemOpen
  \bibfield  {author} {\bibinfo {author} {\bibfnamefont {W.}~\bibnamefont
  {Zhou}}, \bibinfo {author} {\bibfnamefont {H.}~\bibnamefont {Wu}}, \ and\
  \bibinfo {author} {\bibfnamefont {T.}~\bibnamefont {Yildirim}},\ }\href
  {\doibase 10.1021/ja807023q} {\bibfield  {journal} {\bibinfo  {journal} {J.
  Am. Chem. Soc.}\ }\textbf {\bibinfo {volume} {130}},\ \bibinfo {pages}
  {15268} (\bibinfo {year} {2008})}\BibitemShut {NoStop}%
\bibitem [{\citenamefont {M\"{a}rcz}\ \emph {et~al.}(2012)\citenamefont
  {M\"{a}rcz}, \citenamefont {Johnsen}, \citenamefont {Dietzel},\ and\
  \citenamefont {Fjellv\r{a}g}}]{Marcz_2012:iron_member}%
  \BibitemOpen
  \bibfield  {author} {\bibinfo {author} {\bibfnamefont {M.}~\bibnamefont
  {M\"{a}rcz}}, \bibinfo {author} {\bibfnamefont {R.~E.}\ \bibnamefont
  {Johnsen}}, \bibinfo {author} {\bibfnamefont {P.~D.}\ \bibnamefont
  {Dietzel}}, \ and\ \bibinfo {author} {\bibfnamefont {H.}~\bibnamefont
  {Fjellv\r{a}g}},\ }\href {\doibase 10.1016/j.micromeso.2011.12.035}
  {\bibfield  {journal} {\bibinfo  {journal} {Micropor. Mesopor. Mat.}\
  }\textbf {\bibinfo {volume} {157}},\ \bibinfo {pages} {62} (\bibinfo {year}
  {2012})}\BibitemShut {NoStop}%
\bibitem [{\citenamefont {Yu}\ \emph {et~al.}(2013)\citenamefont {Yu},
  \citenamefont {Yazaydin}, \citenamefont {Lane}, \citenamefont {Dietzel},\
  and\ \citenamefont {Snurr}}]{Yu_2013:combined_experimental}%
  \BibitemOpen
  \bibfield  {author} {\bibinfo {author} {\bibfnamefont {D.}~\bibnamefont
  {Yu}}, \bibinfo {author} {\bibfnamefont {A.~O.}\ \bibnamefont {Yazaydin}},
  \bibinfo {author} {\bibfnamefont {J.~R.}\ \bibnamefont {Lane}}, \bibinfo
  {author} {\bibfnamefont {P.~D.~C.}\ \bibnamefont {Dietzel}}, \ and\ \bibinfo
  {author} {\bibfnamefont {R.~Q.}\ \bibnamefont {Snurr}},\ }\href {\doibase 10.1039/C3SC51319J} {\bibfield  {journal} {\bibinfo  {journal} {Chem. Sci.}\
  }\textbf {\bibinfo {volume} {4}},\ \bibinfo {pages} {3544} (\bibinfo {year}
  {2013})}\BibitemShut {NoStop}%
\bibitem [{\citenamefont {Caskey}\ \emph {et~al.}(2008)\citenamefont {Caskey},
  \citenamefont {Wong-Foy},\ and\ \citenamefont
  {Matzger}}]{Caskey_2008:dramatic_tuning}%
  \BibitemOpen
  \bibfield  {author} {\bibinfo {author} {\bibfnamefont {S.~R.}\ \bibnamefont
  {Caskey}}, \bibinfo {author} {\bibfnamefont {A.~G.}\ \bibnamefont
  {Wong-Foy}}, \ and\ \bibinfo {author} {\bibfnamefont {A.~J.}\ \bibnamefont
  {Matzger}},\ }\href {\doibase 10.1021/ja8036096} {\bibfield  {journal}
  {\bibinfo  {journal} {J. Am. Chem. Soc.}\ }\textbf {\bibinfo {volume}
  {130}},\ \bibinfo {pages} {10870} (\bibinfo {year} {2008})}\BibitemShut
  {NoStop}%
\bibitem [{\citenamefont {Dietzel}\ \emph {et~al.}(2009)\citenamefont
  {Dietzel}, \citenamefont {Besikiotis},\ and\ \citenamefont
  {Blom}}]{Dietzel_2009:application_metal-organic}%
  \BibitemOpen
  \bibfield  {author} {\bibinfo {author} {\bibfnamefont {P.~D.~C.}\
  \bibnamefont {Dietzel}}, \bibinfo {author} {\bibfnamefont {V.}~\bibnamefont
  {Besikiotis}}, \ and\ \bibinfo {author} {\bibfnamefont {R.}~\bibnamefont
  {Blom}},\ }\href {\doibase 10.1039/B911242A} {\bibfield  {journal} {\bibinfo
  {journal} {J. Mater. Chem.}\ }\textbf {\bibinfo {volume} {19}},\ \bibinfo
  {pages} {7362} (\bibinfo {year} {2009})}\BibitemShut {NoStop}%
\bibitem [{\citenamefont {Linton}\ \emph {et~al.}(1999)\citenamefont {Linton},
  \citenamefont {Martin}, \citenamefont {Ross}, \citenamefont {Russier},
  \citenamefont {Crozet}, \citenamefont {Yiannopoulou}, \citenamefont {Li},\
  and\ \citenamefont {Lyyra}}]{Linton_1999:high-lying_vibrational}%
  \BibitemOpen
  \bibfield  {author} {\bibinfo {author} {\bibfnamefont {C.}~\bibnamefont
  {Linton}}, \bibinfo {author} {\bibfnamefont {F.}~\bibnamefont {Martin}},
  \bibinfo {author} {\bibfnamefont {A.}~\bibnamefont {Ross}}, \bibinfo {author}
  {\bibfnamefont {I.}~\bibnamefont {Russier}}, \bibinfo {author} {\bibfnamefont
  {P.}~\bibnamefont {Crozet}}, \bibinfo {author} {\bibfnamefont
  {A.}~\bibnamefont {Yiannopoulou}}, \bibinfo {author} {\bibfnamefont
  {L.}~\bibnamefont {Li}}, \ and\ \bibinfo {author} {\bibfnamefont
  {A.}~\bibnamefont {Lyyra}},\ }\href {\doibase 10.1006/jmsp.1999.7858}
  {\bibfield  {journal} {\bibinfo  {journal} {J. Mol. Spectrosc.}\ }\textbf
  {\bibinfo {volume} {196}},\ \bibinfo {pages} {20} (\bibinfo {year}
  {1999})}\BibitemShut {NoStop}%
\bibitem [{\citenamefont {Pople}\ \emph {et~al.}(1989)\citenamefont {Pople},
  \citenamefont {Head-Gordon}, \citenamefont {Fox}, \citenamefont
  {Raghavachari},\ and\ \citenamefont
  {Curtiss}}]{Pople_1989:gaussian-1_theory}%
  \BibitemOpen
  \bibfield  {author} {\bibinfo {author} {\bibfnamefont {J.~A.}\ \bibnamefont
  {Pople}}, \bibinfo {author} {\bibfnamefont {M.}~\bibnamefont {Head-Gordon}},
  \bibinfo {author} {\bibfnamefont {D.~J.}\ \bibnamefont {Fox}}, \bibinfo
  {author} {\bibfnamefont {K.}~\bibnamefont {Raghavachari}}, \ and\ \bibinfo
  {author} {\bibfnamefont {L.~A.}\ \bibnamefont {Curtiss}},\ }\href {\doibase 10.1063/1.456415} {\bibfield  {journal} {\bibinfo  {journal} {J. Chem.
  Phys.}\ }\textbf {\bibinfo {volume} {90}},\ \bibinfo {pages} {5622} (\bibinfo
  {year} {1989})}\BibitemShut {NoStop}%
\bibitem [{\citenamefont {Gamo}\ \emph {et~al.}(1997)\citenamefont {Gamo},
  \citenamefont {Nagashima}, \citenamefont {Wakabayashi}, \citenamefont
  {Terai},\ and\ \citenamefont {Oshima}}]{Gamo_1997:atomic_structure}%
  \BibitemOpen
  \bibfield  {author} {\bibinfo {author} {\bibfnamefont {Y.}~\bibnamefont
  {Gamo}}, \bibinfo {author} {\bibfnamefont {A.}~\bibnamefont {Nagashima}},
  \bibinfo {author} {\bibfnamefont {M.}~\bibnamefont {Wakabayashi}}, \bibinfo
  {author} {\bibfnamefont {M.}~\bibnamefont {Terai}}, \ and\ \bibinfo {author}
  {\bibfnamefont {C.}~\bibnamefont {Oshima}},\ }\href {\doibase 10.1016/S0039-6028(96)00785-6} {\bibfield  {journal} {\bibinfo  {journal}
  {Surf. Sci.}\ }\textbf {\bibinfo {volume} {374}},\ \bibinfo {pages} {61}
  (\bibinfo {year} {1997})}\BibitemShut {NoStop}%
\bibitem [{\citenamefont {Varykhalov}\ \emph {et~al.}(2008)\citenamefont
  {Varykhalov}, \citenamefont {S{\'a}nchez-Barriga}, \citenamefont {Shikin},
  \citenamefont {Biswas}, \citenamefont {Vescovo}, \citenamefont {Rybkin},
  \citenamefont {Marchenko},\ and\ \citenamefont
  {Rader}}]{Varykhalov_2008:electronic_magnetic}%
  \BibitemOpen
  \bibfield  {author} {\bibinfo {author} {\bibfnamefont {A.}~\bibnamefont
  {Varykhalov}}, \bibinfo {author} {\bibfnamefont {J.}~\bibnamefont
  {S{\'a}nchez-Barriga}}, \bibinfo {author} {\bibfnamefont {A.~M.}\
  \bibnamefont {Shikin}}, \bibinfo {author} {\bibfnamefont {C.}~\bibnamefont
  {Biswas}}, \bibinfo {author} {\bibfnamefont {E.}~\bibnamefont {Vescovo}},
  \bibinfo {author} {\bibfnamefont {A.}~\bibnamefont {Rybkin}}, \bibinfo
  {author} {\bibfnamefont {D.}~\bibnamefont {Marchenko}}, \ and\ \bibinfo
  {author} {\bibfnamefont {O.}~\bibnamefont {Rader}},\ }\href {\doibase 10.1103/PhysRevLett.101.157601} {\bibfield  {journal} {\bibinfo  {journal}
  {Phys. Rev. Lett.}\ }\textbf {\bibinfo {volume} {101}},\ \bibinfo {pages}
  {157601} (\bibinfo {year} {2008})}\BibitemShut {NoStop}%
\bibitem [{\citenamefont {Dedkov}\ and\ \citenamefont
  {Fonin}(2010)}]{Dedkov_2010:electronic_magnetic}%
  \BibitemOpen
  \bibfield  {author} {\bibinfo {author} {\bibfnamefont {Y.~S.}\ \bibnamefont
  {Dedkov}}\ and\ \bibinfo {author} {\bibfnamefont {M.}~\bibnamefont {Fonin}},\
  }\href {\doibase 10.1088/1367-2630/12/12/125004} {\bibfield  {journal}
  {\bibinfo  {journal} {New J. Phys.}\ }\textbf {\bibinfo {volume} {12}},\
  \bibinfo {pages} {125004} (\bibinfo {year} {2010})}\BibitemShut {NoStop}%
\bibitem [{\citenamefont {Canepa}\ \emph
  {et~al.}(2013{\natexlab{c}})\citenamefont {Canepa}, \citenamefont {Chabal},\
  and\ \citenamefont {Thonhauser}}]{Canepa_2013:when_metal}%
  \BibitemOpen
  \bibfield  {author} {\bibinfo {author} {\bibfnamefont {P.}~\bibnamefont
  {Canepa}}, \bibinfo {author} {\bibfnamefont {Y.~J.}\ \bibnamefont {Chabal}},
  \ and\ \bibinfo {author} {\bibfnamefont {T.}~\bibnamefont {Thonhauser}},\
  }\href {\doibase 10.1103/PhysRevB.87.094407} {\bibfield  {journal} {\bibinfo
  {journal} {Phys. Rev. B}\ }\textbf {\bibinfo {volume} {87}},\ \bibinfo
  {pages} {094407} (\bibinfo {year} {2013}{\natexlab{c}})}\BibitemShut
  {NoStop}%
\bibitem [{\citenamefont {Wang}\ \emph {et~al.}(2006)\citenamefont {Wang},
  \citenamefont {Maxisch},\ and\ \citenamefont
  {Ceder}}]{Wang_2006:oxidation_energies}%
  \BibitemOpen
  \bibfield  {author} {\bibinfo {author} {\bibfnamefont {L.}~\bibnamefont
  {Wang}}, \bibinfo {author} {\bibfnamefont {T.}~\bibnamefont {Maxisch}}, \
  and\ \bibinfo {author} {\bibfnamefont {G.}~\bibnamefont {Ceder}},\ }\href
  {\doibase 10.1103/PhysRevB.73.195107} {\bibfield  {journal} {\bibinfo
  {journal} {Phys. Rev. B}\ }\textbf {\bibinfo {volume} {73}},\ \bibinfo
  {pages} {195107} (\bibinfo {year} {2006})}\BibitemShut {NoStop}%
\bibitem [{\citenamefont {Bloch}\ \emph {et~al.}(2011)\citenamefont {Bloch},
  \citenamefont {Murray}, \citenamefont {Queen}, \citenamefont {Chavan},
  \citenamefont {Maximoff}, \citenamefont {Bigi}, \citenamefont {Krishna},
  \citenamefont {Peterson}, \citenamefont {Grandjean}, \citenamefont {Long},
  \citenamefont {Smit}, \citenamefont {Bordiga}, \citenamefont {Brown},\ and\
  \citenamefont {Long}}]{Bloch_2011:Selective_Binding}%
  \BibitemOpen
  \bibfield  {author} {\bibinfo {author} {\bibfnamefont {E.~D.}\ \bibnamefont
  {Bloch}}, \bibinfo {author} {\bibfnamefont {L.~J.}\ \bibnamefont {Murray}},
  \bibinfo {author} {\bibfnamefont {W.~L.}\ \bibnamefont {Queen}}, \bibinfo
  {author} {\bibfnamefont {S.}~\bibnamefont {Chavan}}, \bibinfo {author}
  {\bibfnamefont {S.~N.}\ \bibnamefont {Maximoff}}, \bibinfo {author}
  {\bibfnamefont {J.~P.}\ \bibnamefont {Bigi}}, \bibinfo {author}
  {\bibfnamefont {R.}~\bibnamefont {Krishna}}, \bibinfo {author} {\bibfnamefont
  {V.~K.}\ \bibnamefont {Peterson}}, \bibinfo {author} {\bibfnamefont
  {F.}~\bibnamefont {Grandjean}}, \bibinfo {author} {\bibfnamefont {G.~J.}\
  \bibnamefont {Long}}, \bibinfo {author} {\bibfnamefont {B.}~\bibnamefont
  {Smit}}, \bibinfo {author} {\bibfnamefont {S.}~\bibnamefont {Bordiga}},
  \bibinfo {author} {\bibfnamefont {C.~M.}\ \bibnamefont {Brown}}, \ and\
  \bibinfo {author} {\bibfnamefont {J.~R.}\ \bibnamefont {Long}},\ }\href
  {\doibase 10.1021/ja205976v} {\bibfield  {journal} {\bibinfo  {journal} {J.
  Am. Chem. Soc.}\ }\textbf {\bibinfo {volume} {133}},\ \bibinfo {pages}
  {14814} (\bibinfo {year} {2011})}\BibitemShut {NoStop}%
\bibitem [{\citenamefont {Bader}(1990)}]{Bader_1990:atoms_molecules}%
  \BibitemOpen
  \bibfield  {author} {\bibinfo {author} {\bibfnamefont {R.~F.~W.}\
  \bibnamefont {Bader}},\ }\href@noop {} {\emph {\bibinfo {title} {Atoms in
  Molecules: A Quantum Theory}}}\ (\bibinfo  {publisher} {Oxford University
  Press},\ \bibinfo {address} {New York},\ \bibinfo {year} {1990})\BibitemShut
  {NoStop}%
\bibitem [{\citenamefont {Henkelman}\ \emph {et~al.}(2006)\citenamefont
  {Henkelman}, \citenamefont {Arnaldsson},\ and\ \citenamefont
  {J{\'o}nsson}}]{Henkelman_2006:fast_robust}%
  \BibitemOpen
  \bibfield  {author} {\bibinfo {author} {\bibfnamefont {G.}~\bibnamefont
  {Henkelman}}, \bibinfo {author} {\bibfnamefont {A.}~\bibnamefont
  {Arnaldsson}}, \ and\ \bibinfo {author} {\bibfnamefont {H.}~\bibnamefont
  {J{\'o}nsson}},\ }\href {\doibase 10.1016/j.commatsci.2005.04.010} {\bibfield
   {journal} {\bibinfo  {journal} {Comp. Mater. Science}\ }\textbf {\bibinfo
  {volume} {36}},\ \bibinfo {pages} {354} (\bibinfo {year} {2006})}\BibitemShut
  {NoStop}%
\bibitem [{\citenamefont {Sabatini}\ \emph {et~al.}(2013)\citenamefont
  {Sabatini}, \citenamefont {Gorni},\ and\ \citenamefont
  {de~Gironcoli}}]{Sabatini_2013:nonlocal_van}%
  \BibitemOpen
  \bibfield  {author} {\bibinfo {author} {\bibfnamefont {R.}~\bibnamefont
  {Sabatini}}, \bibinfo {author} {\bibfnamefont {T.}~\bibnamefont {Gorni}}, \
  and\ \bibinfo {author} {\bibfnamefont {S.}~\bibnamefont {de~Gironcoli}},\
  }\href {\doibase 10.1103/PhysRevB.87.041108} {\bibfield  {journal} {\bibinfo
  {journal} {Phys. Rev. B}\ }\textbf {\bibinfo {volume} {87}},\ \bibinfo
  {pages} {041108(R)} (\bibinfo {year} {2013})}\BibitemShut {NoStop}%
\bibitem [{\citenamefont {Garrity}\ \emph {et~al.}(2014)\citenamefont
  {Garrity}, \citenamefont {Bennett}, \citenamefont {Rabe},\ and\ \citenamefont
  {Vanderbilt}}]{Garrity_2014:pseudopotentials_high-throughput}%
  \BibitemOpen
  \bibfield  {author} {\bibinfo {author} {\bibfnamefont {K.~F.}\ \bibnamefont
  {Garrity}}, \bibinfo {author} {\bibfnamefont {J.~W.}\ \bibnamefont
  {Bennett}}, \bibinfo {author} {\bibfnamefont {K.~M.}\ \bibnamefont {Rabe}}, \
  and\ \bibinfo {author} {\bibfnamefont {D.}~\bibnamefont {Vanderbilt}},\
  }\href {\doibase 10.1016/j.commatsci.2013.08.053} {\bibfield  {journal}
  {\bibinfo  {journal} {Comp. Mat. Science}\ }\textbf {\bibinfo {volume}
  {81}},\ \bibinfo {pages} {446} (\bibinfo {year} {2014})}\BibitemShut
  {NoStop}%
\bibitem [{dse()}]{dsepDFTdisc}%
  \BibitemOpen
  \href@noop {} {}\bibinfo {note} {Caution must be used when using DFT to
  determine the Li dimer triplet binding. We find that all tested local,
  semilocal, and nonlocal spin-density functionals underestimate the
  experimental binding separation of 4.2~\AA\
  \cite{Linton_1999:high-lying_vibrational} by 0.5~--~0.7~\AA, with svdW-DF-cx
  and LDA giving the smallest and largest deviation, respectively.}\BibitemShut
  {Stop}%
\bibitem [{\citenamefont {Konowalow}\ and\ \citenamefont
  {Fish}(1984)}]{Konowalow_1984:molecular_electronic}%
  \BibitemOpen
  \bibfield  {author} {\bibinfo {author} {\bibfnamefont {D.~D.}\ \bibnamefont
  {Konowalow}}\ and\ \bibinfo {author} {\bibfnamefont {J.~L.}\ \bibnamefont
  {Fish}},\ }\href {\doibase 10.1016/0301-0104(84)85195-2} {\bibfield
  {journal} {\bibinfo  {journal} {Chem. Phys.}\ }\textbf {\bibinfo {volume}
  {84}},\ \bibinfo {pages} {463} (\bibinfo {year} {1984})}\BibitemShut
  {NoStop}%
\bibitem [{\citenamefont {Poteau}\ and\ \citenamefont
  {Spiegelmann}(1995)}]{Poteau_1995:calculation_electronic}%
  \BibitemOpen
  \bibfield  {author} {\bibinfo {author} {\bibfnamefont {R.}~\bibnamefont
  {Poteau}}\ and\ \bibinfo {author} {\bibfnamefont {F.}~\bibnamefont
  {Spiegelmann}},\ }\href {\doibase 10.1006/jmsp.1995.1120} {\bibfield
  {journal} {\bibinfo  {journal} {J. Mol. Spect.}\ }\textbf {\bibinfo {volume}
  {171}},\ \bibinfo {pages} {299} (\bibinfo {year} {1995})}\BibitemShut
  {NoStop}%
\bibitem [{\citenamefont {Jasik}\ and\ \citenamefont
  {Sienkiewicz}(2006)}]{Jasik_2006:calculation_adiabatic}%
  \BibitemOpen
  \bibfield  {author} {\bibinfo {author} {\bibfnamefont {P.}~\bibnamefont
  {Jasik}}\ and\ \bibinfo {author} {\bibfnamefont {J.~E.}\ \bibnamefont
  {Sienkiewicz}},\ }\href {\doibase 10.1016/j.chemphys.2005.10.025} {\bibfield
  {journal} {\bibinfo  {journal} {Chem. Phys.}\ }\textbf {\bibinfo {volume}
  {323}},\ \bibinfo {pages} {563} (\bibinfo {year} {2006})}\BibitemShut
  {NoStop}%
\bibitem [{\citenamefont {Grossman}(2002)}]{Grossman_2002:benchmark_quantum}%
  \BibitemOpen
  \bibfield  {author} {\bibinfo {author} {\bibfnamefont {J.~C.}\ \bibnamefont
  {Grossman}},\ }\href {\doibase 10.1063/1.1487829} {\bibfield  {journal}
  {\bibinfo  {journal} {J. Chem. Phys.}\ }\textbf {\bibinfo {volume} {117}},\
  \bibinfo {pages} {1434} (\bibinfo {year} {2002})}\BibitemShut {NoStop}%
\bibitem [{\citenamefont {Chase}(1998)}]{Chase_1998:nist-janaf_thermochemical}%
  \BibitemOpen
  \bibfield  {author} {\bibinfo {author} {\bibfnamefont {M.~W.}\ \bibnamefont
  {Chase}},\ }\href@noop {} {\bibfield  {journal} {\bibinfo  {journal}
  {NIST-JANAF Thermochemical Tables, 4th ed., J. Phys. Chem. Ref. Data
  Monogr.}\ }\textbf {\bibinfo {volume} {9}},\ \bibinfo {pages} {Suppl.~1}
  (\bibinfo {year} {1998})}\BibitemShut {NoStop}%
\bibitem [{\citenamefont {Shao}\ \emph {et~al.}(2006)\citenamefont {Shao},
  \citenamefont {Molnar}, \citenamefont {Jung}, \citenamefont {Kussmann},
  \citenamefont {Ochsenfeld}, \citenamefont {Brown}, \citenamefont {Gilbert},
  \citenamefont {Slipchenko}, \citenamefont {Levchenko}, \citenamefont
  {O{'}Neill}, \citenamefont {DiStasio~Jr}, \citenamefont {Lochan},
  \citenamefont {Wang}, \citenamefont {Beran}, \citenamefont {Besley},
  \citenamefont {Herbert}, \citenamefont {Yeh~Lin}, \citenamefont
  {Van~Voorhis}, \citenamefont {Hung~Chien}, \citenamefont {Sodt},
  \citenamefont {Steele}, \citenamefont {Rassolov}, \citenamefont {Maslen},
  \citenamefont {Korambath}, \citenamefont {Adamson}, \citenamefont {Austin},
  \citenamefont {Baker}, \citenamefont {Byrd}, \citenamefont {Dachsel},
  \citenamefont {Doerksen}, \citenamefont {Dreuw}, \citenamefont {Dunietz},
  \citenamefont {Dutoi}, \citenamefont {Furlani}, \citenamefont {Gwaltney},
  \citenamefont {Heyden}, \citenamefont {Hirata}, \citenamefont {Hsu},
  \citenamefont {Kedziora}, \citenamefont {Khalliulin}, \citenamefont
  {Klunzinger}, \citenamefont {Lee}, \citenamefont {Lee}, \citenamefont
  {Liang}, \citenamefont {Lotan}, \citenamefont {Nair}, \citenamefont {Peters},
  \citenamefont {Proynov}, \citenamefont {Pieniazek}, \citenamefont {Min~Rhee},
  \citenamefont {Ritchie}, \citenamefont {Rosta}, \citenamefont
  {David~Sherrill}, \citenamefont {Simmonett}, \citenamefont {Subotnik},
  \citenamefont {Lee Woodcock~III}, \citenamefont {Zhang}, \citenamefont
  {Bell}, \citenamefont {Chakraborty}, \citenamefont {Chipman}, \citenamefont
  {Keil}, \citenamefont {Warshel}, \citenamefont {Hehre}, \citenamefont
  {Schaefer~III}, \citenamefont {Kong}, \citenamefont {Krylov}, \citenamefont
  {Gill},\ and\ \citenamefont {Head-Gordon}}]{Shao_2006:advances_methods}%
  \BibitemOpen
  \bibfield  {author} {\bibinfo {author} {\bibfnamefont {Y.}~\bibnamefont
  {Shao}}, \bibinfo {author} {\bibfnamefont {L.~F.}\ \bibnamefont {Molnar}},
  \bibinfo {author} {\bibfnamefont {Y.}~\bibnamefont {Jung}}, \bibinfo {author}
  {\bibfnamefont {J.}~\bibnamefont {Kussmann}}, \bibinfo {author}
  {\bibfnamefont {C.}~\bibnamefont {Ochsenfeld}}, \bibinfo {author}
  {\bibfnamefont {S.~T.}\ \bibnamefont {Brown}}, \bibinfo {author}
  {\bibfnamefont {A.~T.}\ \bibnamefont {Gilbert}}, \bibinfo {author}
  {\bibfnamefont {L.~V.}\ \bibnamefont {Slipchenko}}, \bibinfo {author}
  {\bibfnamefont {S.~V.}\ \bibnamefont {Levchenko}}, \bibinfo {author}
  {\bibfnamefont {D.~P.}\ \bibnamefont {O{'}Neill}}, \bibinfo {author}
  {\bibfnamefont {R.~A.}\ \bibnamefont {DiStasio~Jr}}, \bibinfo {author}
  {\bibfnamefont {R.~C.}\ \bibnamefont {Lochan}}, \bibinfo {author}
  {\bibfnamefont {T.}~\bibnamefont {Wang}}, \bibinfo {author} {\bibfnamefont
  {G.~J.}\ \bibnamefont {Beran}}, \bibinfo {author} {\bibfnamefont {N.~A.}\
  \bibnamefont {Besley}}, \bibinfo {author} {\bibfnamefont {J.~M.}\
  \bibnamefont {Herbert}}, \bibinfo {author} {\bibfnamefont {C.}~\bibnamefont
  {Yeh~Lin}}, \bibinfo {author} {\bibfnamefont {T.}~\bibnamefont
  {Van~Voorhis}}, \bibinfo {author} {\bibfnamefont {S.}~\bibnamefont
  {Hung~Chien}}, \bibinfo {author} {\bibfnamefont {A.}~\bibnamefont {Sodt}},
  \bibinfo {author} {\bibfnamefont {R.~P.}\ \bibnamefont {Steele}}, \bibinfo
  {author} {\bibfnamefont {V.~A.}\ \bibnamefont {Rassolov}}, \bibinfo {author}
  {\bibfnamefont {P.~E.}\ \bibnamefont {Maslen}}, \bibinfo {author}
  {\bibfnamefont {P.~P.}\ \bibnamefont {Korambath}}, \bibinfo {author}
  {\bibfnamefont {R.~D.}\ \bibnamefont {Adamson}}, \bibinfo {author}
  {\bibfnamefont {B.}~\bibnamefont {Austin}}, \bibinfo {author} {\bibfnamefont
  {J.}~\bibnamefont {Baker}}, \bibinfo {author} {\bibfnamefont {E.~F.~C.}\
  \bibnamefont {Byrd}}, \bibinfo {author} {\bibfnamefont {H.}~\bibnamefont
  {Dachsel}}, \bibinfo {author} {\bibfnamefont {R.~J.}\ \bibnamefont
  {Doerksen}}, \bibinfo {author} {\bibfnamefont {A.}~\bibnamefont {Dreuw}},
  \bibinfo {author} {\bibfnamefont {B.~D.}\ \bibnamefont {Dunietz}}, \bibinfo
  {author} {\bibfnamefont {A.~D.}\ \bibnamefont {Dutoi}}, \bibinfo {author}
  {\bibfnamefont {T.~R.}\ \bibnamefont {Furlani}}, \bibinfo {author}
  {\bibfnamefont {S.~R.}\ \bibnamefont {Gwaltney}}, \bibinfo {author}
  {\bibfnamefont {A.}~\bibnamefont {Heyden}}, \bibinfo {author} {\bibfnamefont
  {S.}~\bibnamefont {Hirata}}, \bibinfo {author} {\bibfnamefont {C.-P.}\
  \bibnamefont {Hsu}}, \bibinfo {author} {\bibfnamefont {G.}~\bibnamefont
  {Kedziora}}, \bibinfo {author} {\bibfnamefont {R.~Z.}\ \bibnamefont
  {Khalliulin}}, \bibinfo {author} {\bibfnamefont {P.}~\bibnamefont
  {Klunzinger}}, \bibinfo {author} {\bibfnamefont {A.~M.}\ \bibnamefont {Lee}},
  \bibinfo {author} {\bibfnamefont {M.~S.}\ \bibnamefont {Lee}}, \bibinfo
  {author} {\bibfnamefont {W.}~\bibnamefont {Liang}}, \bibinfo {author}
  {\bibfnamefont {I.}~\bibnamefont {Lotan}}, \bibinfo {author} {\bibfnamefont
  {N.}~\bibnamefont {Nair}}, \bibinfo {author} {\bibfnamefont {B.}~\bibnamefont
  {Peters}}, \bibinfo {author} {\bibfnamefont {E.~I.}\ \bibnamefont {Proynov}},
  \bibinfo {author} {\bibfnamefont {P.~A.}\ \bibnamefont {Pieniazek}}, \bibinfo
  {author} {\bibfnamefont {Y.}~\bibnamefont {Min~Rhee}}, \bibinfo {author}
  {\bibfnamefont {J.}~\bibnamefont {Ritchie}}, \bibinfo {author} {\bibfnamefont
  {E.}~\bibnamefont {Rosta}}, \bibinfo {author} {\bibfnamefont
  {C.}~\bibnamefont {David~Sherrill}}, \bibinfo {author} {\bibfnamefont
  {A.~C.}\ \bibnamefont {Simmonett}}, \bibinfo {author} {\bibfnamefont {J.~E.}\
  \bibnamefont {Subotnik}}, \bibinfo {author} {\bibfnamefont {H.}~\bibnamefont
  {Lee Woodcock~III}}, \bibinfo {author} {\bibfnamefont {W.}~\bibnamefont
  {Zhang}}, \bibinfo {author} {\bibfnamefont {A.~T.}\ \bibnamefont {Bell}},
  \bibinfo {author} {\bibfnamefont {A.~K.}\ \bibnamefont {Chakraborty}},
  \bibinfo {author} {\bibfnamefont {D.~M.}\ \bibnamefont {Chipman}}, \bibinfo
  {author} {\bibfnamefont {F.~J.}\ \bibnamefont {Keil}}, \bibinfo {author}
  {\bibfnamefont {A.}~\bibnamefont {Warshel}}, \bibinfo {author} {\bibfnamefont
  {W.~J.}\ \bibnamefont {Hehre}}, \bibinfo {author} {\bibfnamefont {H.~F.}\
  \bibnamefont {Schaefer~III}}, \bibinfo {author} {\bibfnamefont
  {J.}~\bibnamefont {Kong}}, \bibinfo {author} {\bibfnamefont {A.~I.}\
  \bibnamefont {Krylov}}, \bibinfo {author} {\bibfnamefont {P.~M.~W.}\
  \bibnamefont {Gill}}, \ and\ \bibinfo {author} {\bibfnamefont
  {M.}~\bibnamefont {Head-Gordon}},\ }\href {\doibase 10.1039/B517914A}
  {\bibfield  {journal} {\bibinfo  {journal} {Phys. Chem. Chem. Phys.}\
  }\textbf {\bibinfo {volume} {8}},\ \bibinfo {pages} {3172} (\bibinfo {year}
  {2006})}\BibitemShut {NoStop}%
\end{thebibliography}
\end{document}